\documentstyle[preprint,aps,harvard,floats]{revtex}

\citationstyle{dcu}

\newcommand{\nc}{\newcommand}

\nc{\cit}{\citeasnoun}

\nc{\beq}{\begin{equation}} \nc{\eeq}{\end{equation}}
\nc{\beqa}{\begin{eqnarray}} \nc{\eeqa}{\end{eqnarray}}
\nc{\lsim}{\begin{array}{c}\,\sim\vspace{-21pt}\\< \end{array}}
\nc{\gsim}{\begin{array}{c}\sim\vspace{-21pt}\\> \end{array}}

\def\kahler{K\"ahler\ }
\def\dsb{dynamical supersymmetry breaking\ }
\def\susy{supersymmetry\ }

\nc{\ra}{\rightarrow}

\nc{\al}{\alpha}
\nc{\ald}{{\dot{\alpha}}}
\nc{\be}{\beta}
\nc{\bed}{{\dot{\beta}}}
\nc{\lam}{\lambda}
\nc{\Lam}{\Lambda}
\nc{\Lamp}{\Lambda_{N-2}}
\nc{\Qbar}{\overline Q}
\nc{\qbar}{\overline q}
\nc{\Fbar}{\overline F}
\nc{\Bbar}{\overline B}
\nc{\bbar}{\overline b}
\nc{\nud}{{\dot{\nu}}}
\nc{\lamd}{{\dot{\lam}}}
\nc{\sig}{\sigma}
\nc{\Pf}{{\rm Pf}}
\nc{\Tr}{{\rm Tr}}
\nc{\abs}[1]{\left | #1 \right |}
\nc{\vev}[1]{\langle #1 \rangle}
\nc{\cntr}[1]{\begin{center} #1 \end{center}}

\def\etal{{\it et al}}
\def\eg{{\it e.g.\ }}
\def\sun{SU(N_c)}

\newcommand{\drawsquare}[2]{\hbox{%
\rule{#2pt}{#1pt}\hskip-#2pt
\rule{#1pt}{#2pt}\hskip-#1pt
\rule[#1pt]{#1pt}{#2pt}}\rule[#1pt]{#2pt}{#2pt}\hskip-#2pt
\rule{#2pt}{#1pt}}

\newcommand{\Yfund}{\raisebox{-.5pt}{\drawsquare{6.5}{0.4}}}
\newcommand{\Ybfund}{\bar{\Yfund}}
\newcommand{\Ysymm}{\raisebox{-.5pt}{\drawsquare{6.5}{0.4}}\hskip-0.4pt%
        \raisebox{-.5pt}{\drawsquare{6.5}{0.4}}}
\newcommand{\Yasymm}{\raisebox{-3.5pt}{\drawsquare{6.5}{0.4}}\hskip-6.9pt%
        \raisebox{3pt}{\drawsquare{6.5}{0.4}}}

\begin{document}
\preprint{hep-th/9907225, PUPT-1876, WIS-99/26/07-DPP}
\title{Dynamical Supersymmetry Breaking}
\author{Yael Shadmi}
\address{Department of Particle Physics\\
         Weizmann Institute of Science, Rehovot 76100,
	 Israel\\
         and\\
         Physics Department\\ 
         Princeton University, Princeton, NJ 08544, USA}
\author{Yuri Shirman}
\address{Physics Department\\
         Princeton University, Princeton, NJ 08544, USA\\}

\maketitle

\begin{abstract}
\noindent
Supersymmetry is one of the most plausible and theoretically motivated 
frameworks for extending
the Standard Model. However, any supersymmetry in Nature 
must be a broken symmetry.
Dynamical supersymmetry breaking (DSB) is an attractive idea
for incorporating supersymmetry into a successful description of Nature. 
The study of DSB has recently enjoyed dramatic progress,
fueled by advances in our understanding of the dynamics of supersymmetric
field theories.
These advances have allowed for direct analysis of DSB in strongly coupled
theories, and for the discovery of new DSB theories,
some of which contradict early criteria for DSB.
We review these criteria, emphasizing recently discovered exceptions. 
We also describe, through many examples, various techniques
for directly establishing DSB by studying the infrared theory,
including both older techniques in regions of weak coupling, 
and new techniques in regions of strong coupling.
Finally, we present a list of representative DSB models,
their main properties, and the relations between them.
\end{abstract}

\bigskip
\centerline {{\em Submitted to Reviews of Modern Physics}}

\newpage

\tableofcontents

\newpage

\section{Introduction}   \label{introduction}

Supersymmetry (SUSY), which rotates bosons
into fermions and vice versa,
is a beautiful theoretical idea.
But nature is certainly not supersymmetric. If it were,
we would see a fermionic partner for each known gauge
boson, and a scalar partner for each known fermion,
with degenerate masses. But experimentalists have been
looking for ``superpartners'' long and hard, and so
far in vain, pushing the limits on superpartner masses
to roughly above a 100~GeV.
Thus, any discussion of supersymmetry in nature is
necessarily a discussion of {\it broken} supersymmetry. 

Still, even broken supersymmetry is theoretically more
appealing than no supersymmetry at all.
First, supersymmetry   
provides a solution to the gauge hierarchy problem.
Without supersymmetry, the scalar Higgs mass is quadratically
divergent, so that the natural scale
for it is the fundamental scale of the theory, e.g.,
the Planck scale, many orders of magnitude above
the electroweak scale.
In a supersymmetric theory, the mass of the scalar Higgs
is tied to the mass of its fermionic superpartner.
Since fermion masses are protected by chiral symmetries,
the Higgs mass can naturally be around the electroweak
scale, 
and radiative corrections do not destabilize this hierarchy.
This success is not spoiled even when explicit supersymmetry-breaking
terms are added to the Lagrangian of the theory,
as long as these terms are ``soft'', that is, they only introduce
logarithmic divergences, but no quadratic divergences, into
scalar masses.
The appearance of explicit supersymmetry-breaking terms
in the low-energy effective theory can be theoretically 
justified if the underlying theory is supersymmetric,
yet the vacuum state breaks supersymmetry spontaneously.

While {\it spontaneously} broken supersymmetry would explain
the stability of the gauge hierarchy against radiative
corrections, it still does not
explain the origin of the hierarchy,
that is, the origin of the small 
mass ratios
in the theory.
Indeed if supersymmetry
were broken at 
the classical level (tree level), 
the scale of the soft terms
would be determined by explicit mass parameters in the
supersymmetric Lagrangian, and one would still have to
understand why such parameters are so much smaller than the
Planck scale.
However,
the origin of the hierarchy can be
understood if supersymmetry is broken {\it dynamically}~\cite{wittendsb}.
By ``dynamical supersymmetry breaking'' (DSB) we mean that
supersymmetry is broken spontaneously in a theory
that possesses supersymmetric vacua at the tree level,
with the breaking triggered by dynamical effects.
The crucial point about DSB
is that if supersymmetry is unbroken at tree-level,
supersymmetric non-renormalization 
theorems~\cite{wznr,fiz,grisigroc}
imply that
it remains unbroken to all orders in perturbation theory,
and can 
therefore
only be broken by non-perturbative effects, which
are suppressed by roughly 
$e^{-8\pi^2/g^2}$, where $g$ is
the coupling.
The electroweak scale is related
to the size of the soft supersymmetry breaking terms, 
and thus it is proportional 
to the supersymmetry breaking scale. The latter is suppressed
by the exponential above, and can easily be of the correct
size, about 17 orders of magnitude below the Planck scale.

In addition, supersymmetry, or more precisely, local supersymmetry, 
provides the only known framework 
for a consistent description of gravity,
in the context of string theory.
If indeed the underlying fundamental physics is described by
string theory, 
one can contemplate two qualitatively different scenarios.
One is that SUSY is directly broken by stringy effects.
Then, however, the SUSY breaking scale is generically
around the string scale (barring new and better understanding of string
vacua), and thus the gauge hierarchy problem is not 
solved by supersymmetry.
Therefore, we shall focus here on a second possible scenario,
namely, that in the low-energy limit, string theory
gives rise to an effective field theory, and supersymmetry
is spontaneously broken by the dynamics of this low-energy effective theory.

The aim of this review is then
to describe the phenomenon of dynamical
supersymmetry breaking in field theories 
with ${\cal N}=1$ global supersymmetry.
(${\cal N}$ counts the number of supersymmetries.
For  ${\cal N}=1$
there are 
four
supersymmetry charges and this
is the smallest amount of supersymmetry allowed in 
four
dimensions.)

The restriction on ${\cal N}$ comes from the fact that
only ${\cal N}=1$ supersymmetry has chiral matter,
which we need in the low-energy theory if it is
to contain the standard model.
Moreover, theories with ${\cal N} >1$ supersymmetry are believed
to have an exact moduli space and thus are not expected to exhibit
dynamical supersymmetry breaking.

The restriction to global supersymmetry still
allows us to answer most of the
questions we would be interested in.
This situation is quite analogous to studying the 
breaking of a gauged bosonic symmetry
in say, a theory with scalar matter.
In that case one can determine the pattern of symmetry breaking
just by studying the scalar potential.
Similarly, we shall be able to determine
whether supersymmetry is broken, and, if the theory is weakly coupled,
what the vacuum energy and the light spectrum are. 
From our perspective, the most relevant
consequence of ``gauging'' supersymmetry is the analogue
of the Higgs mechanism by which the massless fermion accompanying
supersymmetry breaking, the Goldstino, is eaten by the gravitino.

As we shall see, supersymmetry is broken if and only if
the vacuum energy is non-zero.
Furthermore, as we mentioned above,
if supersymmetry is unbroken
at tree level, it can only be broken by non-perturbative 
effects.
Thus studying supersymmetry breaking
requires understanding the non-perturbative dynamics
of gauge theories
in the infrared.
Fortunately, in recent years, there has been tremendous progress
in understanding the dynamics of supersymmetric field
theories.

The potential of a supersymmetric theory is determined by two quantities,
the \kahler potential, which contains the kinetic terms for
the matter fields, and the superpotential, a holomorphic function of
the matter fields which controls their Yukawa interactions.
Holomorphy, together with the symmetries of the theory,
may be used 
to determine the physical degrees of freedom and the 
superpotential of the infrared theory~\cite{seibergexact,seibergduality}. 
Since the latter two
are precisely the ingredients
needed for studying supersymmetry breaking,
this progress has fueled the discovery of many new
supersymmetry breaking
theories, as well as new techniques for establishing supersymmetry
breaking.

The structure of this article is as follows.
We start by 
describing  general properties of supersymmetry breaking and 
studying examples of tree level breaking in 
section~\ref{generalities}.
In section~~\ref{indirect},
we 
discuss  indirect methods for finding
theories with dynamical supersymmetry breaking,
and for establishing supersymmetry
breaking.
As we shall see, these methods direct
the search for supersymmetry breaking towards chiral
theories with no flat directions, preferably possessing
an anomaly free R-symmetry. 
These criteria do not amount to necessary conditions for
supersymmetry breaking, and we shall point out ``loopholes''
in the indirect methods which allow the possibility of
supersymmetry breaking in
theories which violate all of the above requirements.
Recent developments
have led to the discovery of such SUSY breaking theories
and we shall 
postpone the discussion of
representative examples to 
section~\ref{exceptions}.
Still, some of the indirect methods we shall describe,
most notably, the breaking of a global symmetry
in a theory with no flat directions, provide
the most convincing evidence for supersymmetry breaking
in theories that cannot be directly analyzed.

In sections~\ref{calculable} and~\ref{strong}
we turn to theories that can be directly
analyzed in the infrared. In the early 80's,
such studies were 
limited to a semi-classical analysis
in regions of weak coupling, and we shall describe
such analyses in section~\ref{calculable}. 
The main development in recent years has been
the better understanding of supersymmetric theories
in regions of strong coupling,
and we shall study 
supersymmetry breaking in
such theories in section~\ref{strong}.
In both cases, the analysis of supersymmetry breaking
involves
two ingredients.
The first is 
identifying the correct degrees
of freedom of the theory, 
in terms of which the \kahler potential is non-singular.
In all the theories we shall study, 
in the interior of the moduli space, 
these are either the confined variables, 
or the variables of a dual 
theory.~\footnote{At the boundary of
moduli space, the microscopic degrees of freedom will
often be more appropriate.}
Indeed, we shall see that 
duality~\cite{seibergduality}---the fact that different UV theories
may lead to the same infrared physics---can be a useful tool
for establishing supersymmetry breaking,
as one can sometimes pick a more convenient theory in which to
study whether supersymmetry is broken or not.
A second, related  ingredient
is finding the exact superpotential
in terms of the light, physical degrees of freedom.

Having established these two ingredients,
at low energies one then typically 
has a theory of chiral superfields, with all gauge dynamics
integrated out, with a known superpotential.
The problem of establishing supersymmetry breaking
is then reduced to solving a system of equations
to check whether or not the superpotential can be extremized.

Although the results we shall use on the infrared degrees of freedom
and the exact superpotential apply to supersymmetric theories,
they may still be used to argue for supersymmetry breaking, since the
theories we shall study in this way are
obtained by perturbing a supersymmetric theory.
For a sufficiently small perturbation, the scale of supersymmetry
breaking can be made sufficiently small, so that we can work
above this scale and still use known results on
the infrared supersymmetric theory.
If supersymmetry is indeed broken in the theory,
the breaking should persist even as the perturbation is increased.
Otherwise, the theory undergoes a phase transition
as some coupling is varied, being supersymmetric in some
region and non-supersymmetric in another.
However, one does not expect supersymmetric theories
to undergo phase transitions as couplings are 
varied~\cite{seiwit1,seiwit2,intseiphase}.

Having learned various techniques for the analysis
of DSB, we use these in section~\ref{exceptions}
to study a few examples of theories that break supersymmetry 
dynamically even though they violate some of the criteria
described in section~\ref{indirect}.

Perhaps the most disappointing aspect of the recent progress
in our understanding of supersymmetry breaking
is that it still has not yielded any organizing principle
to the study and classification of supersymmetry-breaking
theories. 
In section~\ref{known} we shall describe one method for
generating new supersymmetry-breaking theories from known theories.
However, this is far from a full, systematic classification.
Nor can we tell immediately, without detailed analysis, whether
a specific theory breaks supersymmetry or not.
For these reasons we find it useful to present
a rough survey of known models in section~\ref{known}, 
pointing out their main features,
the relations between them, and where applicable,
their relevant properties for model-building purposes. 
While we shall see many different mechanisms by which supersymmetry
is broken in these examples, the breaking is almost always the 
consequence of the interplay between 
instanton effects and a tree-level superpotential.

Another area requiring further study is 
the analysis of
supersymmetry-breaking
vacua, their symmetries and light spectra,
in strongly coupled theories.
One may hope that 
recently discovered
realizations of supersymmetric gauge theories as extended brane
configurations in string theory will lead to further
progress in 
this direction, as well as to some organizing principle for DSB.
Indeed, several DSB models have
been realized as D-brane configurations in string/M-theory,
see for example \cite{bhoo,lpt}. 
Moreover, in some cases  
the dynamical effects leading
to supersymmetry breaking were understood in stringy
language~\cite{bhoo}. 
However, this approach has yet to lead to results 
which 
cannot be directly
obtained in a field theory analysis.

We limit ourselves in this review to the theoretical
analysis of supersymmetry breaking in different models.
We do not discuss the questions of whether, and how, this
breaking can feed down to the standard model.
Ideally, a simple extension of
the standard model would break supersymmetry
by itself, generating an acceptable superpartner spectrum.
Unfortunately, this is not the case.
In 
simple supersymmetry breaking extensions of the 
Standard Model without new gauge interactions,
non-perturbative effects would probably
be too small to generate soft terms of the 
correct size~\cite{adsdsb}.
Moreover, unless some of the scalars
obtained their masses either radiatively, or from non-renormalizable
operators, some superpartners would be lighter than the lightest
lepton or quark~\cite{dimgeo}.
Thus, supersymmetry must be broken 
by a new, strongly-interacting sector,
and then communicated to the SM either by supergravity
effects, in which case the soft terms are generated
by higher-dimension operators or at the loop level,
or by gauge-interactions, in which case the soft terms
occur at the loop-level.
These different possibilities introduce different requirements
on the SUSY breaking sector. For example, gravity
mediation often requires singlet fields which participate
in the SUSY breaking. Simple models of gauge mediation
require a large unbroken global symmetry at the minimum
of the SUSY breaking theory, in which the standard
model 
gauge group
can be embedded. 
Several of the supersymmetry breaking models discovered recently
have some of these desired properties, and thus allow
for improved phenomenological models for the communication
of supersymmetry breaking.

Another issue which is important for phenomenological applications of DSB
that we shall not address is the cosmological constant problem. 
In globally supersymmetric theories, fermionic and bosonic contributions
to the vacuum energy cancel each other and the cosmological
constant vanishes. Upon supersymmetry breaking this is no
longer true, and the cosmological constant is comparable to the
scale of supersymmetry breaking. While in a framework of local
supersymmetry a further cancellation is possible, 
significant fine tuning is required. Eventually, a
microscopic understanding of such a fine tuning
is needed in any successful
phenomenological application of dynamical supersymmetry
breaking.

\bigskip
\noindent
{\it How to Use This Review}

In the body of 
this review we assume that the reader is familiar with
the general properties of supersymmetric field theories
and rely heavily on symmetries and a number of exact
non-perturbative results obtained in recent years.  
The reader who is just beginning the study of
supersymmetry should first consult the Appendix,
where we briefly present 
basic facts about supersymmetry and
relevant results to make our presentation self-contained.
For a more complete introduction to SUSY see for example,
\cit{WB} and \cit{nillesrev}. 
Several excellent reviews of the recent progress in the
study of strongly coupled SUSY gauge theories exist, see for
example \cit{ISreview}, \cit{peskinreview},
and \cit{shifmanreview}.

In Appendix~\ref{notations} we introduce notations and
basic formulae  for the Lagrangians of supersymmetric theories. 
The knowledge of these results is necessary in every section of the
review.
In~\ref{Dflatsection} we discuss a method for finding the D-flat directions
of a SUSY gage
theory (directions along which the gauge interaction terms in the scalar
potential vanish),
and the parametrization of D-flat directions
in terms of gauge-invariant operators.
These results are necessary for the study of supersymmetry
breaking in non-abelian gauge theories which we discuss starting
in Section~\ref{examples}.
In \ref{psun} through \ref{duality} we turn to SUSY QCD with different numbers
of flavors. While the results we present are used directly in
various places in Sections~\ref{calculable}--\ref{known}, the discussion in 
these Appendixes also illustrates techniques in the
analysis of the dynamics of  general SUSY gauge theories
and are applicable to models with matter transforming in general
representations of the gauge group.
The discussion also provides simple examples of phenomena such as
theories with no quantum moduli spaces, deformed quantum moduli spaces,
confinement without chiral symmetry breaking, and duality.
We shall encounter these phenomena in various theories
throughout the review.
References to analyses of the dynamics of theories other than
$SU(N)$ are collected in~\ref{othermodels}.

We also note that the reader who is only interested
in a general knowledge of DSB can skip sections~\ref{exceptions}
and \ref{known}. Section~\ref{modellist} can be used independently
of the rest of the article as a guide to DSB models.

Finally, we note that the interested reader
can find several useful reviews of dynamical supersymmetry breaking
which have appeared in the past couple of years. 
A short introduction to recent developments can be found
in \cite{skiba,nelson,poppitz,thomas}. 
Most notably, the review by~\cit{poptrirev},
although smaller in scope than the present review,
emphasizes recent developments in the field.
It also contains a discussion of supersymmetry breaking
in quantum mechanical systems.
\cit{SVreview} give an excellent introduction
to instanton techniques and discuss their application
for supersymmetry breaking. 
The review by~\cit{GRreview}, 
focuses on applications of DSB
to building models of gauge mediated supersymmetry breaking.

\section{Generalities}  \label{generalities}
\subsection{Vacuum Energy -- The Order Parameter of SUSY Breaking,
and $F$ and $D$ flatness}
\label{vacuum}

A positive vacuum energy is a necessary and sufficient condition
for spontaneous SUSY breaking. This follows from the fact that
the Hamiltonian of the theory is related to the 
absolute
square of the SUSY generators (see Appendix~\ref{notations})
\beq\label{hamiltonian}
H= \frac{1}{4} (\Qbar_1 Q_1 +Q_1 \Qbar_1 +\Qbar_2 Q_2 +
Q_2 \Qbar_2 ) \ .
\eeq
The energy is then either positive or zero. 
Furthermore, a state that is annihilated by 
$Q_\al$ has zero energy, and conversely, a zero-energy
state is annihilated by  $Q_\al$.
Thus, the vacuum energy serves as an order parameter 
for supersymmetry breaking.

Therefore the study of supersymmetry breaking requires the
knowledge of the scalar potential of the theory. It is
convenient to formulate the theory in ${\cal N}=1$
superspace, where space-time (bosonic) coordinates are
supplemented by anti-commuting (fermionic)
coordinates\footnote{For more detail see Appendix
\ref{notations}.}. In this formulation fields of different
spins related by supersymmetry are combined in the
supersymmetry multiplets, superfields. Matter fields form
chiral superfields, while gauge bosons and their spin $1/2$
superpartners form (real) vector superfields. In the
superspace formulation, physics, and in particular, the scalar
potential, is determined by two functions of the superfields,
the superpotential and the \kahler potential. The superpotential encodes
Yukawa-type interactions in the theory, in particular it
contributes to the scalar potential. The superpotential is
an analytic function of the superfields. This fact together
with symmetries and known (weakly coupled) limits often
allows to determine the exact non-perturbative superpotential of
the theory. The \kahler potential, on the other hand, is a
real function of superfields, and can only be reliably
calculated when a weakly couled description of the theory
exists. From our perspective, the \kahler potential is
important in two respects. 
First, it gives rise to gauge-interaction terms 
in the scalar potential.
Second, it determines the kinetic terms of the matter fields and
thus modifies scalar interactions arising from the superpotential.
Assuming a canonical (quadratic in the fields) \kahler
potential, the scalar potential is
\beq
\label{Vgeneralities}
V = \sum_a (D^a)^2 + \sum_i F_i^2 \ ,
\eeq
where the sum runs over all gauge indices $a$ and all matter
fields $\phi_i$.
In (\ref{Vgeneralities}), the D-terms and F-terms are auxiliary
components of vector and chiral superfields
respectively. D-terms and F-terms are not dynamical and one
should solve their equations of motion. In particular,
F-terms are given by  derivatives of the superpotential 
(for more details, and the analogous expressions for D-terms,
see Appendix~(\ref{notations}).)
$$ F_i = \frac{\partial W}{\partial \phi_i} \ .$$
For supersymmetry to remain unbroken, there has to be
some field configuration for which both the $F$-terms and the 
$D$-terms vanish.\footnote{ Here we implicitly assumed that the
\kahler potential is a regular function of the fields and
has no singularities. For example, if the derivatives of the
\kahler potential
vanish, the scalar potential can be non-zero even
if all $F$ and $D$ terms vanish, see Eq.~(\ref{susypotential}).} 
In fact, generically, such configurations exist not only at
isolated points but on a subspace of the field space. 
This subspace is often referred to as the moduli
space of the theory.

Classically, one could set all superpotential couplings to
zero. Then the moduli space of the theory is the
set of ``$D$-flat directions",
along which the $D$-terms vanish.
A particularly useful parameterization of $D$-flat directions,
which we discuss in Appendix~\ref{Dflatsection},
can be given in terms of the gauge invariant operators of the
theory~\cite{lutytaylor}.
Even when small tree-level
superpotential couplings are turned on, the
vacua  will lie near the $D$-flat directions. 
It is convenient, therefore, to
analyze SUSY gauge theories in two stages. First find 
the $D$-flat directions,
then analyze the $F$-terms along these directions.
The latter have classical contributions from the tree-level
superpotential, and may ``lift" some, or all, of the $D$-flat
directions. Since a classical superpotential is a polynomial
in the fields, $F$-terms typically grow for large scalar
vacuum expectation values (vev),
and vanish at the origin.\footnote{An important exception are superpotentials
terms that are linear in the fields, which lead to potentials that are
nonzero even at the origin. Such  terms  necessarily involve 
gauge-singlets, and require the introduction of some mass scale.
However,  linear term can be generated dynamically. We shall encounter
examples of trilinear, or higher, superpotential terms that become
linear after confinement.}

As mentioned in the Introduction,
a key point in the study of SUSY breaking is the fact that,
due to the supersymmetric non-renormalization 
theorems~\cite{wznr,fiz,grisigroc,seibergnr},
the moduli space remains unmodified in perturbation theory.
If the classical potential vanishes for some choice of vevs, it
remains exactly zero to all orders in perturbation theory.
Thus, only non-perturbative effects may generate a non-zero potential,
and lift the classical zeros.
Indeed, non-perturbative effects can
modify the moduli space;
they can lift the moduli space completely; 
or, finally,
the quantum moduli space may coincide with the classical one.

There are numerous possibilities then
for the behavior of the theory.
If a theory breaks supersymmetry, it has some ground state
of positive energy at some point in field space (or it may,
in principle, have several ground states at different 
points\footnote{In this latter case the ground states are
non-degenerate even if they appear to have the same energy
in a certain approximation. This is because the 
low energy physics is
non-supersymmetric, and the vacuum energy receives quantum
corrections (on top of the non-perturbative effects which
led to the non-vanishing energy in the first place). Since
different non-supersymmetric vacua are non-equivalent, these
quantum corrections lift the degeneracy.}).
Alternatively, the theory may remain supersymmetric, 
with either one ground
state of zero energy at some point in field space, or a few
ground states, at isolated points, or with a continuum
of ground states, corresponding to completely flat directions
that are not lifted either classically or
non-perturbatively. 
It is also possible that the theory does
not have a stable vacuum state. In such a case, while
a supersymmetric vacuum does not exist, the energy can become
arbitrarily small along some direction on the moduli
space.\footnote{We shall call such directions in the moduli
space ``runaway'' directions, and study them carefully
in Sections \ref{nonchiral}--\ref{flat}. }
While such a theory can still be given a cosmological
interpretation~\cite{adsqcd}, we shall not consider it a
SUSY breaking theory for our
purposes.\footnote{We also note that the runaway
moduli may be stabilized, and supersymmetry broken, due to 
\kahler potential effects
when the theory is coupled to gravity \cite{dk}. However, since the
typical vevs in this case will be of Planck size, the vacuum
will be determined by the details of the microscopic theory
at $M_P$, and the dynamical supersymmetry breaking is not 
calculable in the low energy effective field theory.}

Note that supersymmetric theories are very different, in this respect,
from other theories. In a non-supersymmetric theory, multiple
ground states are usually related by a symmetry, and are therefore
physically equivalent. 
On the other hand, different ground states of a supersymmetric theory
may describe completely different physics.
For example, classically, one 
flat direction of an $SU(3)$ gauge theory
with two flavors is
(see Appendix~\ref{adsW} for details)
\beq
\label{nonequivalent}
f=\overline{f}=
\left( \begin{array}{cc}
v&0\\[-5pt]
0&0\\[-5pt]
0&0
\end{array}
\right),
\eeq
where $f$ and $\overline{f}$ are the scalar components of the
$SU(3)$  fundamentals and antifundamentals respectively.
For any given choice of $v$, the low-energy theory is an
$SU(2)$ gauge theory, whose gauge coupling depends on
$v$.\footnote{Here we consider the classical vacua of the
theory. Quantum mechanically, the flat directions~(\ref{nonequivalent}) are
lifted, and the theory does not have a stable vacuum. Yet, in
many models non-equivalent quantum vacua exist.}
Note also that the flat directions of SUSY theories may extend
to infinity, unlike the customary compact flat directions
of bosonic global symmetries.
Furthermore, in the case of other global symmetries, the existence
of a flat direction is usually associated with spontaneous
breaking of the symmetry, with the massless Goldstone bosons
corresponding to motions along the flat direction.
This is not the case with SUSY. The reason, of course, is
that the SUSY generators do not correspond to motions in field space.
Theories with unbroken SUSY may have degenerate vacua precisely
because SUSY is unbroken.
And, as we shall see in the next section,
theories with spontaneously broken SUSY have the analog 
of Goldstone particles even when they only have a single
ground state.

\subsection{The Goldstino}    \label{goldstino}

The breaking of any bosonic global symmetry is accompanied by
the appearance of massless Goldstone bosons that couple linearly 
to  the symmetry current.
Similarly, a theory with broken supersymmetry contains a massless
fermion, which is usually referred to as a ``Goldstone fermion",
or in short, ``Goldstino", that couples linearly to the SUSY 
current~\cite{salam,wittendsb}.

The Goldstino coupling to the SUSY current can be expressed as 
\beq
\label{goldst}
J^\mu_\al = f\, {\sig^\mu}_\al^\bed\, \psi^G_\bed\ +\ldots\ ,
\eeq
where $\psi^G_\bed$ is the Goldstino, and as we shall see momentarily,
$f$ is a constant which is non-zero when SUSY is broken.
The ellipsis in~(\ref{goldst}) stand for terms quadratic in the fields
and for potential derivative terms.
Conservation of the SUSY current then implies that the Goldstino 
is massless.
To justify~(\ref{goldst}), note that, for broken SUSY~\cite{wittendsb},
\beq
\label{proofa}
\int d^4x\,\partial_\eta\, 
\langle 0\vert
T\, J^\eta_\al(x)\, {J^\nu}_\bed(0)\vert0\rangle \ =\
\langle 0\vert
\{Q_\al\, , {J^\nu}_\bed(0)\}\vert0\rangle \ \neq 0\ .
\eeq
If indeed there is a massless fermion coupling to the current as
in~(\ref{goldst}), then the LHS of~(\ref{proofa})
is equal to
\beq
\label{proofb}
f^2\, {\sig^\eta}_\al^\ald \, {\sig^\nu}_\bed^\be
\int d^4 x\partial_\eta\,
\langle 0\vert
T\, \psi^G_\ald(x)\, \psi^G_\beta(0)\vert0\rangle\ =\
f^2\, {\sig^\eta}_\al^\ald \, {\sig^\nu}_\bed^\be
[-ip_\eta\, G(p)_{\ald\be}]_{p\rightarrow 0}\ =\
f^2\, {\sig^\nu}_{\al\bed}\ ,
\eeq
where $G(p)_{\ald\bed}$ is the Goldstino propagator.
So indeed $f$ is nonzero. 
Note that a fermion with derivative coupling to the current
would not contribute to the RHS of~(\ref{proofb})
because of the additional factors of the momentum in the numerator.  
In fact, since from the SUSY algebra
$\langle 0\vert\{Q_\al\, , {J^\nu}_\bed(0)\}\vert0\rangle \ = 
2E\,{\sig^\nu}_{\al\bed}$, with $E$ the energy,
we have $f^2 = 2E$.

To see the appearance of the Goldstino more concretely, 
consider  the SUSY current
\beq
\label{j}
J^\mu_\al \sim \sum_\phi {\delta {\cal L}\over 
\delta(\partial_\mu\phi)}\, {(\delta\phi)}_\al\ , 
\eeq
where the sum is over all  fields, and ${(\delta\phi)}_\al$
is the shift of the field  $\phi$ under a SUSY transformation.
Because of Lorentz symmetry, the only linear terms 
in~(\ref{j}) come from vacuum expectation values 
of ${(\delta\phi)}$. 
Examining the SUSY transformations of the
chiral and vector multiplets of 
${\cal N}=1$ SUSY
(see for example \citename{WB}, \citeyear{WB}),
we see from Lorentz invariance that
the only fields whose SUSY transformations contain Lorentz-invariant objects,
which can develop vevs, are the matter fermion $\psi_i$, 
whose SUSY transformation gives $F_i$, and the gauge fermion $\lambda^a$,
whose transformation gives $D^a$.
One then finds,
\beq
\label{jfd}
\psi^G \sim \sum \langle F_i\rangle\, \psi_i 
+ {1\over\sqrt{2}}\,\sum \langle D^a\rangle\, \lambda^a \ ,
\eeq
so that the Goldstino is a linear combination of the
chiral and gauge fermions whose auxiliary fields $F$ and $D$
acquire vevs.
Note that Eq.~(\ref{jfd}) actually only holds with a canonical
(quadratic) \kahler potential, otherwise derivatives of the \kahler potential
enter as well.
We can use this to argue that a non-vanishing
$F$-vev or $D$-vev is a necessary condition for
SUSY breaking. When SUSY is broken, there is a massless fermion,
the Goldstino, that transforms inhomogeneously under the action of
the SUSY generators. 
But the only Lorentz-invariant objects that appear in the SUSY variations
of the ${\cal N}=1$ multiplets, and therefore may obtain vevs, are
the auxiliary fields $F$ and $D$. 
Thus, for SUSY breaking to occur, some $F$- or $D$- fields should
develop vevs.

In light of the above, it would first seem
that if SUSY is relevant to nature, we should observe the massless
Goldstino. 
However, we ultimately need to promote  SUSY to a local symmetry
to incorporate gravity into the full theory.
In the framework of local supersymmetry,
the massless Goldstino becomes the 
longitudinal component of the
Gravitino, much like in the case of the Higgs mechanism.
The Gravitino then has a coupling to ordinary matter other than
the gravitational interaction, by virtue of its Goldstino component.
It should come as no surprise then, that the Gravitino mass as
well as its coupling to matter fields,
are related to the SUSY breaking scale. 
This has important phenomenological implications.
In particular, in models with low-scale SUSY breaking, the gravitino
is very light, and the decay of other superpartners into the gravitino
may be observed in collider searches 
for supersymmetry~\cite{ddrt,swy}.

\subsection{Tree-Level Breaking} \label{tree} 

\subsubsection{O'Raifeartaigh Models} \label{oraf}

One of the simplest models of spontaneous supersymmetry
breaking was proposed by \cit{oraf}, and is based on a theory
of chiral superfields. Supersymmetry in the model is
broken at tree level: while the lagrangian of
the model is supersymmetric, even the classical potential is
such that a supersymmetric vacuum state does not exist.

In addition to giving the simplest example of
spontaneously broken supersymmetry, the study of O'Raifeartaigh
models will be useful for our later studies of  DSB, as the 
low-energy description of 
many 
dynamical models we shall encounter will be given by an 
O'Raifeartaigh-type model.

Before writing down the simplest example of an O'Raifeartaigh
model, let us describe the general properties 
of such models.
First, we shall restrict
our attention to superpotentials with only positive
exponents of the fields. We shall later analyze a number of
models where the low-energy description involves 
superpotentials containing negative
exponents of the fields. Such terms, however, are generated by
the non-perturbative dynamics of the underlying 
(strongly coupled) microscopic theory, and are not
appropriate in the tree level superpotential we consider here.

Second, since the superpotential is a polynomial in the fields,
at least one of the fields in this
model needs 
to appear linearly
in the superpotential, or
there will be a supersymmetric vacuum at the origin of the
moduli space.

It would prove useful, for future purposes, to pay special attention
to the R-symmetry of the model.
(For the definition of an R-symmetry, see Appendix~(\ref{notations})). 
If this symmetry is unbroken by the 
superpotential, there is then at least one field of R-charge 2.
More generally,
consider a model containing  fields $\phi^c_i,~
i=1, \ldots, k$ with R-charge $2$,  and
fields $\phi^n_a,~ a=1, \ldots, l$ with R-charge 0.
(For convenience, we shall call them charged and
neutral respectively, even though various components of the
superfields transform differently under R-symmetry.) 
The most general superpotential respecting
the R-symmetry can be written as
\beq
\label{orafgeneral}
W =\sum_{i=1}^k \phi^c_i f_i(\phi^n_a)\ ,
\eeq 
and, for supersymmetry to break, 
at least one of the $f_i$'s, say $f_1$, 
contains a constant term, independent of the fields.
The equations of
motion for the R-charged fields $\partial W /\partial \phi^c_i =
f_i(\phi^n_a) = 0$ give $k$ equations for $l$ unknowns
$\phi_a^n$. If $k>l$ there are no solutions for generic
functions $f_i$, the F-term conditions can not all be satisfied and
supersymmetry is broken. 

One can modify these models by
adding fields with R-charges $0<Q_R<2$. Since such fields
can not couple to the fields $\phi^c_i$ while preserving 
the R-symmetry,
they will not change the above discussion, and supersymmetry
remains broken. If, on the other hand, fields with negative
R-charges are added to the model, the total number of
variables on which the $f_i$'s depend increases, and in general
supersymmetry is unbroken. 
Finally, we should note that adding to the superpotential
explicit R symmetry violating couplings which do not involve
fields of R charge 2 will not modify the above discussion.
On the other hand,
R symmetry violating terms which include fields of R charge
2 will generically lead to the restoration of supersymmetry.

It is also useful to look at the equations of motion for
the R-neutral fields. First, note that their vev's are fixed by
minimizing the part of the scalar potential arising from the
F-terms of the R-charged fields
(the remaining terms in the
scalar potential vanish at least when $\vev{\phi_i^c}=0$ for
all $i$). 
Therefore, there are 
$l$ F-term equations
depending on $k$ independent variables
$\phi^c_i$. In a SUSY breaking model $k>l$, 
so there are
$k-l$ linear combinations of the fields $\phi^c_i$
that are left undetermined. 
Thus O'Raifeartaigh models necessarily posses
directions of flat (non-zero) potential
in the tree level approximation \cite{zumino,einjones,pol}. 
As we shall discuss later this
is not a generic situation in models of DSB.

The simplest example of an O'Raifeartaigh model requires two
fields with R-charge 2, one field with R-charge 0 ($k=2$, $l$=1), 
and has the superpotential
\beq
\label{orafw}
W=\phi_1 (M_1^2 -  \lam_1 \phi^2) + m_2 \phi \phi_2\ .
\eeq
It is easy to see that the F-term conditions for $\phi_1$
and $\phi_2$ are incompatible. 
We can directly minimize the scalar potential
\beq
\label{orafv}
V=\abs{M_1^2 - \lam_1 \phi^2}^2  + \abs{m_2 \phi}^2
+ \abs{m_2 \phi_2 - 2 
\lam_1 \phi_1 \phi}^2 \ .
\eeq
The first two terms in (\ref{orafv})
determine the value of $\phi$ at the minimum.
In the limit $m_2^2/(\lam_1 M_1^2) \gg 1$ the minimum is found
at $\phi=0$, while for small $m_2^2 /(\lam_1 M_1^2)$ we find
$\phi= (M_1^2 - m_2^2/2)^{1/2}/\lam_1$. 
Note that $\phi_1$ and $\phi_2$ only appear in the last term
in (\ref{orafv}). This term should be set to zero 
for the potential to be extremal with respect to
$\phi_1$ and $\phi_2$.
This is achieved when  $\phi_2 = - 2 \lam_1 \phi\phi_1 /m_2$, and      
therefore, at tree level the linear combination
$m_2 \phi_1-2 \lam_1 \vev{\phi} \phi_2$
is arbitrary, as was expected from the previous
discussion. Equivalently, we can parameterize different vacua by
the expectation values of $\phi_1$.
Note that different vacua are physically
non-equivalent, in particular the spectrum depends 
on $\vev{\phi_1}$.

It is easy to find the tree level 
spectrum of the model. For any choice of parameters
it contains a massless fermion, the Goldstino. 
The spectrum also contains the scalar field associated with the 
flat direction whose mass arises entirely due to radiative corrections
(however, there are no quadratic divergences since the action of
the theory is supersymmetric although supersymmetry is 
not realized linearly). All other states are massive. 
Another important feature of this spectrum is
that the supertrace of the mass matrix squared, ${\rm STr}\,
m_i^2$, vanishes. This property of the spectrum holds for any
model with  tree level supersymmetry breaking \cite{FGP}.

Since supersymmetry is broken, the vacuum degeneracy 
is lifted in perturbation theory. \cit{huq}
calculated the one loop corrections to the potential of this model.
They are given by
\beq
\label{orafCW}
\Delta V(\phi_1) = \sum_i \frac{(-1)^F}{64 \pi^2}
m_i(\phi_1)^4 \ln \left(\frac{m_i(\phi_1)^2}{\mu^2} \right),
\eeq
where the sum is over all massive fields (and the masses
depend on the  $\phi_1$ vev). \cit{huq} found that the corrections
generate a positive mass for $\phi_1$, and that the non-supersymmetric
vacuum is located at $\phi_1=0$ with unbroken R-symmetry.
He also analyzed a model with an $SU(3)$ global symmetry and
a model constructed by \cit{fayetchiral} with $SU(2) \times
U(1)$ symmetry, and in both cases found that the tree-level
modulus acquires positive mass due to one-loop corrections
to the \kahler potential, leading to the unique vacuum with
unbroken R-symmetry. In fact, this conclusion is not  surprising. 
In the model discussed above, 
quantum corrections to the vacuum energy come
from the renormalization of the mass parameter $M^2$. 
Due to the holomorphy of the superpotential, these are
completely determined by the wave function renormalization
of $\phi_1$ and, since the model is infrared free, 
necessarily generate positive
contribution to the scalar potential. 
It is important to note that in
modifications of the model that include gauge fields, there
may be a negative contribution to the potential. The balance
of the two perturbative effects may produce a stable minimum
at large values of the modulus vev \cite{witteninverted}.

\subsubsection{Fayet-Iliopoulos Breaking} \label{fi}

Another useful example of tree level supersymmetry breaking 
is given by a model with $U(1)$ gauge interactions
\cite{FI}. In this model supersymmetry breaking is driven by
D-term contributions to the potential, but depending on
the parameters of the Lagrangian, the non-zero vacuum energy
either comes entirely from D-term contributions, or from
both D- and F-terms. 
To understand how D-terms can drive supersymmetry
breaking, we recall that the \kahler potential can be written as
a function of the gauge invariant combination of fields
\beq
K=f(\phi^\dagger e^{V} \phi, \cal W^\dagger \cal W, S),
\eeq
where $\phi$ represents matter fields transforming in some
representation of the gauge group, $V$ is a vector
superfield whose supersymmetric field strength is $\cal W$, and
$S$ represents gauge singlet fields. 
In a nonabelian theory this is the only possible form 
of field dependence in the
\kahler potential. In an abelian theory, 
however, the D-term of the vector superfield $V$ 
is invariant under the gauge and supersymmetry
transformations by itself. Thus, if one does not require
parity invariance, the lowest order \kahler potential 
of a $U(1)$ gauge theory can be
written as\footnote{We restrict our attention to two matter
multiplets with charges $\pm 1$.}
\beq
K = Q^\dagger e^{V} Q + \Qbar^\dagger e^{-V} \Qbar +
\xi_{FI} V.
\eeq

This \kahler potential together with 
superpotential mass terms for the matter fields
leads to the following
scalar potential
\beq
\label{FIscalar}
V=\frac{g^2}{2}(\abs{Q}^2 - \abs{\Qbar}^2 + \xi)^2 + 
m^2 (\abs{Q}^2 +\abs{\Qbar}^2)\ .
\eeq
It is easy to see that the vacuum energy determined by this
potential is necessarily positive and supersymmetry is
broken. When $g^2 \xi < m^2$ both scalar fields have
positive mass and their vevs vanish. The positive
contribution to the vacuum energy comes entirely from
the D-term in the potential. The scalar mass matrix has
eigenvalues $m^2_\pm = m^2 \pm g^2\xi$. The gauge symmetry
is unbroken, and thus the gauge boson remains massless. The
matter fermions retain their mass $m$, while the gaugino remains
massless and 
plays the role of the Goldstino.
(In accord with with the fact that here $\langle F_i\rangle=0$
and $\langle D\rangle\neq 0$, see Eq.~\ref{jfd}.)

When $g^2 \xi > m^2$, the field $\Qbar$ has negative mass and
acquires a vev. At the minimum of
the potential $Q=0$ and $\Qbar=v$, where $v=(2\xi-4
m^2/g^2)^{1/2}$. We see that both the gauge symmetry and
supersymmetry are broken. Moreover,
both D-term and F-term are
non-vanishing and supersymmetry breaking is
of the mixed type.
One can easily find that the spectrum of the model
contains one vector field and one real scalar field of mass
squared $\frac{1}{2} g^2 v^2$, one complex scalar of mass
squared $2m^2$, two fermions of mass 
 $(m^2 + \frac{1}{2} g^2 v^2)^{1/2}$, and a massless
Goldstone fermion which is a linear combination of the
Goldstino and the positively charged fermion
\beq
\label{FIgoldstone}
\tilde \lam =
\frac{1}{\sqrt{m^2+\frac{1}{2}g^2v^2}}\left(m\lam +
\frac{igv}{\sqrt{2} }\psi_Q\right)\ .
\eeq

\section{Indirect Criteria for DSB}
\label{indirect}

As we have seen in Section \ref{vacuum}, 
the fact that the vacuum energy is the relevant 
order parameter immediately points the way
in our quest for SUSY breaking: we should study   
the zeros of the scalar potential.
This, indeed, is what we shall undertake to do in 
sections~\ref{calculable}
and onward. 
Unfortunately, directly studying the zeros of the potential will not always
be possible, or easy.
In this Section we review several alternate ``indirect"
methods that are useful in the search for supersymmetry breaking.

\subsection{The Witten Index}
\label{witten}

Supersymmetry breaking is related to the existence of zero-energy
states. Rather than looking at the total number of zero energy
states, it is often useful to consider the Witten index~\cite{index},
which measures the difference between the number of 
bosonic and fermionic states of zero energy,
\beq\label{wittenindex}
{\rm Tr} (-1)^F \equiv n_B^0 - n_F^0 \ .
\eeq

If the Witten index is nonzero, there is at least
one state of zero energy, and supersymmetry is unbroken.
If the index vanishes, supersymmetry may either
be broken, with no states of zero energy, or it may be unbroken,
with identical numbers of fermionic and bosonic states of zero energy.

The Witten index is a topological invariant of the
theory. In this lies its usefulness. It may be calculated
for some convenient choice of the parameters of the theory,
and in particular, for weak coupling,
but the result is valid generally.
To see this, note that 
in a finite volume,
fermionic and bosonic states 
of positive energy are paired
by the action of the SUSY generator:
\beq\label{pair}
Q\vert b_E\rangle \sim \sqrt{E}\vert f_E\rangle\ \ \ \  
Q\vert f_E\rangle \sim \sqrt{E}\vert b_E\rangle\ , 
\eeq
where $\vert b_E\rangle$ ($\vert f_E\rangle$) is a bosonic (fermionic)
state of energy $E$.\footnote{This in fact justifies including only 
zero-energy states in~(\ref{wittenindex}). States of non-zero energy
do not contribute.
} 
(Recall that states of zero energy are annihilated by $Q$, and are therefore
not paired). 
Thus, under ``mild'' variations of the parameters of the theory, states
may move to zero energy and from zero energy, but they always
do so in Bose-Fermi pairs, leaving the Witten index unchanged. 

Let us be a bit more precise now about what is meant by ``mild''
variations above. As long as a parameter of the theory,
which is originally nonzero, is varied to a different nonzero
value, we do not expect the Witten index to change, since different states
can only move between different energy levels in pairs. The danger
lies in the appearance of new states of zero energy. 
This can happen if the asymptotic (in field space) 
behavior of the potential changes, which may happen if some parameter
of the theory is set to zero, or is turned on.
In that case, states may ``come in'' from infinity or ``move out''
to infinity.

The index of several theories was calculated by \cit{index}.
In particular, Witten found that the index of a pure supersymmetric
Yang-Mills (SYM) theory is 
non-zero.~\footnote{
For $SU$ and $SP$ groups the index is
equal to  $r+1$, where $r$ is the rank of the group.
This is the same as the number of gaugino condensates for these
groups.
More generally, the index equals
the dual coxeter number of the group, which is different from 
$r+1$ for some groups, notably some of the $SO$ 
groups~\cite{wittennew}.
The fact that the number of gaugino condensates does not
always equal $r+1$, which was believed to be the value of the index,
remained a puzzle until its resolution by Witten recently~\cite{wittennew}.
This puzzle partly motivated the conjecture
that SYM theories have a vacuum with no gaugino condensate~\cite{nocondensate}.
This possibility would have far-reaching consequences for supersymmetry
breaking. The vast majority of theories that break SUSY do so by virtue
of a superpotential generated by gaugino condensation.
A vacuum with no gaugino condensate would mean an extra ground state,
or an entire branch of ground states, with zero energy and unbroken SUSY.} 
Thus, these theories do not break supersymmetry spontaneously.
An important corollary is that SYM theories with massive matter 
(and no massless matter) do not break supersymmetry either. 
The reason is that, at least in weak coupling,
one can take all masses large, so that
there are no massless states in these theories beyond those of the
pure SYM theory, and so the value of the index is the same
as in the pure SYM theory.

What happens when the mass of the matter fields is taken to zero?
The theory with zero mass has flat directions, along which the 
potential is classically zero (away from these flat directions
the potential behaves as the fourth power of the field strength).
In contrast, the theory with mass for all matter fields has no 
classical flat directions, with
the potential growing at least quadratically for large fields.
Thus, as the mass is taken to zero, the asymptotic behavior
of the potential changes, and the Witten index may change too.
In fact, the index is ill-defined in the presence of flat directions,
since zero modes associated with the flat directions lead to a continuous
spectrum of states. (Indeed, to calculate the index of any theory one
needs to consider the theory in a finite volume so that the resulting
spectrum is discrete.)
We therefore cannot say anything about supersymmetry breaking in massless,
non-chiral theories based on the Witten index of the pure SYM theory. 

Consider for example SQCD with $N$ colors and $F$ flavors of mass $m$,
which we discuss in Appendix \ref{adsW}--\ref{duality}.
As explained there, in the presence of mass terms 
$m^i_j Q_i \cdot {\bar Q}^j$, the theory has $N$ vacua at
\beq\label{vac}
M_i^j \equiv Q_i\cdot {\bar Q}^j = \Lambda^{(3N-F)/N}\,
({\rm det} m)^{1/N} {(m^{-1})}_i^j \ ,
\eeq 
corresponding to the $N$ roots of unity.
This is in agreement with the Witten index $N$ of pure 
$SU(N)$ gauge theory. 
Consider now the massless limit, $m^i_j \rightarrow 0$.
For $F < N$, the vacua~(\ref{vac}) all tend to infinity.
The theory has no ground state at finite field vevs. The potential
is nonzero in any finite region of field space and 
slopes to zero at infinity.
The massless limit of the theory is therefore not well-defined.
For $F > N$, by taking  $m^i_j\rightarrow 0$ in different ways,
any value of $M_i^j$ may be attained.
The massless theory has an entire moduli space of vacua, 
parameterized by $M_i^j$.
The $N = F$ case is more subtle,
but in this case too, the theory has a moduli space of vacua.
Thus, the ground states of the massless $SU(N)$ theory with $F$ flavors
are drastically different from those of the massive theory.
In these examples we explicitly see how zero-energy states can disappear to
infinity, or come in from infinity. Again, this is possible
because the asymptotic behavior of the potential changes
as the mass tends to zero.
The theory including mass terms has no flat
directions. Asymptotically the potential
rises at least quadratically.
The massless theory has classical flat directions.
Quantum mechanically, they are completely lifted for $F < N$, and the
potential  asymptotes to zero as a fractional power of the field.
For $F \geq N$  flat directions remain even quantum mechanically.
In any case, adding mass terms changes 
the asymptotic behavior of the potential.

In the examples above, the massive theory was supersymmetric (with
zero energy states at finite fields vevs) and the massless theory
was either supersymmetric (with a continuum of vacua) or not
well defined (with no ground state).
It is natural to ask whether there exist  vectorlike (parity-conserving)
theories that break SUSY as the relevant masses are taken to zero. 
The answer to this question is affirmative as
we shall see in an explicit example in Section~\ref{ITIY}.
It is useful to understand the general properties of the
potential in such a theory.
As before, we expect the fully massive theory to have
a non-zero Witten index. 
The only way to obtain supersymmetry breaking as the masses
are taken to zero, is if the masses change the asymptotic 
behavior of the potential. 
Suppose that for any finite value of the mass parameter $m$, 
the theory possesses a supersymmetric vacuum at some vev $v_0(m)$, 
which moves away to
infinity as some of the masses are taken to zero. Clearly,
for small finite masses, the directional derivative 
of the potential with respect to the modulus is negative
for large values of $v < v_0(m)$.
(The theory may have various minima for finite values of $v$,
but we are interested in the asymptotic behavior
of the potential at large $v$.)    
In the absence of a phase
transition at zero mass, such a directional derivative
will remain non-positive in the limit $m \ra 0$. However,
there are still two possibilities. First, it is possible
that the directional derivative is negative for any finite vev,
and only vanishes in the double limit, $m\ra 0$, $v \ra
\infty$. In such a case the theory does not have a stable
vacuum. However, it is also possible that the derivative
vanishes in the limit $m\ra 0$ for sufficiently large, but
otherwise arbitrary $v$. If this is the case, the asymptotic
behavior of the potential changes, and it becomes a non-zero constant
asymptotically far along the flat direction, 
so supersymmetry is broken.
However, because the potential is flat, running effects cannot 
be neglected.
Indeed, as we shall argue in Section~\ref{ITIY}, 
such effects may 
lift the vacuum degeneracy and determine the true
non-supersymmetric vacuum.

To summarize, pure SYM theories, as well as 
vectorlike theories with masses for all  matter
fields, have a non-zero index, and do not break supersymmetry. 
When some masses are taken to zero, the resulting theories have classical
flat directions, and therefore, the asymptotic behavior of the potential
is different from that of the massive theory. The Witten index
may then change discontinuously and differ, if it is well
defined, from that of the massive theory.

What about chiral (parity-violating) 
theories? In such theories, at least some of the
matter fields cannot be given mass. Thus, these theories cannot be obtained
by deforming a massive vectorlike theory, 
and there is no a priori reason to expect,
based on existing computations of
Witten index, that these theories are supersymmetric.
Indeed, most known examples of supersymmetry breaking are chiral.

\subsection{Global Symmetries and Supersymmetry Breaking}
\label{global}

In this section we shall discuss the connection between
global symmetries and supersymmetry breaking,
which motivates two criteria for supersymmetry breaking.
While these are useful guidelines for finding supersymmetry-breaking
theories, they are not strict rules, and we shall encounter
several exceptions in the following.

Consider first a theory with an exact, non-anomalous global symmetry, 
and no flat directions. If the global symmetry is spontaneously
broken, there is a massless scalar field, the Goldstone boson,
with no potential. 
With unbroken supersymmetry,  
the Goldstone boson is part of a chiral supermultiplet that
contains an additional massless scalar, again with no potential. 
This scalar describes 
motions along a flat direction of zero potential.
But this contradicts our initial assumption that there are no 
flat directions. 
To avoid the contradiction we should drop
the assumption of unbroken supersymmetry. 
This gives a powerful tool for establishing supersymmetry 
breaking~\cite{adssufive,adsdsb}: If a theory  has
a spontaneously broken global symmetry and no flat directions, 
the theory breaks supersymmetry.

We have assumed here that the additional massless
scalar corresponds to motions along a {\it non-compact} flat direction.
This is often the case, since, in supersymmetric theories,
the superpotential is invariant under the complexified global
symmetry, with the Goldstone boson corresponding to the imaginary part
of the relevant order parameter, and its supersymmetric partner
corresponding to the real part of the order 
parameter.\footnote{The possibility that the low-energy theory is a theory
of Goldstone bosons and massless chiral fields such 
that the supersymmetric scalar
partner of any Goldstone is a Goldstone too is ruled out.
Such theories can only be coupled to gravity for discrete
values of Newton's constant~\cite{bagwit}, and so 
cannot describe the low-energy
behavior of renormalizable gauge theories.}

In general, deciding whether a global symmetry is broken 
requires detailed knowledge of the potential of the theory,
and is at least as hard as 
determining whether the vacuum energy vanishes.
However, if a theory is strongly coupled at the scale at which
supersymmetry might be broken, one cannot directly answer
either of these questions.\footnote{The most obvious example is a theory
which does not possess any adjustable parameters, and has
only one scale, like an $SU(5)$ model 
of Section~\ref{examples}. }
Still, in some cases, one may argue, based on 't Hooft anomaly matching
conditions~\cite{thooft}, that a global symmetry is broken.

If a global symmetry is unbroken in the ground-state, then the
massless fermions of the low-energy theory should reproduce
the global triangle anomalies of the microscopic theory~\cite{thooft}.
Thus there should be a set of fields, with appropriate charges under 
the global symmetry, that give a solution to the anomaly-matching
conditions. 
This fact may be used when trying to determine
whether a theory confines, and how its global symmetries
are realized in the vacuum.
For example, if the gauge-invariants which can be constructed
out of the microscopic fields of the theory saturate the
anomaly-matching conditions for some subgroup of the global
symmetry of the microscopic theory, it is plausible that
the theory confines, and that the relevant symmetry subgroup
remains unbroken in the vacuum.
In contrast,
if all possible solutions to the anomaly-matching
conditions are very complicated, that is, they require
a large set of fields, it is plausible to conclude
that the global symmetry is spontaneously broken.

In the case of an R-symmetry there is another way to
determine whether it is spontaneously broken.
In many theories, the scale of supersymmetry breaking
is much lower than the strong-coupling scale, so that
supersymmetry breaking can be studied in a low-energy effective
theory involving chiral superfields only, with all gauge dynamics 
integrated out. In fact, the low-energy theory is an O'Raifeartaigh-like
model (with possibly negative exponents of the fields in the
superpotential arising from non-perturbative effects in the
microscopic description). 
In some cases it is easy to see that the origin is excluded
from the moduli space, because, for example,  the potential 
diverges there.
Then typically, some terms appearing in the superpotential
obtain vevs. 
Since all terms in the superpotential have R-charge 2 this
implies that R-symmetry is broken.
Obviously such an argument is not applicable to other
global symmetries because the superpotential is necessarily
neutral under non-R symmetries.
So it is often easier to prove that
an R-symmetry is broken, than to prove that a non-R symmetry
is broken. If the theory has no flat directions one can 
then conclude that supersymmetry is broken.

It is not surprising that R-symmetries, which do not  
commute with supersymmetry, should play a special role
in supersymmetry breaking. Let us discuss this role further,
following \cit{NS}.

In what follows we shall assume that the gauge dynamics was
integrated out.
Suppose we have a low-energy theory with a superpotential $W(\{X_i\})$,
where $X_i$ are chiral fields and $i=1\ldots n$. 
For supersymmetry to remain unbroken, the superpotential should be
extremal with respect to all fields,
$${\partial W\over \partial X_i} =0 \ .$$
If the theory has no symmetries, the number of unknowns, $X_i$, equals the
number of equations. Similarly, if the theory has a global symmetry that
commutes with supersymmetry, the number of equations equals the
number of unknowns.
To see this note that in this case, the superpotential
can only depend on chiral field combinations that are
invariant under the symmetry. Therefore, if there are $k$
symmetry generators, the superpotential depends on $n-k$
invariant quantities (for example, for a $U(1)$ symmetry
these could be $X_i/X_1^{q_i\over q_1}$, where $i=2\ldots n$,
and $q_i$, $q_1$ are the $U(1)$ charges of $X_i$, $X_1$ respectively)
while the remaining $k$ fields do not appear in the superpotential.
Thus, for supersymmetry to remain unbroken the superpotential
should be extremal with respect to $n-k$ variables, leading
to $n-k$ equations in $n-k$ unknowns.
Thus, generically, there is a solution and supersymmetry is unbroken. 

In contrast, suppose the theory has an $R$ symmetry 
that is spontaneously broken.
Then there is a field, $X$, with $R$ charge  $q \ne 0$,
which gets a non-zero vev.
The superpotential then can be written as
$$W = X^{2/q} f(Y_i=X_i^q/X^{q_i})\ ,$$
where $q_i$ is the charge of $X_i$. 
For supersymmetry to be unbroken we need
$${\partial f\over \partial Y_i} =0 \ ,$$
and 
$$f=0\ .$$
Thus there is one more equation than unknowns, and generically
we do not expect a solution. Roughly speaking, what we mean
by ``generically'' is that the superpotential is a generic
function of the fields, that is, it contains
all terms allowed by the symmetries. We shall return to this point
shortly.

If the extremum of the superpotential were determined by a
system of homogeneous linear equations, the above discussion
would lead us to conclude that an R-symmetry is a necessary
condition for supersymmetry breaking, and a spontaneously
broken R-symmetry is a sufficient condition. While this conclusion
generally holds
for theories in which the superpotential is a
generic function consistent with all the symmetries, 
there may be
exceptions to this rule. This is because the F-flatness
conditions are given by a system of non-linear equations
which may contain negative powers of fields
(arising from the dynamical
superpotential) as well as terms
independent of fields (arising from linear terms, either
generated dynamically or included in the tree level
superpotential).
Such a system of equations is not guaranteed to have solutions. 
In fact, we have already argued  in Section \ref{oraf} that
one can add explicit R-symmetry breaking terms to an
O'Raifeartaigh model without restoring supersymmetry. Later
we shall encounter other examples of supersymmetry
breaking models without
R-symmetry.

Let us make a bit more precise what we mean by a generic
superpotential. The superpotential contains two parts.
One is generated dynamically, 
and certainly does not contain
all terms allowed by the 
symmetries~\cite{seibergnr}. In particular,  such terms
could involve arbitrarily large negative powers of the fields.
The other part is the classical superpotential, which is
a polynomial (of some degree $d$) in the fields, that preserves
some global symmetry. 
Here what we mean by ``generic''
is that no term, with dimension smaller or equal to $d$
that is allowed by this global symmetry was omitted
from the superpotential.
On the other hand, the 
tree-level
superpotential can still be considered
generic if the operators with dimension higher than $d$ are
omitted. 
Indeed, in a renormalizable Lagrangian with a stable vacuum  
we do not expect Planck scale vevs. The analysis of
\cit{NS} shows that the inclusion of 
non-renormalizable operators can only produce additional
minima with Planck scale vevs, that is in a region of 
field space where our approximation of  global
supersymmetry is not sufficient anyway. On the other hand, in
models where a stable vacuum appears only after the
inclusion of  non-renormalizable terms of dimension $d$, 
the typical
expectation values
will depend on the Planck scale (or other large scale) 
but often will remain
much smaller than it.  
As long as the expectation values are small compared to the
Planck scale, these minima will remain stable local
minima even if operators of dimension higher than $d$ are added.

We shall encounter several examples of 
theories that break supersymmetry even 
though they do not possess an $R$ symmetry. 
In some cases, while the microscopic
theory does not have an $R$ symmetry, there is an effective, 
spontaneously broken  $R$ symmetry in the low energy theory.
In other cases, there is not even an effective $R$ symmetry.
In one example, SUSY will be broken even though the tree-level
superpotential is generic and does not preserve any $R$-symmetry,
and there is no effective $R$-symmetry.

\subsection{ Gaugino Condensation}
\label{gauginocond}

Let us now introduce a criterion for SUSY breaking which is based on 
gaugino condensation~\cite{mv,arv}.  
Suppose that a certain chiral superfield 
(or a linear combination of chiral
superfields) does not appear in the superpotential, yet all
the moduli are stabilized. In such a case the Konishi
anomaly \cite{konishi,cps} implies
\beq
\label{konishi1}
\bar {\cal D}^2 ({\bar \Phi} e^V \Phi) \sim {\Tr} W^2 \ ,
\eeq   
where $\cal D$ is a supersymmetric covariant derivative (see
Appendix \ref{notations}),
$\Phi$ is a chiral superfield
and $V$ is the vector superfield.
It is instructive to consider Eq. (\ref{konishi1}) in 
component form. It is given by an anomalous commutator with
the supersymmetry generator $Q$,
\beq
\label{konishi2}
\{Q, \psi_\Phi \phi\} \sim \lambda \lambda \ ,
\eeq
where $\psi_\Phi$ and $\phi$ are the fermionic
and scalar components of $\Phi$ respectively,
and $\lambda$ is the gaugino.
From this equation we see that the vacuum energy is
proportional to the lowest component of  $W^2$, that is,
to  $<\Tr \lam \lam >$. Therefore, if the gaugino
condensate forms one can conclude that supersymmetry is
broken. 
Note that if the fields $\Phi$ and ${\bar\Phi}$ appear in the
superpotential, the right-hand side of Eqs. (\ref{konishi1})
and (\ref{konishi2}) can be modified and the gaugino
condensate may form without violating supersymmetry. For
example, if there is a 
superpotential mass term, Eq. (\ref{konishi2})
becomes
\beq
\label{konishi3}
\{Q, \psi_\Phi \phi\} = m {\bar\phi} \phi 
+\frac{1}{32\pi^2} \lambda
\lambda \ .
\eeq
The latter equation is compatible with supersymmetry and
determines the vevs of the scalar fields in terms of the gaugino
condensate.

This criterion is related to the global symmetry
arguments of \cit{adsdsb},
since if a gaugino condensate develops in a theory possessing an R-symmetry,
this symmetry is spontaneously broken. In the absence
of flat directions, the Affleck-Dine-Seiberg argument leads
to the conclusion that SUSY is broken.

\subsection{Examples}    \label{examples}

We shall now demonstrate the techniques described in 
\ref{global} and \ref{gauginocond} by a few examples.

\subsubsection{Spontaneously Broken Global Symmetry: the
$SU(5)$ Model}

Consider an $SU(5)$ gauge theory with one antisymmetric tensor 
(${\bf 10}$) $A$,
and one anti-fundamental 
${\bar F}$~\cite{adssufive,mv}.
The global symmetry of the
theory is  $U(1)\times U(1)_R$, under which we can take the charges
of the fields to be $A(1,1)$ and ${\bar F}(-3,-9)$.
There are no gauge invariants one can make out of $A$ and ${\bar F}$.
Thus there are no flat directions, and classically the theory has a
unique vacuum at the origin. The theory is strongly coupled near the origin,
and we have no way to determine the behavior of the quantum theory. 
Because there are no chiral gauge invariants, the theory does not admit any 
superpotential. If supersymmetry is broken, the only possible scale for its
breaking is the strong coupling scale of $SU(5)$.

Following \cit{adssufive} we shall now use R-symmetry to argue
that supersymmetry is indeed broken 
(we shall consider a gaugino condensation argument due to \citename{mv},
\citeyear{mv} in the following subsection).
Assuming the theory confines, the massless gauge-invariant
fermions of the confined theory should  reproduce the  triangle
anomalies generated in the microscopic theory.
\cit{adssufive} showed that the minimal number of fermions required,
with $U(1)$ and $U(1)_R$  charges under 50, is five. This makes it
quite implausible that the full global symmetry remains unbroken.   
But if the global symmetry is spontaneously broken and there are no flat
directions, the theory breaks supersymmetry by the arguments of~\ref{global}.

\subsubsection{Gaugino Condensation: the $SU(5)$ Model}

We  now  would like to apply the gaugino condensate argument
to the $SU(5)$
model discussed in the previous subsection. 
Since the gaugino condensate serves as an order parameter for
supersymmetry breaking, we need to establish
that it is non-zero. To do that 
we follow \cit{mv}
and consider the correlation function
\beq
\Pi(x,y,z) =  \left < T(\lam^2(x),\lam^2(y),
\chi(z))\right> \ ,
\eeq
where 
\beq
\chi =  \epsilon^{abcde}\lam_a^{a^\prime}
\lam_{a^\prime}^{b^\prime} \Fbar^{c^\prime} A_{b^\prime
c^\prime}A_{b c} A_{d e} \ ,
\eeq
and $\Fbar$ and $A$ are scalar components of $\bar 5$ and $10$
respectively.

In the limit $x,y,z \ll \Lam^{-1}$ one can show that the
condensate $\Pi \sim \Lam^{13}$. Then one can take the limit
of large $x$, $y$, and $z$ and using cluster decomposition
properties argue that $<\lam \lam> \ne 0$. As a result
supersymmetry must be broken.

\subsubsection{R-symmetry and SUSY breaking}

In the example of the $SU(5)$ model we could not 
explicitly
verify
the spontaneous breaking of the global symmetry, and had to rely on 
the use of 't~Hooft anomaly matching conditions to establish
supersymmetry breaking.
Now we consider examples in which 
one can explicitly show that R-symmetry is spontaneously broken,
and to 
use that 
to establish supersymmetry breaking. 
While we shall be content to only consider models
with tree level breaking, similar arguments can be applied
to a number of dynamical models discussed later.

Consider first the O'Raifeartaigh model of Eq.~(\ref{orafw}).
There is an $R$ symmetry under which $\phi_1$, $\phi_2$ have
$R$-charge 2, and $\phi$ has $R$ charge zero. There is also a
discrete $Z_2$ symmetry under which $\phi_1$ is neutral, while
$\phi_2$ and $\phi$ change sign.
The superpotential is the generic one consistent with the symmetries
and supersymmetry is broken.
We can add to the superpotential $\phi$ to some even power so
that the $R$-symmetry is broken. Still, supersymmetry is
broken. However, the superpotential is no longer generic,
and we can add the term $\phi_1^2$ without breaking the remaining $Z_2$
symmetry. This term will restore supersymmetry. 

If we do not wish to impose a discrete symmetry the most
general superpotential is
\beq
W= \sum_{i=1}^2 M_i^2\phi_i +m_i\phi_i \phi +
\lam_i \phi_i \phi^2.
\eeq
The model with this superpotential still breaks supersymmetry
unless $M_1/M_2 = m_1/m_2 = \lam_1/\lam_2$. Note that if the
parameters are chosen so that this latter equality is satisfied
the superpotential is independent of one linear combination of the
fields, $\tilde \phi = (m_2 \phi_1 - m_1 \phi_2)/(m_1^2+m_2^2)$,
and thus is not generic.

As another example, consider the superpotential
\beq 
W = P_1 X_1 + P_2 X_2 + A(X_1 X_2 - \Lambda^2) + \alpha X_1 X_2\ .
\eeq
For $\alpha=0$ the theory has a $U(1)\times U(1)_R$ symmetry
with $A(0,2)$, $X_1(1,0)$, $X_2(-1,0)$, $P_1(-1,2)$, $P_2(1,2)$.
Supersymmetry is broken whether or not $\alpha=0$. For  $\alpha\neq0$
the $R$-symmetry is broken and the potential is no longer generic.
Terms such as $P_1 P_2$, $A^2$, which respect the $U(1)$ could restore
supersymmetry.
This example is a simple version of the low energy
theory of the example we shall study in section~\ref{noR}.

\subsubsection{ Generalizations of the $SU(5)$ Model}
\label{antisym-general}

To conclude our presentation of the basic examples of 
supersymmetry breaking we construct an infinite class of
models generalizing the $SU(5)$ model discussed earlier
\cite{adsdsb,mv}.
These models have an $SU(2N+1)$ gauge group, 
with matter transforming as an antisymmetric tensor $A$ and
$2N-3$ antifundamentals $\Fbar_i$, $i=1,\ldots, 2N-3$, \cite{adsdsb}.
For $N>2$, all these models have
D-flat directions, 
which are all lifted by the most general R-symmetry preserving superpotential
\beq
W= \lam_{ij} A \Fbar_i \Fbar_j\ ,
\eeq
for the appropriate choice of the matrix of coupling constants.

First, note that for small superpotential
coupling the model possesses almost flat directions, and as a
result, part of the dynamics can be analyzed directly. 
Yet  the scale of the unbroken gauge dynamics in the low
energy theory is
comparable to the scale of SUSY breaking and the model is
not calculable. 
To analyze  supersymmetry breaking it is convenient to start
from the theory without the tree level superpotential.
In this case the models have
classical flat directions along which the effective theory 
reduces to the $SU(5)$ theory with an antisymmetric and an
antifundamental as well as the light modulus parameterizing
the flat direction. 
We already know
that in this effective theory SUSY is broken with a vacuum energy
$\sim~\Lam_L^4$. The low energy scale depends on the vev of the
modulus as in Eq.~\ref{scalematching},
and thus there is a potential
\beq
\label{strongrun}
V(\phi) \sim \Lam_L^4 \sim \phi^{-\frac{4}{13}(4N-8)}\ .
\eeq
When small Yukawa couplings are turned on, the flat directions
are stabilized by the balance between the tree level contribution of
order $\lam^2 \phi^4$ and the dynamical potential (\ref{strongrun}).
The minimum of the potential then occurs for
\beq
\phi\sim \lam^{-\frac{13}{2(4N+5)}} \Lam~~~{\rm with}~~~~
E_{vac}\sim \lam^{8\frac{N-2}{4N+5}} \Lam^4\ .
\eeq
We found that the potential is stabilized at finite value of the
modulus and the effective low energy description is given in terms of
the SUSY breaking theory. Thus  supersymmetry must be broken in
the full theory.
We also note that at the minimum, R-symmetry is broken, giving us
additional evidence for supersymmetry breaking.

Before concluding this section
we comment on the analogous theories with $SU(2N)$ gauge
groups~(\cit{adsdsb}).
In this case the tree level superpotential allowed by
symmetries (including R symmetry) does not lift all the classical flat
directions. On the other hand, a dynamical superpotential is
generated, pushing the theory away from the origin. Thus the model does
not have a stable ground state. It is possible to lift all flat
directions by adding 
R-symmetry breaking terms to the
tree level superpotential. While lifting the flat directions, these
terms lead to the appearance of a stable supersymmetric vacuum.

\section{Direct Analysis: Calculable Models} \label{calculable}

As we have seen, SUSY breaking is directly related to the
zero-energy properties of the theory, namely, the ground-state
energy and the appearance of the massless Goldstino.
Fortunately, then, to establish SUSY breaking, 
we only need to understand the low-energy behavior of the theory
in question. 
As we shall see, many models can be described, in certain regions 
of the moduli space,
by a low-energy O'Raifeartaigh-like effective theory.
The question of whether SUSY is  broken simply amounts to the question
of whether all $F$-terms can vanish simultaneously. 
The tricky part, of course, is obtaining the correct low-energy
theory.
This involves a number of related ingredients: establishing the 
correct degrees of freedom, and determining the superpotential
and the \kahler potential.
In many cases, holomorphy and symmetries indeed determine the
superpotential, but the same is not true for the \kahler
potential. 
However, if all we care about is whether the energy vanishes or
not, it suffices to know that the \kahler potential is not singular
as a function of the fields that make up the low-energy theory.
This, in turn, is related to whether or or not we chose
the correct degrees of freedom of our low-energy theory.

We shall divide our discussion into two parts.
In this section we shall consider weakly-coupled theories.
By tuning some parameters in the superpotential to be very small,
we can typically drive some of the fields to large expectation
values, with the gauge symmetry completely broken, so
that  all gauge bosons are very heavy.
We can then neglect gauge interactions, and write down, as advertised,
a low-energy O'Raifeartaigh-type model.
Since the theory is weakly-coupled, we shall also be able
to calculate the \kahler potential, and thus, completely
determine the low-energy theory including the ground-state energy,
the composition of the Goldstino, and the masses of low-lying
states.

But we can also, in many cases, analyze the theory near the origin
in field space, where the theory is very strongly coupled.
If the theory confines, then below the confinement scale,
we are again left with an O'Raifeartaigh-type model.
In Section~(\ref{strong}), we shall see examples of this kind.
In some cases, though we shall not be able to analyze the
theory in question, we shall be able to analyze a dual theory,
that, as we just described, undergoes  confinement.
In all these cases, however, we shall not be able to calculate
the \kahler potential. 
Thus, while we shall ascertain that SUSY is broken, the details 
of the low-energy theory, and in particular the vacuum energy,
the unbroken global symmetry, and the masses of low-lying states
remain unknown.

In the examples we encounter, supersymmetry is broken due to a 
variety of effects.
Still, it is 
always  
the consequence
of the interplay between, on the one hand, 
 a tree-level superpotential, which
gives rise to a non-zero potential everywhere except at the
origin in field space, 
and, on the other hand, non-perturbative effects, either 
in the form of instantons or gaugino condensation, that generate
a potential that is non-zero at the origin.

\subsection{The 3--2 Model} 
\label{threetwo}

Probably the simplest model of \dsb is the 3--2 model of~\cit{adsdsb}. 
Here we shall 
choose the parameters of the model so that the low-energy
effective theory is weakly coupled, and thus the model is calculable. 
In this weakly coupled regime, the main ingredient
leading to supersymmetry breaking in the model is an instanton
generated superpotential.
In Section \ref{ITIY},
we shall analyze the same model in a strongly coupled
regime where SUSY is broken through the quantum deformation
of the moduli space
(which is again the result of instanton effects).
We shall also discuss numerous generalizations
of the 3--2 model.

The model is based on an $SU(3)\times SU(2)$ gauge group with
the following matter content (we also show charges under
the global $U(1)\times U(1)_R$ symmetry of the model):
\beq
\begin{array}{c|cccc}
&SU(3)  &SU(2) & U(1) & U(1)_R\\
\hline
Q &3       & 2& 1/3 & 1\\[-5pt]
{\bar u} & \bar 3 & 1& -4/3 & -8\\[-5pt]
{\bar d} & \bar 3 & 1& 2/3 & 4\\[-5pt]
L & 1      & 2& -1 & -3
\end{array}
\eeq
Using methods described in Appendix \ref{Dflatsection}
one can easily determine the classical moduli space of the
model.
In the absence of a tree level superpotential, it
 is given by
\beq
Q_{if}=\overline{Q}^{if}=
\left( \begin{array}{cc}
a&0\\[-5pt]
0&b\\[-5pt]
0&0
\end{array}
\right),
\ \ \ L=(0,\sqrt{a^2-b^2})\ ,
\eeq
where  $\overline{Q}=(\bar u, \bar d)$, and $i,f$ are
$SU(3)$ color and flavor indices respectively. For generic
values of $a$ and $b$ the gauge group is completely broken, and
there are three light chiral fields. While it is not difficult to
diagonalize the mass matrix and find the light degrees of freedom, it
is convenient to use
an alternative parameterization of
the classical moduli space in terms of
the composite operators
\beq
\label{threetwoinv}
X_1= Q_{i\alpha}\overline{d}^{i}L_\beta\epsilon^{\alpha\beta}, \ \
X_2= Q_{i\alpha}\overline{u}^{i}L_\beta\epsilon^{\alpha\beta}, \ \
Y=\det(Q\overline{Q})\ ,
\eeq
where greek indices denote $SU(2)$ gauge indices.
The most general renormalizable superpotential
which preserves the $U(1)\times U(1)_R$ global symmetry is
\beq
\label{threetwotree}
W_{tree}=\lam Q\bar d L\ = \lam X_1 \ .
\eeq
This superpotential lifts all classical
flat directions.

Let us now analyze the quantum theory. To do that we choose
$\lam^2 \ll g_2^2 \ll g_3^2$. The former inequality implies that
the minimum of the scalar potential lies very close to the D-flat
direction. To simplify the analysis we shall, in fact, impose
D-flatness conditions.~\footnote{However, 
in this and in other calculable models it is
easy to take D-term corrections to the scalar potential into
account in numerical calculations.} 
The latter inequality
guarantees that effects due to the $SU(2)$ non-perturbative
dynamics are exponentially suppressed compared to those due to the
$SU(3)$ dynamics. In particular, at scales below $\Lam_3$ and
much bigger that $\Lam_2$,
the $SU(2)$ gauge theory is weakly coupled and its dynamics
can be neglected.
$SU(3)$ on the other hand, confines, so
we can write down an effective theory in terms of its mesons
$M^f_\alpha = Q_\alpha\cdot {\bar Q}^f$,
subject to the non-perturbative superpotential,
\beq
\label{threetwonp}
W_{np}=\frac{2 \Lam_3^7}{\det(Q\overline{Q})}\ ,
\eeq
which is generated by an $SU(3)$ instanton.
The reader may now note that in this effective theory, $SU(2)$
appears anomalous; it has three doublets, $M^{f=1,2}$, $L$.
However, this is not too surprising, because the 
superpotential~(\ref{threetwonp}) drives the fields $Q$, ${\bar Q}$
away from the origin, so that $SU(2)$ is broken everywhere.
Indeed, as discussed in the appendix, an $SU(3)$ gauge theory with two
flavors has no moduli space.

In fact, we can already conclude that  supersymmetry is broken.
Since the  superpotential~(\ref{threetwonp}) drives the fields $Q$, $\bar Q$
away from the origin, the $R$-symmetry of the model is spontaneously broken.
Combining this with the fact the model has no flat directions,
we see, based on the arguments of section~\ref{global}, 
that the theory breaks supersymmetry.

Let us go on to analyze supersymmetry breaking in the theory in
more detail. As we saw above, in the absence of a tree-level superpotential,
the theory has a ``runaway'' vacuum with $Q, {\bar Q}\to\infty$.
This is precisely what allows us to find a calculable minimum
in this model. The tree-level superpotential lifts all classical 
flat directions. Any minimum would result from a balance between $W_{tree}$,
which rises at infinity, and $W_{np}$, which is singular at the origin.
If we choose $\lambda$ to be very small, the minimum would occur for large
$Q$, ${\bar Q}$ vevs, so that the gauge symmetry is completely broken,
and gauge interactions are negligible. 
Thus for a small $\lam \ll 1$
we can conclude that the light degrees of freedom can still
be described by the gauge invariant operators
$X_1$, $X_2$ and $Y$ (in the following we shall
see additional arguments supporting the
fact that $X_1$, $X_2$ and $Y$ are indeed the 
appropriate degrees of freedom).
Furthermore, the superpotential
in this limit is given by
\beq
\label{threetwofull}
W = 
\frac{2 \Lam_3^7}{Y} \ + \lam X_1\ .
\eeq
We now see explicitly that supersymmetry is broken, since
$W_{X_1} =\lam\neq 0$. Note that this conclusion depends crucially
on the fact the we have the full list of massless fields. In
general, one should be careful in drawing a conclusion about 
supersymmetry breaking based on the presence
of a linear term for a composite field in the superpotential.
If at some special points additional fields become massless,
the \kahler\ metric is singular, and the potential 
$V = W_i K^{-1}_{i j*} W_{j*}$ may vanish even if all $W_i$
are non-zero. Moreover, if the theory has classical flat
directions, it is possible that the \kahler potential
(written in terms of composites) has
singularities at the boundaries of moduli space, with some fields
going to infinity, and possibly others to the origin. As a result
supersymmetry may be restored at the origin.

As we saw above, for the choice of parameters $\Lambda_3 \gg \Lambda_2$,
$\lam \ll1$, the theory is weakly coupled. The \kahler\ potential
of the low energy theory 
is therefore the canonical \kahler\ potential in terms of the elementary 
fields, $Q$, ${\bar u}$, ${\bar d}$ and $L$, projected on the 
$D$-flat direction. In terms of the gauge invariant operators it
is given by~\cite{adsdsb,bpr}
\beq
\label{threetwokahler}
K=24 \frac{A+B x}{x^2},
\eeq
where $A=1/2 (X_1^\dagger X_1 + X_2^\dagger X_2)$, $B=1/3
\sqrt{Y^\dagger Y}$, and
\beq
x\equiv 4\sqrt{B} \cos\left(\frac{1}{3}
\rm{Arccos}
\frac{A}{B^{3/2}} \right).
\eeq 
We therefore have all the ingredients of the low energy theory,
including the superpotential and the \kahler\ potential.
This allows one to explicitly minimize the scalar potential.
For details of the analysis, we refer the reader to  \cit{adsdsb}
and \cit{bpr}. Here we just give some qualitative results.
It is possible to work in terms of either the elementary fields, or
 the gauge invariants fields. Simple dimensional analysis shows
that the minimum occurs for elementary field vevs 
$v \sim \lam^{-1/7} \Lambda_3$, and that the vacuum energy is 
of order $\lam^{5/14} \Lambda_3$. Explicitly one finds that
at the minimum $X_2=0$, so that the global $U(1)$ symmetry is unbroken
(not surprising, as points of higher symmetry are extremal).
The massless spectrum contains the Goldstino, a massless fermion 
of $U(1)$ charge $-2$, which saturates the $U(1)$ anomaly,
as well as a massless scalar which is the Goldstone boson of the broken
$R$-symmetry (usually known as the $R$-axion).

This concludes our discussion of the calculable minimum of the
3--2 model, but let us make a few more comments.

First, the above analysis of supersymmetry breaking did not
involve the strong dynamics of $SU(2)$. It is interesting to
see therefore the effect of turning off the $SU(2)$ gauge interactions.
We then have an $SU(3)$ gauge theory with two flavors,
plus two singlets $L_{\alpha=1,2}$, and with the 
superpotential~(\ref{threetwotree}).
Classically, this superpotential leaves a set of flat directions.
Up to global symmetry transformations
(which now include an $SU(2)$ global symmetry), these flat directions are
parameterized by $L_1$ and $Q_2 \bar u$.
The nonperturbative dynamics leads to runaway
towards a supersymmetric vacuum at infinity along this
direction.~\footnote{We shall discuss the quantum behavior of SUSY QCD
coupled to singlet fields in detail in Section \ref{sqcdsinglets}.}
This dangerous direction is no longer
$D$-flat when the $SU(2)$ is turned on.

Second, even though so far we concentrated on the limit 
$\Lam_3 \gg \Lam_2$,
it is possible to derive the exact superpotential of the 3--2 model
for any choice of couplings, and to use it to establish supersymmetry breaking.
Note first that the complete list of independent gauge invariants 
is $X_1$, $X_2$, $Y$ and $Z\equiv Q^3 L$ (we suppress all
indices). The latter vanishes classically, or more precisely
in the limit $\Lam_2 \ra 0$.
In the limit $\Lam_3\gg\Lam_2$, $\lam =0$, the superpotential is 
given by~(\ref{threetwonp}). In the limit $\Lam_2\gg\Lam_3$, $\lam =0$,
the theory is an $SU(2)$ gauge theory with two flavors and a
quantum constraint which can be 
implemented
in the
superpotential using a Lagrange multiplier $A$, as
$A (Z - \Lam_2^4)$.~\footnote{Actually,
with $SU(3)$ turned off, $Z$ in this superpotential should be understood
as the Pfaffian of the $SU(2)$ mesons.}
The most general superpotential  that respects all the symmetries of 
the theory is
\beq
\label{wderiv}
W = 
\frac{2 \Lam_3^7}{Y}\, f(t,z^\prime) \ + A (Z - \Lam_2^4)\, g(t,z^\prime) \ ,
\eeq
where $t \equiv \lam X_1 Y/\Lambda_3^7$, $z^\prime\equiv Z/\Lambda_2^4$
are the only dimensionless field combinations neutral under all symmetries.
In the limit $\Lambda_3, \lambda\to 0$ (for which any value of $t$
can be attained) we find an $SU(2)$ theory with 4 doublets (and a
set of non-interacting singlet fields). Since the exact
superpotential for this theory is known we find $g(t, z^\prime)
\equiv 1$. Similarly, in the limit $\Lam_2 \ra 0$ we find $f(t,z^\prime)=1+t$.
The exact superpotential is then
\beq
\label{wexact}
W = 
\frac{2 \Lam_3^7}{Y}\, + \,  A (Z - \Lam_2^4)\, + \, \lam X_1 \ .
\eeq
It is clear from this superpotential that $Z$ obtains a
mass. To see if the mass is large one needs to know the \kahler
potential for this field. In the limit of weakly coupled $SU(2)$
and $\lam \ll 1$
both gauge groups are strongly broken,
and the \kahler potential is close to the classical one. 
Since classically $Z$ vanishes, the projection of the
classical \kahler potential on it also vanishes 
(see Eq. (\ref{threetwokahler})).
For small, but non-vanishing $\Lam_2$ the
\kahler potential of $Z$ is suppressed by
some function of $\Lam_2/v$. Restoring the canonical
normalization for the kinetic term we find that the mass of $Z$
is enhanced by the inverse of this function. 
We were therefore justified in keeping only $X_1$, $X_2$ and $Y$ as the 
light fields.

Looking at~(\ref{wexact}), we can conclude that supersymmetry is broken for
any choice of the parameters of the theory.
However, unlike in the limit $\Lam_3\gg\Lam_2$, in general the 
theory is strongly coupled, and we have no control over the \kahler\
potential. 
Therefore, while we may be able to estimate the scale of supersymmetry breaking
we cannot say anything about the vacuum, {\it e.~g.} we cannot establish
the pattern of symmetry breaking.

In the above, we first showed that supersymmetry is broken in a specific
limit, and later realized that it is always broken. Indeed, we do not
expect a theory to break supersymmetry for some choice of parameters,
and to develop a supersymmetric minimum for other choices.
The reason is that no phase transitions are expected to occur 
in supersymmetric theories  as their parameters are 
varied~\cite{seiwit1,seiwit2,intseiphase}.
If a theory is supersymmetric for some choice of parameters, it remains
supersymmetric for any choice. This allows us to establish supersymmetry
breaking by considering a convenient limit.
In some theories, including the 3--2 model, we can establish supersymmetry
breaking in different limits.
The details of supersymmetry breaking, such as the vacuum energy
and the source of the breaking, may be very different in the different limits.
 
Finally, another interesting feature of the 3--2 model is the possibility
of gauging the  global $U(1)$ symmetry, provided that a new
field $E^+$, with  $U(1)$ charge $+2$, is added to cancel the
$U(1)^3$ anomaly. With the addition of this field, the  analysis of  \dsb
does not change since neither new classical flat directions appear
nor are new tree level superpotential terms allowed. This
possibility proved to be useful in phenomenological model
building \cite{dns}. For our purposes, however, the importance of
this $U(1)$ is in the observation \cite{dnns}
that with the addition of $E^+$, the matter content of the model
falls into  complete $SU(5)$ representations, and in
fact they are the same representations which are required for 
DSB in the $SU(5)$ model of Section~\ref{examples}. In the
following section we shall discuss another simple and calculable
model of DSB based on an $SU(4)\times U(1)$ gauge group and again see
that the matter fields form complete $SU(5)$ representations.
In Section~\ref{discarded} we shall introduce a method of
constructing large classes of DSB models based on this
observation. This method will lead us to an infinite class of
models generalizing 3--2 model. Many other calculable and
non-calculable generalizations will be discussed in 
Section~\ref{known}.

\subsection{The 4--1 Model}
\label{fourone}

Another example of a calculable DSB model is the 4--1 model
constructed by \cit{dnns} and \cit{PT1}. Consider an
$SU(4)\times U(1)$ gauge group with matter transforming as
an antisymmetric tensor of $SU(4)$, $A_2$ (where
the subscript indicates $U(1)$ charge), a fundamental, $F_{-3}$,
an anti-fundamental, $\overline F_{-1}$, and an $SU(4)$ singlet,
$S_4$.

For a range of parameters of the model, the scale of
the gauge dynamics will be below the SUSY breaking scale. Thus
one could analyze supersymmetry breaking in terms of
the microscopic variables \cite{dnns}. Indeed,  in terms of the microscopic
variables the \kahler potential of the light degrees of
freedom is nearly canonical and it is easy to calculate the
vacuum energy. We shall, however, 
analyze this model in terms of
the gauge invariant polynomials. 
Again for convenience we shall
work in a regime where the couplings are arranged hierarchically, with
the superpotential Yukawa coupling  the smallest,
and with the $U(1)$ coupling weak at the $SU(4)$ strong coupling
scale\footnote{Since both Yukawa and $U(1)$ couplings become weaker
in the infrared 
we can just
choose them to be 
sufficiently small in the ultraviolet.}.
The $SU(4)$ moduli space is given
by the fields $M=F\overline F$, 
$X={\Pf} A$,
and $S$. The model
possesses a non-anomalous R-symmetry, and the unique superpotential
allowed by the symmetries is
\beq
\label{Wfourone}
W=\frac{\Lam_4^5}{\sqrt{MX}}+\lam\, S\, M\ .
\eeq
The tree level term in the superpotential 
lifts all
classical flat directions (note that one more condition on the $SU(4)$
moduli is imposed by the $U(1)$ D-term). Due to the
non-perturbative superpotential the vacuum cannot lie in the
origin of the moduli space of the theory. As a result, the
R-symmetry is spontaneously broken at the minimum of the potential
and supersymmetry is broken.

Let us argue that the non-perturbative term in the superpotential
(\ref{Wfourone}) is indeed generated. We would also like to
establish that the model is calculable, namely, that for some
choices of parameters, the corrections to the classical \kahler
potential are small near the minimum.
To this end, neglect the tree level superpotential
and consider a region
of the classical moduli space with $M,S^2 \gg X$. In this region
the gauge group is broken down to an $SU(3)$ subgroup. Apart from
the light modulus which controls the scale of the unbroken gauge
group there is one $SU(3)$ flavor in the fundamental representation
coming from the components of $A$.
In this effective theory the non-perturbative superpotential is
generated by gaugino condensation
\beq
W=\frac{\Lam_3^4}{\sqrt{q\bar q}}\ ,
\eeq
where $q$ and $\bar q$ denote light $SU(3)$ fields, and
$\Lam_3^4$ is an $SU(3)$ scale. Using the scale matching condition
\beq
\Lam_3^4=\frac{\Lam_4^5}{\sqrt{M}}\ ,
\eeq
we easily recognize the superpotential (\ref{Wfourone}).

Furthermore, we see that the effective $SU(3)$ also possesses a
flat direction along which $q$ and $\bar q$ acquire vevs
and the gauge group is broken down to $SU(2)$. The strong scale of
this $SU(2)$ tends to zero along the flat direction. While the
tree level superpotential stabilizes the theory at finite vevs,
the corrections to the classical scalar potential which scale as
$\Lambda_2/v$ (where $v$ is the typical vev) are negligible for
sufficiently small $\lam$ and therefore the model is calculable. The vacuum
energy in this model was explicitly calculated by
\cit{dnns}.

It is worth noting that one can add an R-symmetry
breaking (and non-renormalizable) term, 
$M \Pf A$,
to the superpotential
(\ref{Wfourone})~\cite{PT1}. 
We shall discuss DSB models without R-symmetry in more
detail in Section \ref{noR}.

\section{Direct Analysis: Strongly-Coupled Theories} \label{strong}

\subsection{Supersymmetry Breaking Through Confinement}
\label{isssection}

In  previous sections we have seen that, at the classical level,
 supersymmetric gauge
theories without explicit mass terms possess a zero energy minimum
at least at the origin of  field space.
Classical tree level superpotentials may lift the moduli space, but
the supersymmetric vacuum at the origin survives.
In traditional calculable models of DSB (such as the 3--2 model)
the vacuum at the origin is lifted due to a dynamical superpotential
generated by non-perturbative effects. On the other hand, in
the $SU(5)$ model, no superpotential can be generated, and
supersymmetry is broken by the confining dynamics. Unfortunately,
the low energy spectrum of the $SU(5)$ model is not known, and thus
the main arguments for DSB are based on the complexity of the
solutions to the 't Hooft anomaly matching conditions. It would be
very useful to investigate a model in which supersymmetry breaking
is generated by the confining dynamics, with a known low energy
spectrum. In fact a very simple and instructive model of this type
was constructed by \cit{ISS}. 
This model clearly illustrates the fact that a crucial ingredient
in studying supersymmetry breaking is the knowledge of the
correct degrees of freedom of the low-energy theory. Supersymmetry 
breaking in this theory hinges on whether the theory confines,
or has an interacting Coulomb phase at the origin. It seems 
very plausible that the theory indeed confines, and that supersymmetry
is  broken as a result.

The model is based on an $SU(2)$ gauge
theory with a single matter field, $q_{\alpha\beta\gamma}$, in a
three-index symmetric representation. The model is chiral, since
the quadratic invariant, $q^2$, vanishes by the Bose statistics of
the superfields. It also possesses an R-symmetry
under which $q$ has the charge $3/5$. Moreover, the model is
asymptotically free, and thus non-trivial infra-red dynamics may
lift the supersymmetric vacuum at the origin of  field space.
It is, therefore,
a candidate model of DSB according to traditional criteria for
supersymmetry breaking.

The only non-trivial gauge invariant polynomial
which can be constructed out of $q$ is $u=q^4$ (with appropriate
contraction of indices). This composite parametrizes the only flat
direction of the theory along which the gauge group is completely
broken.
R-symmetry and holomorphy restrict any effective 
superpotential to be of the form $W=a \Lam^{-1/3} u^{5/6}$, where
$\Lam$ is the dynamical scale of the theory. This
superpotential, however, does not have a sensible behavior as
$\Lam\ra 0$, since for large $u/\Lam^4$ the moduli space should
be close to the classical one with $W\ra 0$. 
Therefore, $a=0$.
This means that the quantum theory also has a moduli space of
degenerate vacua.
The moduli space may be lifted
by the tree level non-renormalizable superpotential
\beq
\label{ISStree}
W=\frac{\lam}{M} u\ ,
\eeq
where $\lam$ is a constant of order 1.
In the presence of the nonrenormalizable term,
the model can be thought of
as a low energy effective description of a more fundamental
theory, which is valid below the scale $M$.
Choosing $\Lam \ll M$, there is a region of moduli space,
with $\Lam^4\ll u\ll M^4$, in which the
gauge dynamics is weak 
and we have a good description of the physics in terms
of an effective theory of chiral superfields.

In the presence of the non-renormalizable term,
holomorphy and symmetries restrict the exact superpotential 
to be
\beq
\label{ISSexactW}
W=\frac{\lam}{M} u f(t=\Lam^2 u /M^6)\ ,
\eeq
where the function $f$ is given by the sum of instanton
contributions. In the allowed region,
$\abs{t} \ll 1$, $f \approx 1$ and we can use the classical superpotential.

Naively, the linear superpotential for $u$ leads to  supersymmetry
breaking since $F_u\ne 0$. One should remember, however, that as $u$
is a composite field, its \kahler potential may be quite
complicated. In particular, the \kahler potential may be 
singular at some points in the moduli space, potentially 
leading to supersymmetry restoration.

To determine the behavior of the \kahler potential, consider
first the theory for
large expectation values of $u$. In this regime
the description of the
model should be semiclassical and thus the \kahler potential
scales as
\beq
\label{ISSKclass}
{\cal K} \sim Q^\dagger Q \sim (u^\dagger u)^{1/4}\ .
\eeq
Indeed this \kahler potential is singular at $u=0$.
The singularity reflects the fact that at $u=0$ the gauge bosons
become massless and must be included in the effective description.
There are two plausible alternatives for the nature of the
singularity in the quantum theory. It is possible that the theory
is in a non-abelian Coulomb phase. On the other hand,
it is possible that the singularity is
smoothed out quantum mechanically. \cit{ISS} argued that this
latter option is 
probably realized
since the massless
composite field $u$ satisfies the 't Hooft anomaly matching
conditions, which is quite 
non-trivial.\footnote{See, however, \cite{misleading}, for a class of theories
in which the existence of simple solutions to the anomaly matching conditions
suggests that the theories confine, yet the theories in fact do not
confine.}
We shall assume that this is indeed the case. 
Then R-symmetry, smooth behavior near
the origin and semiclassical
behavior at infinity imply that the \kahler potential is a
(smooth) function of $u^\dagger u/\abs{\Lam}^8$ satisfying
\beq
\label{ISSKexact}
K=\abs{\Lam}^2 k(u^\dagger u/\abs{\Lam}^8) \sim \cases
{u^\dagger u /\abs{\Lam}^6,~~~u^\dagger u \ll \Lam^8\cr 
(u^\dagger u)^{1/4}, ~~~~u^\dagger u \gg \Lam^8\ .\cr}
\eeq

Combining this form of the \kahler potential with the
superpotential of~Eq.(\ref{ISSexactW}) (with $f\equiv 1$)
 we find that the scalar potential
\beq
\label{ISSscalar}
V= (K_{u^\dagger u})^{-1}\abs{W_u}^2 = (K_{u^\dagger u})^{-1}
\abs{\frac{\lam}{M}}^2 \ ,
\eeq
necessarily breaks supersymmetry with a vacuum energy of order
\beq
\label{ISSenergy}
E\sim \frac{\abs{\Lam}^6}{M^2}\ .
\eeq

At this point we should comment on several other interesting
properties of the model. 
Before adding the tree-level superpotential~(\ref{ISStree}),
the effective description of the confined
theory in terms of the $u$ modulus possesses an accidental global
$U(1)$ symmetry (as the \kahler potential does not depend on the
phase of $u$).
This $U(1)$ is anomalous in terms of the elementary
degrees of freedom. The tree level superpotential explicitly
breaks the R-symmetry of the model, 
as well as the accidental $U(1)$.
However in the low energy
description there is an effective R-symmetry which is a
combination of the $U(1)_R$ and accidental $U(1)$ symmetries.
Since R-symmetry in the macroscopic description is explicitly
broken by the tree level superpotential, 
higher order terms can generically correct~(\ref{ISStree}).  These terms
explicitly violate the effective R-symmetry of the low energy
description. According to our analysis in section~\ref{global}
this leads to the appearance of  supersymmetric vacua, but
these vacua will lie outside the region of validity $\abs{u} <
M^4$ of our analysis, and the non-supersymmetric minimum will remain a
(metastable) local minimum of the potential.

\subsection{Establishing Supersymmetry Breaking through
a Dual Theory}
\label{dual}

In this section we shall encounter a class of theories,
with $SU(N)\times SU(N-2)$ gauge symmetry, that break
supersymmetry for odd $N$. By directly studying these
theories, we can show that they have
calculable, supersymmetry-breaking minima for a certain
choice of parameters. But we cannot show that there is
no supersymmetric minimum, simply because we cannot analyze
the low-energy theory
in all regions of the moduli space.
However, we shall be able to construct a Seiberg-dual of
the original theory, that can be reliably analyzed at
low-energy. As we shall see, the dual theory breaks
supersymmetry. We can then conclude that the original
theory breaks supersymmetry as well. The reason is, that
at least as long as supersymmetry is unbroken, the two duals
should have the same physics at zero energy. It is therefore
impossible for one of them to be supersymmetric, with a
vacuum at zero energy, and for the other one to break supersymmetry,
with non zero energy vacuum.

Furthermore, it is possible that the  two dual theories actually
agree not just at zero energy, but in a small, finite
energy window. 
In the theories at hand,
the scale of supersymmetry breaking is proportional
to some superpotential coupling, and can be tuned to be
small enough so that it is within this energy window.
It is important to stress however 
that we only use the dual
to establish that supersymmetry is broken. 
The details
of supersymmetry breaking may be different between 
the original theory and its dual.

While we mostly concentrate on the application of duality to
establish supersymmetry breaking, one could adopt a
different point of view, and use duality to construct
new models of DSB starting with known models. 
Generically new models constructed in such a way will
describe completely different, yet non-supersymmetric
infrared physics.

Let us turn now to our example. The  theory we start with
is an $SU(N)\times SU(N-2)$ ($N\geq 5$) gauge theory with 
fields
$Q_{i \alpha}$, transforming as $(N,N-2)$ under the
gauge groups, $N-2$ fields, $\bar{L}^i_I$, transforming as
$(\overline{N}, {\bf 1})$, and $N$ fields,  $\bar{R}^{\al}_A$,
that  transform as  $({\bf 1}, {\overline{N-2}})$.   
We denote the gauge indices of  $SU(N)$
and $SU(N-2)$  by $i$ and $\al$, respectively, while
$I = 1\ldots N-2$ and $A = 1 \ldots N$ are  flavor indices. Note that
these theories  are  chiral---no mass terms can be added for any of the matter
fields.

In the following, we shall only outline the main stages
of the analysis. For details we refer the reader to~\cite{PST2}. 
In particular, we omit numerical factors and some scale factors
throughout this section.

The classical moduli space of the theory is given by
the gauge invariants
$Y_{IA} = \bar{L}_I \cdot Q \cdot \bar{R}_A$,
$\bar{b}^{A B} = (\bar{R}^{N - 2})^{ A B }$
and
$\bar{\cal{B}} = Q^{N - 2} \cdot \bar{L}^{N - 2}$ (when appropriate,
all indices are contracted with  $\epsilon$-tensors),
subject to the classical constraints $Y_{I A} \bar{b}^{A B} = 0$
and
$\bar{b}^{A B}  \bar{\cal{B}}  \sim (Y^{N-2})^{A B}$. 

To lift all classical flat directions, we can add the superpotential,
\beq
\label{wtree}
W_{tree} = \lambda^{IA} ~ Y_{IA} + \alpha_{AB} ~\bar{b}^{AB}\ ,
\eeq
with $\lam_{IA}= \lam\, \delta_{IA}$ for $A\leq N-2$, and zero otherwise.
$\al_{AB}$ is an antisymmetric matrix, whose non-zero elements
are $\al_{12} = \ldots = \al_{(N-2)\, (N-1)} =\alpha$ for odd $N$,
and $\al_{12} = \ldots = \al_{(N-1) N} =\alpha$
for even $N$.~\footnote{In fact, 
one can lift all flat directions with other choices
for $\lambda^{IA}$ and  $\alpha_{AB}$, see~\cit{PST2}.}
Note that the second term in~(\ref{wtree}) is non-renormalizable
for $N\geq 6$, but has dimension four for $N=5$.

As it turns out, there is an important difference
between the theories with even and odd $N$. For odd $N$,
the superpotential~(\ref{wtree}) preserves an $R$ symmetry,
and one may expect supersymmetry to break. 
For even $N$, there is no $R$ symmetry that is preserved by~(\ref{wtree}),
so supersymmetry is most likely unbroken .
Both of these statements are indeed borne out
by direct analysis, as we shall see.

It is also easy to check that if we set $\al_{AB}=0$ in~(\ref{wtree}),
all flat directions are lifted, except for the ``baryon'' directions
$\bar{b}^{AB}$.

To analyze the quantum theory, we can start with the limit
$\Lam_N \gg \Lamp$, where $\Lam_N$, $\Lamp$ are the strong coupling scales
of $SU(N)$, $SU(N-2)$ respectively.
$SU(N)$ has $N-2$ flavors, so gaugino condensation in an unbroken $SU(2)$
subgroup generates the superpotential 
\beq
\label{wn1}
W \sim \left({\Lam_N^{2N+2}\over \bar{\cal{B}} }\right)^{1/2} \ .
\eeq
Thus there is no moduli space.
Below $\Lam_N$, $SU(N-2)$ appears
anomalous. It is also partially broken. This is reminiscent
of the situation we encountered in the 3--2 model.
However, there, because of the $SU(3)$ superpotential, the $SU(2)$
was completely broken. In contrast, here the $SU(N-2)$ is
not completely broken, so there is some strong dynamics associated with
the unbroken group. It is therefore very hard (or impossible)
to analyze the theory (except for a special choice
of parameters, for which it has a calculable minimum, as we shall see later). 
Fortunately, we can turn to a dual theory, in which the low-energy
dynamics is under control.~\footnote{The 
appearance of the superpotential~(\ref{wn1})
 can be seen in the dual theory as well~\cite{PST2}.} 

Before we do that, one comment is in order. It is already
clear from Eq. (\ref{wn1}) 
that the electric theory has no moduli space.
In addition, with $\alpha=0$, the theory has classical flat directions.
If these are not lifted quantum mechanically,
the superpotential (\ref{wn1}) pushes some fields to large
vevs along these directions.
Then for $\al \ll 1$ we can find a calculable minimum.
This indeed is the case. We shall return to this calculable
minimum towards the end of the section.
First, however, we would like to show that the theory has no supersymmetric
vacua. To do that, we turn to the dual theory.

We construct the dual theory in the limit $\Lamp \gg\Lam_N$.
However, it is expected to give a valid description of
the original theory in the infrared for any $\Lamp/\Lam_N$~\cite{PST1}.
The dual theory is obtained by dualizing the $SU(N-2)$. This can be thought
of as the process of first turning off the $SU(N)$ coupling, so
that we are left with an $SU(N-2)$ with $N$ flavors. Dualizing this
theory we find an $SU(2)$ theory with $N$ flavors. 
Finally, we switch the  $SU(N)$ coupling back on in this dual theory.

The dual theory then has $SU(N)\times SU(2)$ gauge symmetry,
with the following field content:
\beq
\begin{array}{c|cc}
&SU(N)  &SU(2)\\
\hline
q^i_\nu & {\overline N}       & 2\\[-5pt]
{\overline r}^{A\nu} &  1 &  2\\[-5pt]
{1\over\mu}M_{iA} & N &  1\\[-5pt]
{\overline L}^i_I & {\overline N}      & 1
\end{array}
\eeq
The $SU(2)$ singlets $M_{iA}$ correspond to the $SU(N-2)$
mesons $q_i\cdot R_A$, and $\mu$ is a mass scale that relates
the strong coupling scales of $SU(N-2)$ and $SU(2)$:
$\Lamp^{2N-6}~ \Lam_2^{6-N} \sim \mu^N$.

In addition, the dual theory has a Yukawa superpotential:
\beq
\label{wdual}
W = {1 \over \mu} ~ M_{\alpha A}~ \bar{r}^A \cdot q^{\alpha}\ .
\eeq

Note that in this dual theory, $SU(2)$ has $N$ flavors, so
naively it is in the dual regime. We shall soon see however that
the combination of the Yukawa superpotential~(\ref{wdual}) and the
$SU(N)$ dynamics, drives the theory into the confining regime.

To see that, note that $SU(N)$ now has $N$ flavors, and
therefore, a quantum modified moduli space.
Below the $SU(N)$ confining scale~\footnote{This scale is not
$\Lam_N$. Rather, it is a combination of  $\Lam_N$, $\Lamp$, 
and $\mu$~\cite{PST2}.},
we can write down an effective theory in terms of the $SU(N)$
mesons 
$N_{A \nu} \sim  M_{i A}\, q^i_\nu$ and  
$Y_{IA} \sim  M_{i A}\, \bar{L}^i_I$, and the $SU(N)$ baryons
$ \cal{B} \sim{\rm det} (M_{\alpha A})$ and
$$\bar{\cal{B}}^\prime\sim q^2 \cdot \bar{L}^{N-2}
 \sim ~Q^{N - 2} \cdot \bar{L}^{N - 2}
 \sim~\bar{\cal{B}}\ ,$$
where in the last equation we used the baryon map of SQCD,
Eq.~(\ref{mapping}).
Here we omit various scales as well numerical coefficients.

In terms of these variables, the $SU(2)$ still has $2N$ doublets,
$N_A$ and ${\bar r}^A$, but the superpotential~(\ref{wdual}) now turns into
\beq
\label{wdual1}
W \sim N_{A}\cdot \bar{r}^A \ ,
\eeq
which gives masses to all $SU(2)$ doublets. Thus, indeed,
as $SU(N)$ confines, the Yukawa couplings turn into mass terms
and drive $SU(2)$ into the confining regime.

Since $SU(2)$ is now confining, with mass terms
for all its doublets, we should work in terms of its
mesons, all of which obtain vevs. 
A convenient way of keeping track of the correct vevs
is to add the superpotential 
\beq
\label{pf}
W\sim \left( {{\rm Pf} V\over \Lam_{2 L}^{6-N} } \right)^{1\over N-2}\ .
\eeq
Here $\Lam_{2 L}$ is the $SU(2)$ scale after $SU(N)$ confines,
and $V$ stands collectively for the $SU(2)$ mesons
$[N^2]$, $[{\bar r}^2]$, $[N\cdot {\bar r}]$. We use brackets to indicate
that these mesons should be thought of as single fields now.

In addition, recall that the $SU(N)$ dynamics leads to a constraint
that can be implemented through the superpotential
\beq
\label{wn}
A\, \left([N^2]\cdot Y^{N-2} -  {\cal{B}} ~
\bar{\cal{B}} - {\bar\Lambda}_{N L}^
{2 N} \right) \ ,
\eeq
where $A$ is a Lagrange multiplier, and ${\bar\Lambda}_{N L}$ 
is the $SU(N)$ scale.

Combining (\ref{wn}), (\ref{pf}) and (\ref{wdual1}) with
the tree level superpotential, which now has the form
\beq
\label{wtree2}
\lam^{IA}\, Y_{IA} + \al_{AB}\, [{\bar r}^A \cdot {\bar r}^B] \ ,
\eeq
we have the complete superpotential.
Note that in the last step we used the SQCD baryon map
$\bar b^{AB}\sim {\bar r}^A \cdot {\bar r}^B$.

We now have a low energy field theory with all gauge dynamics
integrated out. This low energy theory consists of the
fields  ${\cal{B}}$, $\bar{\cal{B}}$, $Y_{IA}$, $[N^2]$, $[r^2]$
and $[N\cdot r]$, with a superpotential that is given by 
adding~(\ref{wdual1}) through~(\ref{wtree2}).
We can  check then  whether all $F$ terms  vanish
simultaneously. This is a rather tedious task, and we refer
the interested reader to~\cite{PST2}. As the analysis  
shows, for odd $N$, no solution exists, and supersymmetry is broken.
For even $N$ a solution does exist.

It is interesting to see what happens before adding the
tree-level superpotential. In that case, the $F$ equations have
no solution for finite field vevs. Furthermore, for $\al=0$
and $\lam\neq0$, the different F-terms tend to zero as some
of the baryons ${\bar b}^{AB}$ tend to infinity.

One difficulty which we glossed over in the above discussion
is related to the fact that after $SU(N)$ confines, the $SU(2)$
scale $\Lam_{2 L}$ is field dependent. If this scale vanishes,
additional fields may become massless, and the \kahler\ potential
in terms of the degrees of freedom we kept so far may become
singular.
To resolve this issue, one can add a heavy $SU(N)$ flavor.
In this case, no scale is field dependent, and the analysis
confirms the results stated above. In particular, one 
finds that supersymmetry is broken for odd $N$. For further
details see \cite{PST2}.

To summarize, while we could not in general analyze the
original $SU(N)\times SU(N-2)$ theory, we were able to show
that it breaks supersymmetry for odd $N$ by studying
its dual $SU(N)\times SU(2)$ theory.
 
To complete our discussion of this theory, we now
turn to the calculable minimum we mentioned earlier.
This minimum can be studied in the electric theory itself,
so duality plays no role in the analysis.

As we already mentioned, with $\al=0$, all $F$ terms
asymptote to zero along the baryonic flat direction.
Let us now see this in the electric theory.
We choose a particular baryon direction, with $R^i_A = v\, \delta^i_A$.
This corresponds to ${\bar b}^{(N-1) N} \sim v^{N-2}$, with all
other ${\bar b}^{A B} = 0$.
Along this direction, $SU(N-2)$ is completely broken,
so for large $v$ we can neglect its effects.
Furthermore, the first term in~(\ref{wtree})
gives masses $\lam v$ to all $SU(N)$ flavors (in the following we
set $\lam=1$ for convenience).
At low energies we are thus left with a pure $SU(N)$,
whose scale $\Lam_{N L}$ satisfies  
$\Lam_{N L}^{3 N} \sim v^{N-2}\, \Lam_N^{2N+2}$.
Gaugino condensation in this theory then leads
to a superpotential
\beq
\label{ncond}
W \sim  \left(v^{N-2}\right)^{1/N} \sim  \left(\bar b^{(N-1) N}\right)^{1/N}\ .
\eeq
We thus have a low energy theory in terms of the baryons ${\bar b}^{AB}$,
with the superpotential~(\ref{ncond}) (for $\al=0$),
so the $F$ term for $\bar b^{(N-1) N}$ behaves as 
$F\sim {(\bar b^{(N-1) N})}^{{1\over N} -1}$ which goes to zero as
$\bar b^{(N-1) N}\to \infty$.
But whether this is a runaway direction or not depends on the \kahler\
potential. In fact, it can be argued~\cite{shirmanflat} that
the Kahler potential is canonical in terms of the elementary
fields ${\bar R}^A$, up to small corrections.
Thus, if the $F$ terms for ${\bar R}^A$ tend to zero along
this direction, there is a runaway minimum at infinity.
Indeed, these $F$ terms behave as $v^{{N-2\over N}-1} = v^{-2/N}$
and asymptote to zero as $v\to \infty$.

Adding now a small $\al\neq0$, the potential can be stabilized
as $v\to\infty$, with a supersymmetry-breaking minimum
for large values of $v$. Note that for $N\geq 6$, the
baryon term in~(\ref{wtree}) is non-renormalizable,
so $\al$ is naturally small.
In fact, as was shown in~\cite{poptrisigma}, this minimum
can be analyzed using a simple $\sigma$-model approach,
and is interesting for model building purposes, as
there is a large unbroken global symmetry at the minimum,
in which, a priori at least, one can embed the standard-model 
gauge group.

\subsection{Integrating Matter In and Out}
\label{inandout}

In Section~\ref{examples} we discussed the
supersymmetry breaking $SU(5)$ model 
with matter in the antisymmetric tensor and in the antifundamental
representations. This model does not possess any classical or
dynamical superpotential and does not have flat
directions. 
We gave two
arguments establishing DSB.  One was 
based on the complexity of solutions
to 't Hooft anomaly matching conditions, while the other was
based on the formation of the gaugino condensate.
We also discussed generalizations of the $SU(5)$ model.

Here we shall use the same class of theories to illustrate another
method of analysis which is useful in non-calculable
models \cite{murnoncalc,PT1}.
In this method one modifies the model of interest to make it
calculable through the introduction of extra vector-like
matter. When these
vector-like matter fields are
massless, the models
typically posses flat directions along which the gauge group
is broken and the theory is in a weak coupling regime. 
For small masses of these matter fields
the weak coupling
approximation is still reliable and the theory remains
calculable.  
Thus, the modified theory with small masses allows a direct 
analysis of supersymmetry breaking.
Then one considers the limit 
of infinite vector-like matter mass.
In this
limit the vector-like matter decouples and one is left with
the original theory. 
If  supersymmetry is broken in the
modified theory with the additional light fields, 
holomorphy arguments ensure that it is broken for any finite
values of masses. Moreover, since the theories that we have in
mind do not have classical flat directions  both for finite
and infinite masses, we expect that the asymptotic
behavior of the scalar potential (and therefore, the Witten
index) remain unchanged in the infinite mass limit.
Thus the assumption 
that no phase transition occurs when going to the infinite mass
limit, 
leads us to the conclusion that supersymmetry is
broken in the original strongly coupled model. 
We stress that this approach gives strong evidence for
supersymmetry breaking, yet does not help in
understanding the strongly coupled SUSY breaking
vacuum. The reason is that as the mass of the vector-like matter
becomes large, $m \sim \Lambda$,  control of the \kahler
potential is lost and the theory becomes non-calculable.

We now discuss the application of this method to the
models at hand, following \cit{PT1}. We consider models with
an $SU(2N+1)$ gauge group, an antisymmetric tensor
$A_{\alpha\beta}$, $2N-3+N_f$
antifundamentals $\Qbar_i^\alpha$, ($i=1,\ldots, 2N-3+N_f$),
and $N_f$ fundamentals $Q_\alpha^a$, ($a=1,\ldots, N_f$). It
is convenient to start with the case $N_f=3$, and then
to integrate out vector-like matter. The classical moduli
space is described by the following gauge invariant
operators
\beq
\label{asymmod}
\begin{array}{l}
M_i^a = \Qbar_i^\alpha Q_\alpha^a \, , \\
X_{ij}= A_{\alpha\beta} \Qbar_i^\alpha \Qbar_j^\beta \, ,\\
Y^a = \epsilon^{\alpha_1,\ldots,\alpha_{2N+1}}
A_{\alpha_1,\alpha_2}\cdots A_{\alpha_{2N-1}\alpha_{2N}}
Q^a_{\alpha_{2N+1}} \, ,\\
Z= \epsilon^{\alpha_1,\ldots,\alpha_{2N+1}}
A_{\alpha_1\alpha_2} \cdots A_{\alpha_{2N-3,2N-2}}
Q_{\alpha_{2N-1}}^a Q_{\alpha_{2N}}^b Q_{\alpha_{2N+3}}^c
\epsilon_{abc} \, .
\end{array}
\eeq
These moduli overcount by one the number of
the massless degrees of freedom at a
generic point of the moduli space, and thus are related by a
single constraint which easily follows from the Bose
statistics of the superfields
\beq
\label{asymclcon}
Y \cdot M^2 \cdot X^{N-1} - \frac{k}{3} Z \Pf X = 0 \, ,
\eeq
where appropriate contraction of indices is assumed.
Vevs of the moduli (\ref{asymmod}) satisfying the constraint
(\ref{asymclcon}) describe non-equivalent classical vacuum
states. The \kahler potential of the theory written in terms
of the gauge invariant composites is singular at the
origin. As usual this singularity reflects the fact that the gauge
symmetry is restored at the origin of the moduli space and
additional massless degrees of freedom descend into the low
energy theory. In complete analogy with SQCD (see
Eq. (\ref{quantumconstraint})) this constraint is modified by
non-perturbative effects,
\beq
\label{asymqcon}
Y \cdot M^2 \cdot X^{N-1} - \frac{k}{3} Z \Pf X =
\Lam^{4N+2} \, .
\eeq
As a result, the origin of field space where the gauge
symmetry is completely restored does not belong to the
quantum moduli space. The \kahler potential is
non-singular in any finite region of the moduli space, and
we have good control of the physics. We note in passing
that the \kahler potential may still become singular at the
boundaries of the (D-flat) moduli space, where  a subgroup of the
original $SU(2N+1)$ gauge group remains unbroken
(corresponding to the situation with some moduli vevs
vanishing while other  vevs tend to infinity). We shall
carefully  consider models in which the physics in such boundary
regions is important in Section \ref{flat}. For the time being
we note that as long as the
classical superpotential of the theory lifts all flat
directions, such boundary regions are not accessible and we
do not need to worry about them. 

Having understood the properties of the moduli space we turn on
the tree level superpotential. The full
superpotential of the theory with three vector-like flavors
can be written as
\beq
W_{3} = L ( Y \cdot M^2 \cdot X^{k-1} - \frac{N}{3} Z \Pf X -
\Lam^{2(2N+1)}) + m_a^i M_i^a + \lam^{ij} X_{ij} \ ,
\eeq
where $m_a^i$ is a rank three mass matrix, and the matrix of Yukawa
couplings $\lam$ is chosen so that all classical flat directions
are lifted. We can now vary the masses $m_a^i$ to move in the parameter space
between the $N_f=3$ and $N_f=0$ theories. However, it is useful to
first choose only one mass eigenvalue to be large, so that the
effective description is of two light vector-like flavors.
In such a case the superpotential takes
the form
\beq
\label{twoflavor}
W_{2} = \frac{\Lam^{4N+3}_{(2)}}{\epsilon_{ac} Y^a
M^c_{i_1}\epsilon^{i_1\ldots i_{2N-1}} X_{i_2 i_3}\cdot
X_{i_{2N-2} i_{2N-1}}} + m_a^i M_i^a + \lam^{ij} X_{ij} \ ,
\eeq
where the low energy scale $\Lam_{(2)}$ is given by the usual
scale matching condition, $\Lam_{(2)}^{4N+3}= m \Lam^{4N+2}$ and
the tree level terms only include fields of the $N_f=2$
model. We note that the
nonperturbative term in Eq. (\ref{twoflavor}) is generated by a
one instanton term in the gauge theory. By solving the equations of
motion for the mesons $M_i^a$ it is easy to see that the
F-flatness conditions can not be satisfied. Together with the
regularity of the \kahler potential in any finite region of the
moduli space and the absence of classical flat directions, this
implies supersymmetry breaking (\cit{PT1}).
We note that for small masses $m \ll \Lam$ and couplings
$\lam \ll 1$ the theory is in a semiclassical regime and the
low energy theory
is calculable. As the masses are increased,
control of the \kahler potential, and as a result, calculability,
are lost, yet supersymmetry remains broken.
For large masses, $m \gg \Lam$, the effective description
is given by the $N_f=0$ models, so that we have given 
an additional
argument
for supersymmetry breaking in these noncalculable theories.

\section{Violations of Indirect Criteria for DSB}
\label{exceptions}

So far we have concentrated on models satisfying 
the~\cit{adsdsb} criteria for dynamical SUSY breaking.
These criteria restricted model building efforts to chiral
models with $R$ symmetries and with no flat directions.
In recent years a number of 
non-chiral models, models with classical flat directions,
and models with no R-symmetry,
have been shown to break supersymmetry dynamically. 
In this section we shall discuss such examples in turn.

\subsection{Non-chiral Models}
\label{nonchiral}

\subsubsection{SUSY QCD with Singlets}
\label{sqcdsinglets}

We shall start the discussion of  non-chiral models
with SUSY QCD coupled to gauge singlet fields. 
We shall vary the number of
flavors in the theory and analyze the quantum behavior along the
classical flat directions.
We should warn the reader that generically these models do not
break supersymmetry. However, 
this analysis will lead us to the non-chiral
ITIY model of DSB \cite{IT,IY} discussed in the
following subsection. Along the way we shall develop
useful techniques for the analysis of flat directions and
illustrate them in additional examples in section~\ref{flat}.

Consider an $SU(N)$ gauge
theory with $N_f$ flavors coupled to a single gauge
singlet field through the superpotential
\beq
W = S \,Q_i \cdot\Qbar_i.
\eeq
This superpotential lifts one classical flat direction of
SUSY QCD, namely, $M_{ij}=v \delta_{ij}$. On the other
hand, there is a flat direction along which $S$ is
non-vanishing. Along this direction the vev of $S$ plays the 
role of a mass for the quark superfields~\footnote{Note that
unlike the case with a tree level mass term 
the superpotential preserves a nonanomalous R-symmetry,
which is only broken spontaneously
by the $S$ vev.}. 

For large $S$ the effective theory is
pure SYM with an effective strong coupling scale
$\Lam_{SYM}^{3N}=S^{N_f} \Lam^{3N-N_f}$, where $S$ denotes the
expectation value, and $\Lam$ is the original $SU(N)$ scale. 
Gaugino condensation in the effective
theory generates the superpotential
\beq
\label{oneS}
W=\Lam_{SYM}^3=S^{N_f/N} \Lam^{3N-N_f}.
\eeq
The superpotential (\ref{oneS}) gives an effective
description at scales much smaller than $\vev{S}$, yet the
fluctuations of $S$ itself remain massless, and (\ref{oneS})
can be considered as an effective superpotential for this
modulus, leading to the scalar potential
\beq
\label{oneSV}
V=\Lam^{2\frac{3N-N_f}{N}}\abs{S}^{2\frac{N_f-N}{N}}.
\eeq
Note that this effective description is only valid for $S\gg \Lam$.
We see that for $N_f < N$ this potential slopes to zero at
infinity, and the vacuum energy is arbitrarily small for large
$S$, exactly in the region where our effective description
is reliable.
For $N_f \ge N$ the potential
for $S$ is non-vanishing at infinity~\cite{adsdsb}. Of course, the
stabilization of this direction in the case $N_f \ge N_c$ 
does not imply supersymmetry
breaking (or even the existence of a stable vacuum) in the
full model. First, the analysis performed so far is
not valid near the origin of field space. In addition 
there are many unlifted mesonic and
baryonic flat directions. Yet, this suggests a way to
stabilize
other flat directions. Namely, one could couple the quarks
to $N_f^2$ gauge singlet fields~\footnote{In this case,
the singlets transform under the global $SU(N_f)_L \times SU(N_f)_R$
group, and the chiral symmetry is preserved by the
superpotential. It is spontaneously broken by the $S_{ij}$
vevs.}
\beq
\label{manyS}
W=\sum_{ij}^k \lam_{ij} \, S_{ij}\, Q_i\cdot\Qbar_j\ ,
\eeq
where the matrix of Yukawa coupling constants has
maximal rank, and in the following we shall choose it to 
be
$\lam_{ij}=\lam\, \delta_{ij}$.~\footnote{Recall that the only
fact we need to know about
the \kahler potential to establish \susy
breaking is that it is non-singular in the appropriate variables.
As a result one can further rescale
$\lam$ to one by field redefinitions,
but it would be useful for us to keep it explicit.}
The superpotential (\ref{manyS}) lifts all
mesonic flat directions $M_{ij}= Q_i \Qbar_j$. 
If baryonic branches of the moduli
space exist, they can be lifted by introducing additional
nonrenormalizable couplings to singlets, but we shall leave
these directions aside for the moment. 

Along the singlet
flat directions all quark superfields generically become
massive, and the effective theory is pure SYM with the
superpotential
\beq
\label{manySeffective}
W = \lam^{\frac{N_f}{N}}\, \tilde S^{\frac{N_f}{N}}\, 
\Lam^{\frac{3N-N_f}{N}}\ ,
\eeq
where $\tilde S = (\det S)^{\frac{1}{N_f}}$. 
This is just a direct generalization
of Eq.~(\ref{oneS}), and we see that the flat direction is
stabilized quantum mechanically for $N_f \ge N$. 
A somewhat more careful analysis would show that the
stabilization happens for all directions $S_{ij}$ as we
shall see below.

To better understand the
quantum behavior of the model, 
we shall repeat the above analysis in more detail.
We write the scalar
potential in the form~\footnote{This potential, of course, is
further modified by corrections to the \kahler potential.}
\beq
\label{singletQCDV}
V= \sum_i^{N_f}\left( \abs{\frac{\partial W}{\partial Q_i}}^2 +
\abs{\frac{\partial W}{\partial \Qbar_i}}^2 \right)+
\sum_{ij}^{N_f}\abs{\frac{\partial W}{\partial S_{ij}}}^2,
\eeq
where $W$ includes all possible nonperturbative
contributions. 
A supersymmetric minimum in the model exists if all three
terms in (\ref{singletQCDV}) vanish. 
The first two contributions in this
potential reproduce the scalar potential of SUSY QCD
with $N_f$ flavors and with the mass matrix
$m_{ij}= \lam\, S_{ij}$. We can,
therefore use Eq.~(\ref{sqcdsolution}) of the Appendix 
to find the meson expectation values for which these terms vanish
\beq
\label{sqcdsolutionS}
M_{ij} = \left( \det (\lam S)\, \Lam^{3N-N_f}\right)^{1/N} 
\left(\frac{1}{\lam S}\right)_{ij}.
\eeq 
Note that analyticity requires that (\ref{sqcdsolutionS}) is
satisfied for all values of $N_f$.
We can now substitute this solution back into
(\ref{singletQCDV})
\beq
\label{manySV}
V= \sum_{ij}^{N_f}\abs{\frac{\partial W}{\partial S_{ij}}}^2
= \abs{\lam}^2 \,\sum_{ij} \abs{M_{ij}}^2 = 
\abs{\lam}^{2\frac{N_f}{N}} \abs{ \det (S) \,\Lam^{3N-N_f}}^{2/N}
 \sum_{ij} \abs{\left(\frac{1}{S}\right)_{ij}}^2.
\eeq
It is easy to see that this term is minimized by $S_{ij} =
\tilde S \delta_{ij} \equiv 
\left (\det S \right)^{1/N_f} \delta_{ij}$. 
Therefore the scalar potential for the lightest modulus
$\tilde S$
is 
\beq
V =\abs{\lam^{N_f}\, \Lam^{3N-N_f}\,\tilde S^{N_f-N}}^{\frac{2}{N}}.
\eeq
This is just the potential which could be derived from
Eq. (\ref{manySeffective}) and we again see that the flat
direction is lifted if $N_f\ge N$.
In fact now we can make a stronger statement. 
When $N_f=N+1$ the model is {\it s}-confining~\cite{css3}, 
and near the origin has  a weakly coupled
description in terms
of composite (mesonic and baryonic) degrees of freedom. 
As a result the potential~(\ref{manySV}) is reliable near
the origin, and we see that supersymmetry is restored
there. When $N_f > N+1$ the weakly coupled description is
given in terms of the dual gauge theory theory, and it 
is also possible to show that a
supersymmetric vacuum exists at the origin.

The most interesting
case for our purposes is $N_f=N$, where the vacuum energy is
independent of the value of $\tilde S$ in the approximation that
the \kahler potential is classical. This statement is equivalent to
the statement that the energy is constant and non-vanishing
everywhere on the mesonic branch of the moduli space.
So far we have not considered the baryonic flat directions. In
fact, in the model with $N_f=N$ flavors and the
superpotential~(\ref{manyS})  the potential slopes to zero
along the baryonic directions.
However, it is easy to see that a simple modification 
leads to DSB \cite{IT,IY}. This modification requires the
introduction of two additional gauge singlet fields with
nonrenormalizable couplings to the $SU(N)$ baryons (in the $N=2$
case the new couplings are renormalizable).

\subsubsection{The Intriligator-Thomas-Izawa-Yanagida Model}
\label{ITIY}

Let us concentrate on a particular case with $SU(2)$ gauge group
with two flavors of matter fields in the fundamental
representation (four doublets $Q_i$, $i=1,\ldots, 4$). 
Because the matter fields
are in the pseudoreal representation the
superpotential~(\ref{manyS}) with $N_f^2$ singlets does not
lift all the mesonic flat directions.~\footnote{Note that the
superpotential~(\ref{manyS}) does not preserve the global
$SU(4)_F$ symmetry of the $SU(2)$ with four doublets.} 
Two mesonic flat
directions remain and lead to a supersymmetric minimum at
infinity in direct analogy with the baryonic flat
directions for general $N$.
A slight modification
of the theory with $N_f^2+2=6$ singlets lifts all mesonic
flat directions
\beq
\label{ITtree}
W=\sum_{ij}\, \lam S_{ij}\, M_{ij}\ ,
\eeq
where $M_{ij}=Q_i\cdot Q_j$, and $S_{ij}$ transform in the
antisymmetric representation of the global $SU(4)_F$ 
symmetry.~\footnote{For simplicity we choose
$\lam$ to respect global $SU(4)$ symmetry, 
but this is not necessary.}
Furthermore, we notice that near the origin of the moduli space, 
the theory has a weakly coupled
description in terms of the mesons, $M_{ij}$.
Thus our preceding discussion immediately leads us to the conclusion
that supersymmetry is broken.

Let us understand qualitatively the mechanism of
supersymmetry breaking. The nonperturbative dynamics 
generates the quantum constraint
\beq
\label{pfaffian}
\Pf(M)=\Lam_2^4\ .
\eeq
This quantum constraint modifies the moduli space. While
the origin 
$M_{ij}=0$ belongs to the classical moduli space, it does not
lie on the quantum moduli space. On the other hand,
the $S$ $F$-terms
only vanish at
the origin, $M_{ij}=0$. Supersymmetry is therefore broken
because the F-flatness conditions are incompatible with the
quantum moduli space.

It is often convenient to impose the quantum constraint
through a Lagrange multiplier in the
superpotential. Then the full
superpotential is
\beq
\label{ITexact}
W=\lam\, S\, M\, + A\, (\Pf M -\Lam_2^4)\ ,
\eeq
where $A$ is the Lagrange multiplier field. For $\lam \ll 1$
the vacuum will lie close to the 
$SU(2)$ quantum moduli space. Thus one can consider the
superpotential~(\ref{ITtree}) as a small perturbation around the
vacuum of the $N_f=N_c$ SQCD with masses
$m=\lam \vev{S}$.~\footnote{This approximation is equivalent to
satisfying the
equation of of motion for the Lagrange multiplier~$A$.}
In this approximation the scalar potential is given again by
(\ref{manySV}) and is minimized when (up to
symmetry transformations)
\beq
\begin{array}{c}
\tilde S = S_{12}=S_{34}\ ,\\
S_{13}=S_{14}=S_{23}=S_{24}=0\ ,\\
M_{12}=M_{34}=\frac{1}{\lam \tilde
S}\left(\frac{\Lam_2^4}{\Pf(\lam S)}\right)=\Lam_2^4\ .
\end{array}
\eeq
The vacuum energy is then given by
\beq
\label{ITenergy}
V=\abs{\lam}^2 \Lam_2^4\ .
\eeq

Thus, we have a non-chiral (left-right symmetric)
model~\footnote{Practically, what is usually meant by a 
non-chiral model is that all fields can be given masses.
This issue is a bit subtle in the ITIY model, as 
quark mass terms can be absorbed by a redefinition of
the singlets. Still, one may first give masses to the singlets
and integrate them out, and then introduce quark masses, 
so that ultimately all fields become heavy.} 
which breaks supersymmetry!
Indeed, as we mentioned in Section~\ref{witten}
the Witten index can change discontinuously if the asymptotic
behavior of the classical potential changes. Consider 
modifying the ITIY model by turning on a mass term for the
singlet,
$m\, S^2$. For sufficiently large mass, the low-energy effective
theory is pure SYM, and the Witten index ${\rm Tr}(-1)^F\ne0$.
This is therefore true for any non-vanishing value of $m$.
As the limit $m \ra 0$ is taken, the asymptotic behavior of the
potential changes (there is now a 
classical
flat direction with $S\ne 0$)
and the Witten index vanishes.
Note that, in accord with our discussion of section~\ref{witten},
the potential~\ref{ITenergy} is flat along the $S$ flat direction.

At the level of analysis we performed so far, there is a
pseudo-flat direction parameterized by $\tilde S$.
Since $\tilde S$ is the only light field in the low energy
theory and  
the superpotential (\ref{ITexact}) is exact,
this direction would be
exactly flat if the \kahler potential for $S$ were canonical.
However, quantum contributions to the \kahler potential lift the
degeneracy. For sufficiently small $\lam$ and large $\lam\, \vev{S}$
it is possible to show \cite{MA-H,DDGR} by renormalization
group arguments that the quantum
corrections due to the wave-function renormalization of $S$ are
calculable and 
lead to a logarithmic growth of the potential at
large $S$. 
It is possible to construct modifications of the
ITIY models with {\it calculable} (but not necessarily
global) supersymmetry breaking \cite{murinvert,DDGR}. This
is achieved by gauging a subgroup of the global symmetry
under which $S$ transforms. As a result, the wave function
renormalization of $S$ as well as the vacuum energy depend on
both the Yukawa and gauge coupling. For an appropriate
choice of parameters, a local minimum of the potential
exists for a large $S$ vev realizing \possessivecite{witteninverted}
idea of inverted hierarchy in a model with dynamical
supersymmetry breaking.

On the other hand, the exact superpotential of the theory
(\ref{ITexact}) is of an O'Raifeartaigh type. Thus it is natural
to ask whether there exists a region of the parameters of the
model such that near the origin of the moduli space
(where, in particular, the singlet vev is zero),
the strong coupling dynamics
decouples and the potential is calculable. Indeed, 
\cit{clp} have argued that for sufficiently small
coupling $\lambda$, and 
$\tilde S\ll \Lam_2/\lam$, the contributions of the strong dynamics to
the scalar potential are small compared with the contributions of
the light particles of  mass $\lam \vev{S}$. The latter
contribution is calculable, and it was found in \cite{clp} that
there exists a minimum of the potential at
$\tilde S =0$. Moreover, it was argued that the calculability
breaks down only when the Yukawa coupling $\lam$ has
non-perturbative strength.
Finally, another minimum of the potential
may exist at $\tilde S \sim {\cal{O}}(\Lam_2 /\lam)$, however, this
possibility can not be verified at present, since the strong
coupling dynamics is important in this region.

We should also mention 
several obvious but useful generalizations of the ITIY model.
Consider an $SP(N)$ gauge group with $N+1$ flavors of matter
fields in the fundamental representation. This 
theory has an $SU(2N+2)$ flavor symmetry. When the quarks are coupled
to gauge singlet fields transforming in the antisymmetric
representation of the flavor symmetry group, supersymmetry is
broken in exactly the same way as in the $SU(2)$ model. A slightly
more complicated generalization is based on an $SU(N)$ gauge
group with $N_f=N$ flavors. In this case the baryonic operators
$B$ and $\overline B$ are required to parameterize the quantum moduli
space. Therefore, the superpotential~(\ref{ITtree}) will not be
sufficient for \susy breaking. In particular there will be a
supersymmetric solution $M_{ij}=0$, $B\overline B=\Lambda_N^{2N}$.
Supersymmetry is broken if two additional fields, $X$ and
$\overline X$ with superpotential couplings
$\lam_1\overline X B +\lam_2 X \overline B$ are added to the
superpotential. To enforce this structure of the superpotential
one can gauge baryon number. We should note that in the case of the
$SU(N)$ models, the renormalization group argument we used to show
that the potential grows at large singlet vev's is not applicable to
the $X$ and $\overline X$ directions. Other models with quantum
modified moduli spaces can also break supersymmetry when each
invariant appearing in the constraint is coupled to a gauge
singlet \cit{css2}.

\bigskip
\noindent
{\it The 3--2 Model Revisited}

Before closing this section we would like to reanalyze, following
\cit{IT}, 
the familiar 3--2 model of \citename{adsdsb} discussed in 
section~\ref{threetwo} in a different limit,
$\Lam_2\gg\Lam_3$. We shall see that the description of  \susy
breaking is quite different in this limit. First, note that
from the point of view of the $SU(2)$ gauge group we have the matter
content of the ITIY model, namely four $SU(2)$ doublets (three $Q$'s
and $L$) and six singlets ($\bar u$ and $\bar d$).
The superpotential couplings of the 3--2 models are not sufficient
to lift all classical flat directions, and in addition to the
``singlets'' there is an $SU(2)$ meson which can acquire a vev. (Of
course all these flat directions, the including ``singlet'' ones, are
lifted by $SU(3)$ D-terms as we learned in  
section~\ref{threetwo}.) Let us parameterize the $SU(2)$ mesons by
$M_{ij}=Q_iQ_j$, $M_{i4}=Q_iL$, where the summation over $SU(2)$
indices is suppressed. In these variables the superpotential of
the model is
\beq
W={\cal A}(\Pf M - \Lam_2^4) + \lam \bar d_i M_{i4}\ ,
\eeq
where ${\cal A}$ is a Lagrange multiplier. Extremizing the
superpotential with respect to $\bar d$ we find that the scalar
potential contains terms
\beq
\label{32terms}
V=\sum_i^3 \abs{M_{i4}}^2+\ldots \ .
\eeq
By an $SU(3)$ rotation we can set $M_{14}=M_{24}=0$.  Thus
supersymmetry is restored if it is possible that
$M_{34}=\epsilon^2 \ra 0$. 
In turn this requirement and the quantum
constraint (\ref{pfaffian}) mean that
\beq
\label{falserun}
M_{12}=\frac{\Lam_2^4}{\epsilon^2} \ra \infty.
\eeq
At large expectation values the quantum moduli space approaches
the classical one. Thus equation (\ref{falserun}) can only be
satisfied if the model possesses classical flat directions. But as we
already know, when the $SU(3)$ D-flatness conditions are imposed, the
model does not have flat directions. Therefore  supersymmetry
must be broken. The natural expectation values of the (canonically
normalized) fields at
the minimum of the potential are of order ${\cal O}(\Lam_2)$,
therefore the quantum corrections to the \kahler
potential are significant and one can only estimate  the vacuum energy
in this limit, $V \sim \abs{\lam^2 \Lam^4}$.

\subsection{Quantum Removal of Flat Directions}
\label{flat}

In the previous section we encountered the ITIY model which 
breaks supersymmetry even though it has a classical flat direction.
Quantum mechanically, the potential becomes non-zero and flat
(up to logarithmic corrections) far along this flat direction.
It is in fact possible for quantum effects to completely ``lift''
classical flat directions, generating a growing potential
along these directions.
Thus it is possible for theories with classical flat directions
to break supersymmetry, with a stable, supersymmetry-breaking minimum.
We shall now build upon the insights gained in our analysis of of SQCD
of the previous section, to develop a method for determining
when classical flat directions are lifted quantum mechanically.
We shall also discuss some examples in which this happens.

As will become clear from our discussion, a crucial
requirement for the quantum removal of flat directions
is that some gauge dynamics becomes strong along the
flat direction. In many models, the opposite happens,
that is, the gauge group is completely broken along
the flat direction, and the dynamics becomes weaker
as $S$ increases. However, it may be the case that
along the flat direction, some gauge group remains unbroken, 
and fields charged under it obtain masses proportional
to $S$. Then the dynamics associated with this gauge
group becomes strong, and may lift the flat direction.

We should stress that in this section we shall be asking two
separate questions. First, we shall ask if  quantum
effects can stabilize the potential 
along a given flat direction.
It is most convenient to answer this question by finding a set
of degrees of freedom which give a weakly coupled
description of the theory in the region of interest on the
moduli space. However, if the quantum stabilization of the
potential indeed happens, the vacuum may well lie in the
genuinely strongly coupled region. Thus an affirmative
answer to the first question is not sufficient to give an
affirmative answer to our second question, whether
supersymmetry is broken in the model. To answer this second
question we shall need to consider the properties of the
exact superpotential in the strong coupling region.

Following~\cit{shirmanflat},
consider a model with classical
flat directions. Assume for simplicity that 
there is a single modulus $S$. In the approximation of a
canonical \kahler potential, the scalar potential of the
model can be written as
\beq
V=V_r + V_S = \sum \abs{\frac{\partial W}{\partial
\phi_i}}^2 + \abs{\frac{\partial W}{\partial S}}^2.
\eeq
The applicability of this scalar potential is restricted by
the assumption 
that the \kahler potential is canonical.
However, for large
enough $S$ vevs the description of the
physics often simplifies, and in fact, it may be possible
to find a description where the theory (or a sector
of the theory) is weakly coupled. In such a limit, it is
convenient to analyze the theory in two steps. First, one
considers a ``reduced'' theory with the scalar potential
$V_r$, where $S$ plays a role of the fixed
parameter. One then studies the behavior of the scalar potential
of the ``reduced'' theory as a function of $S$ as well as
contributions of $V_S$. Let us consider various
possibilities.

\bigskip
\noindent
{\it 1. A SUSY-breaking reduced theory}

Suppose that the potential $V_r$ in the 
reduced theory along the flat direction
is non-zero, so that the reduced theory breaks \susy.
If $V_r$ is an increasing function of $S$,
it is clear that the flat direction is stabilized. 
Typically, $V_r \ra 0$ as $S\ra 0$, but even if $V_r$
tends to a nonvanishing constant one can not conclude at
this stage that supersymmetry is broken 
in the full theory.
This is because the theory is typically in a strong coupling
regime near the origin of the moduli space and therefore the
assumption of a
canonical \kahler potential as well as  the separation of
the scalar potential into the sum of two terms is not
justified. 
An example of a model  with such  behavior is an 
$SU(4)\times SU(3) \times U(1)$ model of  \cit{crs}, see
Section \ref{discarded}. 
Classically there 
is a flat direction along
which the gauge group is broken down to $SU(4)\times U(1)$
and the matter is the same as in the $4-1$ model discussed
in Section \ref{fourone}. The strong coupling scale of the
effective $SU(4)$ gauge group grows with the modulus, and
the flat direction is stabilized \cite{shirmanflat}. Additional 
analysis performed by \cit{crs} shows that there is no
supersymmetric vacuum at the origin, thus allowing them to
conclude that SUSY is broken. This mechanism of quantum
removal of classical flat directions is quite generic for
discarded generator models (see Section \ref{discarded}). 

Another possibility is that $V_r$ 
is a decreasing function of the
modulus leading to a run-away behavior at moderate values of
$S$. In this case one should include in the analysis
contributions from $V_S$. Since the stable vacuum (if it
exists at all) will be found at large values of $S$, the
separation of the scalar potential into two terms is well
justified. We can therefore conclude that as long as $V_S$
stabilizes the flat direction,  SUSY is broken. If, however, $V_S
\ra 0$ as $S \ra \infty$ the theory does not have a stable
vacuum. A very basic example of such behavior is
the antisymmetric tensor models discussed in
Section \ref{antisym-general}. In these models the effective theory along
the classical flat direction is SUSY breaking $SU(5)$ with the
scale vanishing at the boundary of the moduli space. The
theory does not have a stable vacuum. Introducing a tree level
superpotential 
lifts all classical flat directions, stabilizes the
potential and breaks supersymmetry. Note that there is no
weak coupling description anywhere on the moduli space of
the model. This means that the separation of the scalar
potential into $V_r$ and $V_S$ is not 
apriori
justified. However,
$V_r$ arises from non-perturbative effects in the
\kahler potential while $V_S$ arises from the tree level
superpotential. As a
result there is no interference effects between $V_r$ and
$V_S$ and we can separate the potential into two positive
definite terms.

\bigskip
\noindent
{\it 2. A Supersymmetric Reduced Theory}

Now we would like to consider models where $V_r=0$ has
solutions for all values of $S$ (or for a set of moduli). In
these cases we have to analyze the behavior of $V_S$ subject to
the condition that $V_r=0$ is satisfied. It is instructive to
consider as examples two classes of analogous models with
the gauge groups $SP(N/2)\times SU(N-1)$~\cite{IT} and
$SU(N) \times SU(N-1)$~\cite{PST1}. 

We begin with the model of~\cit{IT}.
The matter content is $Q\sim (N, N-1)$, $L\sim (N,1)$,
$R_a\sim(1,\overline {N-1})$, with the
tree level superpotential:
\beq
\label{sptree}
W_{tree} = \lam\, Q\, L\, {\bar R}_2 +
{1 \over M}\,
\sum_{a,b>2}^{N} \lam_{ab}\, Q^2 \,\bar R_a\, \bar R_b ~.
\eeq
This superpotential leaves classical flat directions
associated with the $SU(N - 1)$ antibaryons
$\bar b^a = (\bar Q^{N - 1})^a = v^{N - 1}$
(we shall denote the $R$ vevs by $v$).
The exact superpotential was found in~\cite{IT},
and was used to show that there is no supersymmetric
vacuum in the finite region of moduli space.
Here we shall confine ourselves to discussing the
physics along the classical baryonic flat directions.

Without loss of generality, we can consider
the flat direction $S\equiv {\bar b}^1 =  v^{N - 1} $.
Along this direction, $SU(N-1)$ is completely broken.
On the other hand, all $SP$ flavors get masses proportional
to $v$ through the tree-level superpotential, so that
one is left with a pure $SP(N/2)$ which gets stronger
for larger $S$. Gaugino condensation in this group then
produces the superpotential
\beq
W_S \sim S^{{2\over N+2}} \ ,
\eeq
leading to the potential
\beq
\label{sprunawaylow}
V_S = \abs {\partial W \over \partial S}^2
\sim S^{-{2N \over N+2}} ~.
\eeq
(Note that here $V_r=0$.) Actually, one can obtain this result
starting from the exact superpotential.
However, at  scales $S \gg \Lam_1$ the relevant degrees of freedom are
the elementary ones, so we should consider the behavior of the potential
in terms of $v$,
\beq
\label{larges}
V_S \sim  v^{ 2 {N-4 \over N+2}} ~.
\eeq
We see that for 
$N > 4$ it increases along the classical flat direction.
Thus the classical flat direction is stabilized
quantum mechanically. The analysis of the theory in the finite
region of the field space \cite{IT} shows that SUSY is broken.

We would like to compare these results with the
behavior of the model of \cit{PST1} 
based on an $SU(N) \times SU(N-1)$
gauge group with matter in the fundamental
representations:
$Q\sim (N, N-1)$, $\bar L_i\sim ({\bar N}, 1)$,
and $\bar R_a\ (1, \overline {N - 1})$,
where $i=1\ldots N - 1$, and $a=1\ldots N$.
The tree level superpotential is given by:
\beq
\label{sutree}
W_{tree} = \sum_{ia}\, \lambda_{ia}\, 
Q\, \bar L_i\, \bar R_a + \alpha_a\, \bar b^a\ ,
\eeq
where $\bar b^a = ({\bar R}^{N-1})^a$ is an antibaryon of $SU(N-1)$. 
This superpotential lifts all flat directions as long as the coupling
are chosen so that~\cite{PST1} $\lam_{ia}$ has maximal rank
and
\beq
\label{PSTcouplings}
\lam_{ia}\,\alpha_a \ne 0\ .
\eeq
Since we are interested 
in understanding the physics along the flat directions we shall
set $\alpha_a=0$. Then there are classical
flat directions parameterized by the $SU(N-1)$ antibaryons, in
complete analogy with the $SU(N-1)\times SP(N/2)$ model discussed
above.
Again, along the direction $S\equiv {\bar b}^N$,
$SU(N-1)$ is broken, and all flavors of $SU(N)$
obtain mass. $SU(N)$ gaugino condensation generates
the potential
\beq
\label{surunawaylow}
V_S \sim S^{- 2 {N-1 \over N}} 
\eeq
Again, this potential can also be obtained from the exact
superpotential of the theory, which was obtained in~\cite{PST1}.
In terms of the vev of the elementary field $\bar R$
we then have 
$$
V_S\sim v^{{-2\over N}} \ .
$$
Unlike in the case of the Intriligator-Thomas model
the runaway behavior persists and the model does not
have a stable vacuum state.
Turning on $\alpha_a$ according to (\ref{PSTcouplings}),
all flat directions are lifted. This, together with
the analysis of~\cite{PST1}, which shows that there
is no supersymmetric minimum in the finite region of moduli space, 
allows one to conclude that supersymmetry is broken.

As we mentioned in the beginning of this subsection,
the key ingredient in quantum lifting of flat directions
is that the dynamics of some gauge group becomes
strong along the flat direction.
This in fact happens in both the $SU\times SP$ and the $SU\times SU$
examples we saw above,
as, along the relevant flat direction, one group factor remains unbroken 
and fields charged under it obtain masses.
However, the numerical factors
are such that the potential grows along the flat direction
in the first example, and slopes to zero in the second.

While general criteria for the determination of the
quantum behavior along classical flat directions do not exist, we
have illustrated several techniques which are useful for the analysis. 
We should also stress that we have
concentrated on the simplest examples with a single modulus. In
more general situations it is not sufficient to perform an analysis
for each flat direction separately, assuming that other moduli are
stabilized. One should do a complete analysis allowing all moduli
to obtain independent vevs consistent with $D$ and $F$-flatness
conditions. In particular one should study the moduli which
do not appear in the tree level superpotential.

\subsection{Supersymmetry Breaking with No R-symmetry}
\label{noR}

In section~\ref{global} we discussed the relation between SUSY breaking
and R-symmetries. We saw that theories with a spontaneously
broken R-symmetry
and no flat directions break SUSY. We also saw that
if R-breaking terms are added to the superpotential,
SUSY is typically restored. We emphasized that both
these statements assume that the superpotential is
generic, that is, all terms allowed by the symmetries
appear. 

In this section we shall encounter a theory with the
most general renormalizable superpotential allowed by symmetries,
which breaks supersymmetry even though it does not have
an R-symmetry. Furthermore, unlike the theory of section~\ref{isssection},
it does not possess an effective R-symmetry in the low-energy
description. As we shall see, the reason supersymmetry is
broken is that the dynamical superpotential is not generic.

The model we describe is an $SU(4)\times SU(3)$ gauge
theory studied by~\cite{PST1},
which is the first in a series of $SU(N)\times SU(N-1)$ models
that we already discussed from a different
perspective in Section~\ref{flat}. Here we only state some
of the results. The matter content is $Q\sim(4,3)$,
$\bar L_i\sim({\bar 4},1)$, and $\bar R_a\sim(1,{\bar 3})$
with $i=1\ldots 3$, $a=1\ldots 4$.
One can add the classical superpotential
\beq
\label{fourthreew}
W = \lam\, \sum_{i=1}^3 Q\cdot \bar L_i \cdot \bar R_i\ +\ 
\lam^\prime\, Q\cdot \bar L_1 \cdot \bar R_4\ +\ 
\alpha {({\bar R}^3)}^1 \ +\ \beta {({\bar R}^3)}^4 \ ,
\eeq
with appropriate contractions of the gauge indices
(in particular, ${({\bar R}^3)}^a$ stands for the $SU(3)$
``baryon'' with the field $\bar R_a$ omitted).
This superpotential does not preserve any R-symmetry.
It is the most general renormalizable superpotential
preserving an $SU(2)$ global
symmetry that rotates $\bar L_2$, $\bar L_3$ together with
$\bar R_2$, $\bar R_3$. In addition, it lifts all the classical
flat directions of the model. If we add nonrenormalizable
terms to this superpotential, (supersymmetric) minima will
appear at Planckian field strength. These extra minima will
not destabilize the non-supersymmetric minima we shall be
discussing.~\footnote{Other models in this class (with $N>4$) are
non-renormalizable, and we do not have a reason to neglect
non-renormalizable operators. For $N\le 6$ the
generalization of~(\ref{fourthreew}) still gives the most
general superpotential up to operators whose dimension is
smaller or equal to the dimension of the baryon. Therefore,
the expected SUSY breaking minimum is still a stable local
minimum. In models with $N>6$, the most general
superpotential with no R-symmetry and 
operators whose dimension does not
exceed the dimension of the baryon operator will generically
preserve supersymmetry.}

As was shown in~\cite{PST1}, the theory breaks supersymmetry.
This can be established by carefully analyzing the
low energy theory. In the limit that the $SU(3)$ dynamics
is stronger, $SU(3)$ confines, giving a low
energy theory in which $SU(4)$ has four flavors.
After $SU(4)$ confines one has an O'Raifeartaigh-like
theory, with the fields
$Y_{ia}=Q\cdot \bar L_i\cdot \bar R_a$, ${\bar b}^a = (\bar R^3)^a$,
${\bar B}= Q^3 L^3$, $P_a = Q^3 (Q\cdot \bar R_a)$,
and $B=\det(Q\cdot \bar R)$ (the last two vanish classically).
Taking into account the dynamically generated superpotential we
find that the full superpotential is given by,
\beq
\label{ftfull}
W = {P_a {\bar b}^a -B \over \Lambda_3^5}\ +\
A\, \left(P\cdot Y^3 - {\bar B}\, B - \Lambda_4^9\, \Lambda_3^5
\right)
\ +\ 
\lam\, \sum_{i=1}^3 Y_{ii}\ +\ 
\lam^\prime\, Y_{14}\ +\ 
\alpha\, \bar b^1 \ +\ \beta\, \bar b^4 \ ,
\eeq
where $A$ is a Lagrange multiplier, and $\Lambda_4$, $\Lambda_3$,
are the scales of $SU(4)$, $SU(3)$ respectively.
This superpotential does not preserve any effective
R-symmetry in terms of the variables of the low-energy theory.
Still, as was shown in~\cite{PST1}, supersymmetry is
broken.
The crucial point is that the Lagrange multiplier $A$
only appears linearly in~(\ref{ftfull}). 
If the superpotential contained terms with higher 
powers of $A$, supersymmetry would be restored.
Note that the superpotential~(\ref{ftfull}) is 
reminiscent of the superpotential~(\ref{orafw}) of the
simplest O'Raifeartaigh model, with $A$ playing the role of $\phi_1$.
In the absence of an R-symmetry, one cannot rule out the
presence of higher powers of $\phi_1$ in~(\ref{orafw}),
whereas in the dynamically generated superpotential~(\ref{ftfull}),
$A$  only appears linearly.

Other examples have been found which break supersymmetry
even though the microscopic theory does not have an
R-symmetry. These include, among others, the 4--3--1 model of~\cite{unity},
and the 4--1 model with the superpotential term~$M \Pf A$~\cite{PT1},
as well as, as we mentioned already, the ISS model~\cite{ISS}.
In most of these examples, either the tree-level superpotential
is not generic, or there is an effective R-symmetry in the
low energy theory.

\section{DSB Models and Model Building Tools} 
\label{known}

So far we have discussed several important models which
illustrate the main methods and subtleties in the analysis of DSB.
Many more models (in fact many infinite classes of models) have
been constructed in recent years. The methods of analysis we have
described can be used for these models. In fact often,
not only the method of analysis but the dynamics itself is analogous
to one or the other models discussed in previous sections. Thus
it is not practical to present a detailed investigation of every
known model of DSB.

On the other hand, in many cases the dynamics is not well understood
beyond the conclusion that SUSY must be broken, and further
investigation of the dynamics as well as the connection between
different mechanisms and models of SUSY breaking
may lead to better understanding of the general
conditions for DSB. Therefore, in this section we shall give a
list of known models briefly discussing how SUSY is broken. We
shall emphasize the  relations between various models, and
give a partial classification. In addition we shall
introduce 
a useful model building method which can be used
to construct new models.

We shall also discuss in this section supersymmetry breaking
in theories with anomalous $U(1)$'s which give an example
of dynamical SUSY breaking through the Fayet-Iliopolous mechanism. 

\subsection{Discarded Generator Models} \label{discarded}

Let us recall the observation that both the 4--1 and the 3--2 models have
a gauge group which is a subgroup of $SU(5)$, while the matter
content (after adding  $E^+$ in the 3--2 model) falls into
antisymmetric tensor and antifundamental representations --
exactly as needed for DSB in $SU(5)$.
Based on this observation, \cit{dnns} proposed the following method
of constructing new DSB models. Take a known model of \dsb without
classical flat directions
and discard some of the group generators. This reduces the number
of D-flatness conditions, and therefore, leads to the appearance of
flat directions. On the other hand, the most general
tree level superpotential allowed by the smaller symmetry may
lift all the moduli. It
is also possible that a unique non-perturbative superpotential will
be allowed in such a ``reduced'' model. This construction is guaranteed to
yield anomaly free chiral models which often possess a non-anomalous
R-symmetry, and thus are good candidates for dynamical
supersymmetry breaking. If a model constructed using this
prescription breaks supersymmetry it is often calculable since
for small superpotential couplings it typically possesses almost
flat directions along which the effective description may be
weakly coupled.

In fact, the 3--2 and 4--1 models are the simplest examples of two
infinite classes of DSB models based on $SU(2N-1)\times
SU(2)\times U(1)$ \cite{dnns} and on $SU(2N)\times U(1)$
\cite{dnns,PT1} gauge groups which can be constructed
by using the discarded generator method.

To construct these theories one starts with an $SU(2N+1)$
theory with matter transforming as an antisymmetric tensor, $A$, and
$2N-3$ anti-fundamentals, $\overline F$.
Then one requires gauge invariance under, for example,
an $SU(2N-1)\times SU(2)\times (1)$ subgroup, with the $U(1)$ generator
being
\beq
T={
\rm diag}(2,\ldots,2, -(2N-1),-(2N-1)).
\eeq
Under this group the matter fields decompose as
\beq
\begin{array}{l}
A (\overline{\Yasymm}, 1, 4), ~F (\Yfund, 2, 3-2N),~
S(1,1,2-4N),\\
\overline{F}^a (\overline{\Yfund}, 1, -2)~, 
\phi^a (1, 2, 2N-1), ~~a=1,\ldots, 2N-3~.
\end{array}
\eeq
The most general superpotential consistent with the symmetries is
\beq
W=\gamma_{ab} A \overline F^a \overline F^b + \eta_{ab} S \phi^a
\phi^b + \lambda_a F \overline F^a \phi^a.
\eeq
This superpotential lifts all classical flat directions. Models of
this class have non-anomalous R-symmetry and supersymmetry is
broken. It is
interesting to observe that the coupling $\eta_{ab}$ in the
superpotential above could be set to zero without restoring 
supersymmetry. While for $\eta=0$ classical flat directions
appear, they are lifted by quantum effects. 

The construction of the $SU(2N)\times U(1)$ DSB series is quite
analogous. The mater fields in this class of models are
\beq
A_2,~~ F_{1-n},~~\overline F^a_{-1},~~ S^a_n,~~a=1,\ldots, 2N-3~,
\eeq
where subscripts denote $U(1)$ charges, and superscripts are
flavor indices. With the most general superpotential allowed by
symmetries the models break \susy.

Clearly one can consider many other subgroups of $SU(2N+1)$,
and in fact, several other classes of
broken generator models were constructed:
$SU(2N-2)\times SU(3)\times U(1)$ models \cite{crs,chou},
$SU(2N-3)\times SU(4)\times U(1)$
and $SU(2N-4)\times SU(5)\times U(1)$ models
\cite{clrs}. While these models are similar by construction to
those we discussed above, the supersymmetry breaking dynamics is
quite different, and various models in this class can have
confinement, dual descriptions and quantum removal of
classical flat directions. Since we have already considered
the simplest and illuminating examples of these phenomena in
DSB models, we shall not give a detailed discussion of all
possible discarded generator models. We shall restrict
ourselves to the mention of the $SP(2)\times U(1)$ model by \cit{css}.
This model is interesting because it is an example of the
discarded generator model in which the
rank of the gauge group is reduced compared to the ``parent''
theory.  The matter fields in this model are 
\beq
\begin{array}{c|cr}
&SP(2)  & U(1)\\
\hline
A &\Yasymm       & 2\\[-5pt]
Q_1 & \Yfund & -3\\[-5pt]
Q_2 & \Yfund& -1\\[-5pt]
S_1 & 1      & 2\\[-5pt]
S_2 & 1      & 4
\end{array}
\eeq
Non-renormalizable couplings 
are required to lift all flat directions. The full superpotential
is
\beq
W=\frac{\Lam^7(Q_1Q_2)}{2(A)^2(Q_1Q_2)^2 - (Q_1AQ_2)^2} +
Q_1Q_2S_2 + Q_1AQ_2S_1~,
\eeq
where the first term is generated dynamically.

The existence of a general method for constructing discarded
generator models suggests that there may exist a unified
description of these models.
In fact, \cit{unity} found exactly such a description.
It is based on the antisymmetric tensor models supplemented
by a chiral field $\Sigma$ in the adjoint representation of
the gauge group. We are interested in finding an effective
description of the discarded generator models, or more
generally of the models with $U(1)^{k-1}\times
\prod_{i=1}^k SU(n_i)$ gauge groups (where $\sum_{i=1}^k
n_i = 2N+1$) and the light matter given by decomposing 
the antisymmetric tensor and antifundamentals of $SU(2N+1)$
under the unbroken gauge group. The adjoint $\Sigma$ needs to be
heavy in such a vacuum. This can be achieved by
introducing the superpotential for the adjoint 
\beq
W_{\Sigma} = \sum_{i=2}^{k+1} \frac{s_i}{i} \Tr \Sigma^i \ .
\eeq
We shall be most interested in the case $k=2$. 
For generic coefficients $s_i$ there are several discrete
vacua where $\Sigma$ is heavy and the model contains matter
in desired representations. Note that in the most symmetric
vacuum $\Sigma=0$, the low energy physics is described by
the antisymmetric tensor model, and SUSY is broken.
For supersymmetry to be broken in other vacua one needs to
lift the classical flat directions associated with the light
fields which requires the following
tree level superpotential
\beq
W= \frac{1}{2} m \Sigma^2 + \frac{1}{3} s_3 \Tr \Sigma^3 +
\lam_1^{ij} \Fbar_i A \Fbar_j + 
\lam_2^{ij} \Fbar_i A \Sigma \Fbar_j +
\lam_3^{ij} \Fbar_i \Sigma A \Sigma \Fbar_j \ .
\eeq
This superpotential is chosen so that in each vacuum of
interest it exactly reproduces the superpotential needed for
supersymmetry breaking. \cit{unity} showed that in the
full model supersymmetry is broken for any value of the
adjoint mass including $m=0$. This latter conclusion at
first seems quite unusual, since the one loop beta function
coefficient of the model is $b_0= 2N+4$. In the $SU(2N+1)$
model this might suggest that at least in the absence of the
superpotential the theory is in a non-Abelian
Coulomb phase. However, the analysis of \cit{unity} showed that
indeed the superpotential is quite relevant and, when certain
requirements on the Yukawa couplings are satisfied, SUSY is
broken.
A similar construction with $k>2$ (that is models with more
than two non-Abelian factors and/or more than one abelian
factor in the gauge group) was shown \cite{unity} not
to break supersymmetry.

\subsection{Supersymmetry Breaking from an Anomalous $U(1)$}

Theories with an anomalous $U(1)$ provide a
simple mechanism for supersymmetry 
breaking~\cite{binetruydudas,dvalipomarol}.
Such theories contain a Fayet-Iliopoulos (FI) term,
so supersymmetry can be broken just as in
the FI model we discussed in Section~\ref{fi}. 
In fact, the anomalous $U(1)$ 
theories discussed below are the only known examples where
the Fayet-Iliopoulos mechanism of supersymmetry breaking can
be realized dynamically. 
In the absence of any superpotential, at least one
field with appropriate $U(1)$ charge develops
a vev to cancel the FI term. One can then introduce
an (effective) superpotential mass term for this field
so that some $F$-term and the D-term cannot
vanish simultaneously and supersymmetry is broken.

In our discussion of the FI model in Section~\ref{fi}
we simply put in a tree-level FI term by hand.
It is well known that a $U(1)$ D-term can be renormalized at
one loop \cite{dtermnr,wittendsb}.
Such renormalization is proportional to the sum of the
charges of the matter fields, and therefore, vanishes unless
the theory is anomalous.
Indeed,
in many string models, the low-energy 
field theory contains an anomalous $U(1)$, 
whose anomalies are canceled by shifts of the 
dilaton-axion superfield, through the 
Green-Schwarz mechanism~\cite{greenschwarz}.
A FI term is generated for
this $U(1)$ by string loops. As far as the low-energy 
field theory is concerned, we can treat this FI
term as if it was put in by hand. The only subtleties
associated with supersymmetry breaking involve
the dilaton superfield.

Consider a theory with an anomalous $U(1)$ gauge symmetry
with 
\beq
\label{dgs}
\delta_{GS} = {1\over 192 \pi^2} \sum_i q_i \ ,
\eeq
where $q_i$ denote the $U(1)$ charges of the different
fields of the theory.
The dilaton superfield $S$ then transforms as
\beq
\label{stransf}
S \ra S + i\, {\delta_{GS}\over 2}\, \alpha\ ,
\eeq
under the $U(1)$ transformation 
$A_\mu\ra A_\mu +\partial_\mu\alpha$,
where $A_\mu$ is the $U(1)$ vector boson.
To be gauge invariant, the dilaton \kahler potential
is then of the form
\beq
\label{skahler}
K = K(S+S^* -\delta_{GS} V) \ ,
\eeq
where $V$ is the $U(1)$ vector superfield.
This then gives the FI term
\beq
\label{fiterm}
\xi^2 = -{\delta_{GS}\over 2}\, {\partial K\over \partial S}  \ .
\eeq

Following~\cite{ahdm} we shall consider the
model of~\cite{binetruydudas}, which has, in addition
to the anomalous $U(1)$, an $SU(N)$ gauge symmetry.
The model contains the field $\phi$, an $SU(N)$
singlet with $U(1)$ charge $-1$ (assuming $\delta_{GS}>0$),
and one flavor of $SU(N)$, that is, fields $Q$
and $\bar Q$ transforming as
$(N,q)$ and $(\bar N, \bar q)$ under $SU(N)\times U(1)$.
Working in terms of the $SU(N)$ meson $M=Q\bar Q$,
the superpotential is given by
\beq
\label{wbd}
W = m M \, \left({\phi\over M_P}\right)^{q+ \bar q} +
(N-1)\left({\Lam^{3N-1}\over M}\right)^{1\over N-1} \ ,
\eeq
where the first term is a tree-level term, and the
second term is generated dynamically by $SU(N)$ instantons.
The potential will also contain contributions from
the $U(1)$ D-term, which is given by,
\beq
\label{dbd}
D = - g^2 \left((q+\bar q) \abs{M} -\abs{\phi}^2
+ \xi^2 \right)\ ,
\eeq
where $g$ is the $U(1)$ gauge coupling.
Minimizing the potential, one finds that supersymmetry
is broken. $\phi$ wants to develop a vev to cancel $\xi^2$,
but because of the $SU(N)$ dynamics, the meson
$t$ develops a vev, which then generates, through the
first term in~(\ref{wbd}), a mass term for $\phi$,
so that the potential does not vanish.

It is important to recall though that the $SU(N)$ scale
depends on the dilaton superfield. This dependence
is most easily fixed by requiring that the second term 
in~(\ref{wbd}) is $U(1)$ invariant, giving,
\beq
\label{scaledep}
\Lam^{3N-1} = M_P^{3N-1}\, 
e^{ {-2(q+\bar q)S\over \delta_{GS}}} \ .
\eeq
One can then minimize the potential in terms of $t$, $\phi$
and the dilaton, $S$. 
At the minimum, the D-term as well as the $t$, $\phi$,
and dilaton $F$-terms are non-zero.

Note that if the dilaton superfield is neglected
in the above analysis, the theory seems to have
no Goldstino, as the gaugino and matter fermions
obtain masses either by the Higgs mechanism
or through the superpotential.
In fact, as was shown in~\cite{ahdm}, the Goldstino
in these theories is a combination of the gaugino,
the matter fermion and the dilatino. In this
basis the Goldstino wave-function is given by
\beq
\left({D\over\sqrt{2} g}, F_i, 
\sqrt{{\partial^2 K\over \partial S^2}}\, F_S \right)\ ,
\eeq  
where $D$, $F_i$, and $F_S$ stand for the D-term,
the $i$-th matter field $F$-term, and the dilaton $F$ term
at the minimum, respectively.

\subsection{List of Models and Literature Guide}
\label{modellist}

Finally, in this section we shall present an extensive (but
certainly incomplete) list of models known to break
supersymmetry with references to the original papers where these
models were introduced. While some of the models discussed
below have been studied in  great detail, frequently it is
only known that a given model breaks supersymmetry, but the low
energy spectrum and the properties of the vacuum have not been
studied. 
In addition to the examples presented below
many other models appeared in the literature. New
supersymmetry breaking theories can
be constructed from the known models in a variety of ways.
Moreover, for phenomenological purposes it is often
sufficient to find a model with a local non-supersymmetric
minimum. While establishing the existence of a local
non-supersymmetric minimum may sometimes be more difficult than
establishing the absence of any supersymmetric vacuum, the
methods involved in the analysis are essentially the same,
and we shall not discuss such models here.

$\bullet$ $SU(5)$ with an antisymmetric tensor and an
antifundamental~\cite{adsdsb,mv}. The
arguments of~\cit{adsdsb} were based on the difficulty of
satisfying 't Hooft anomaly matching conditions, while the
\cit{mv} argument was based on gaugino
condensation. The model is not calculable.
\cit{murnoncalc} and \cit{PT1} have used the method
of integrating in and out vector-like matter to give
additional arguments for DSB in the model. \cit{pouliot}
constructed a dual of the
$SU(5)$ model and showed that the dual breaks SUSY at tree
level. For the
discussion of the model in the present review see Sections
\ref{examples} and \ref{inandout}.

$\bullet$ 
$SU(2N+1)$ with an antisymmetric tensor $A$,
$2N-3$ antifundamentals $\Fbar$ and the superpotential
\beq
W=\lam_{ij} A \Fbar_i \Fbar_j \ ,
\eeq 
where $\lam$ has the maximal rank~\cite{adsdsb}. 
These are generalizations of the $SU(5)$ model. 
The integrating in and out
method was used by \cit{PT1} to further analyze these models. See
Section \ref{examples} and also Section
\ref{discarded}.

$\bullet$ $SO(10)$ with a single
matter multiplet in the spinor 
representation~($\bf 16$ 
of $SO(10)$)~\cite{adssoten}. The analysis of supersymmetry
breaking in this model is very similar to that of the non-calculable
$SU(5)$ model. Indeed, the $SU(5)$ model may be constructed from the
$SO(10)$ model by using the discarded generator
method. \cit{murnoncalc} discussed  DSB in this model in the
presence of an extra field in the vector $\bf 10$ 
representation of $SO(10)$. For a small mass of the extra
field, the theory is calculable, and 
assuming no phase transition,
SUSY remains broken when the vector is integrated out. \cit{PS}
considered the same theory by adding an arbitrary number $N>5$ of
vector fields and constructing the dual $SU(N-5)$
theory. They showed that the dual breaks SUSY when masses
for the vectors are turned on. All these arguments can only be
used as additional evidence of DSB in the $SO(10)$ model, but do
not allow one to analyze the vacuum and low energy spectrum of
the theory.

$\bullet$ The two generation $SU(5)$ model~\cite{mv,adstwogen}: 
$SU(5)$ with two
antisymmetric tensors and two antifundamentals, with the
superpotential 
\beq
W = \lam A_1 \Fbar_1 \Fbar_2 \ .
\eeq
\cit{adstwogen} showed that the model is calculable. In
fact, historically,
this is the
first calculable model with supersymmetry breaking
driven by an instanton-induced superpotential. 
The vacuum
and the low energy spectrum of the model were analyzed in
detail by \cit{veld1}. \cit{veld2} also analyzed
generalizations of this model which include extra
vector-like matter with a mass term in the superpotential.

$\bullet$ The $3-2$ model~\cite{adsdsb}: $SU(3)\times SU(2)$ with 
\beq
Q~(3,2)~, ~~~~~~\bar u~ (\bar 3,1)~, ~~~~~~\bar d (\bar
3,1)~, ~~~~~~L~(1,2) \ ,
\eeq
with the superpotential
\beq
W = \lam\, Q\, L\, \overline{d} \ .
\eeq
This model is calculable.
The analysis of the vacuum and low energy spectrum
can be found in
\cite{adsdsb,bpr}. The model possesses a global $U(1)$
symmetry which can be gauged without restoring SUSY, the
relevant details of the vacuum structure in this case
can be found in~\cite{dns}. See Sections~\ref{threetwo} and~\ref{ITIY}.

$\bullet$ Discarded generator models: These include
$SU(n_1)\times SU(n_2) \times U(1)$ and $SU(2N)\times (1)$ 
subgroup of $SU(2N+1)$
(with $n_1+n_2=2N+1$) with matter given by the decomposition
of the antisymmetric tensor and $2N-3$ antifundamentals
of $SU(2N+1)$ under the appropriate gauge group. This
construction was proposed in~\cite{dnns}. For details
see Section~\ref{discarded}. The two smallest
models in this class are the $3-2$ model (Section
\ref{threetwo}) and the $4-1$ model (Section
\ref{fourone}). The $SU(2N)\times U(1)$ models were first
constructed in~\cite{dnns,PT1}; the $SU(2N-1)\times SU(2)\times
U(1)$ models can be found in~\cite{dnns}; the $SU(2N-2)\times
SU(3) \times U(1)$ models are considered in~\footnote{See
also \cite{chou}.} \cite{crs};
finally, the $SU(2N-3)\times SU(4)\times U(1)$ models are
discussed in \cite{clrs}. A unified description of this
class of models, as well as of the non-calculable $SU(2N+1)$ models,
is given in~\cite{unity}.
Another example of the models in this class is
$SP(2) \times U(1)$ model of \cit{css}.

$\bullet$ 
$SU(2N+1)\times SU(2)$~\cite{dnns} with
\beq
Q \sim (\Yfund, \Yfund),~~~~
L \sim (1, \Yfund),~~~~
Q_i \sim (\Ybfund, 1), ~~~i=1,2,
\eeq
and with a superpotential similar to that of the $3-2$ model.
These models are obvious generalizations of the $3-2$ model.
The dynamics in this class of models is very similar to that of the $3-2$
model. For detailed analysis see~\cite{IT2}.
The low energy physics of the $SU(5)\times SU(2)$ model 
in this class, in the limit of a strong $SU(2)$, is described by
the non-calculable $SU(5)$ model~\cite{IT2} to which we paid so much
attention in this review.

$\bullet$ $SU(7) \times SP(1)$ and $SU(9)\times
SP(2)$~\cite{IT2}:
These models are obtained by dualizing 
the $SU(7)\times SU(2)$ and $SU(9)\times SU(2)$ models
of the previous paragraph.
The matter content is 
\begin{eqnarray}
A~ (\Yasymm,
1) \ , ~~~~F~ (\Yfund,1) \ , ~~~~\overline{P}~
(\Ybfund,\Yfund) \ , \\ \nonumber
L~(1,\Yfund) \ , ~~~~\overline{U}~(\Ybfund,1) \ ,
~~~~\overline{D}~(\Ybfund,1) \ ,
\end{eqnarray}
 and the superpotential
\beq
W= A \overline{P} \overline{P} + F \overline{P}L \ .
\eeq
Note that these models can be constructed starting from
the antisymmetric tensor models of \cit{adsdsb},
by gauging a maximal
global symmetry and adding matter to cancel all
anomalies with the most general superpotential.

$\bullet$ 
$SU(2N+1) \times SU(2)$ with
\beq
A \sim (A, 1),~~~~ F \sim (\Yfund, 2),~~~~
\Fbar_i \sim (\Ybfund, 1), ~~~~ D\sim (1,2),
\eeq
where $i=1,\ldots, N-2$.~\cite{dnns} 
To lift all  flat directions,
a non-renormalizable superpotential is required
\beq
W = \sum_{i,j=1}^{2N-2} \gamma_{ij}A\Fbar_i\Fbar_j
+\lam \Fbar_{2N-1} F D + \frac{1}{M} \sum_{i,j=1}^{2N-2}
\alpha_{ij} \Fbar_i\Fbar_j FF \ .
\eeq

$\bullet$ 
$SU(2N+1)\times SP(M)$,  $N \ge M-1$ with
\beq
\label{suspseriesmatter}
Q \sim(\Yfund,\Yfund), ~~~~\Qbar_i\sim
(\Ybfund,1),~~~~L\sim(1,\Yfund) \,
\eeq
where $i=1,\ldots,2M$ is a flavor index~\cite{dnns}. 
These are generalization of the 3--2 model with a non-renormalizable
superpotential.

The $SU(2N+1)$ dynamics generates a dynamical
superpotential~\cite{dnns}. In addition, a quantum
constraint is generated by the $SP(M)$ dynamics for $N=M$. 
The tree level superpotential
\beq
\label{suspseriesW}
W=\lam \Qbar_{2}QL + \sum_{i,j>2}^{2M} \gamma_{ij} Q^2
\Qbar_j \Qbar_j 
\eeq
lifts all flat directions, and supersymmetry is broken. For
details see~\cite{dnns,IT}. 

For $N=M+1$, the tree level superpotential~(\ref{suspseriesW}) does not
lift all classical flat directions, yet they are lifted by
nonperturbative effects~\cite{IT,shirmanflat} and SUSY is
broken. We discussed this model in Section~\ref{flat}.
(Note that in that section, we used a different
notation for $SP$ and refered to this theory as $SU(N-1)\times
SP(N/2)$.)

It is also useful to  note
that for $M+1 < N$, the $SP(M)$ dynamics can have dual
description. \cit{IT2} argued that the dual description (with
$SU(2N+1)\times SP(N-M-1)$ gauge group and matter content
which includes the symmetric tensor of $SU(2N+1)$ as well as
(anti)fundamentals and bifundamentals breaks SUSY. When
$N>3M+2$ it is  only the dual description which is asymptotically
free and can be interpreted as a microscopic theory.

An interesting modification of these models~\cite{LT} is an 
$SU(2N+1)\times SP(N+1)$ theory with 
\beq
Q \sim(\Yfund,\Yfund), ~~~~\Qbar_i\sim
(\Ybfund,1),~~~~L_a\sim(1,\Yfund) \ ,
\eeq
where $i=1,\ldots,2(N+1)$ and $a=1,\ldots, 2N+1$.
We note that this version of the model
possesses an $SU(2N+1)\times U(1) \times U(1)_R$ global
symmetry. \cit{LT} considered only the renormalizable
superpotential
\beq
W =  \lam \Qbar Q L \ ,
\eeq
where $\lam$ has maximal rank. They showed that while this
model has a large number of classical flat directions, all
of them are lifted quantum mechanically. 
One can now add mass terms for some flavors of the $SP(N+1)$
fields $L_a$. For an appropriately chosen mass matrix
supersymmetry remains broken.
Choosing a mass matrix of maximal rank and integrating out
the massive matter, we recover the non-renormalizable
model discussed above.

$\bullet$ Non-renormalizable $SU(2N)\times U(1)$ model~\cite{dnns} 
with chiral superfields transforming under the
gauge group and a global $SU(2N-4)$ symmetry as
\beq
A\sim (\Yasymm, 2N-4, 1), ~~~~
\Fbar \sim(\Ybfund, -(2N-2), \Ybfund),~~~~
S \sim (1,2N, \Ysymm) \ .
\eeq
The superpotential required to stabilize all flat directions
\beq
W = A\Fbar\Fbar S \ ,
\eeq
explicitly breaks the global symmetry down to a subgroup. 
The anomaly free $SU(N-2)$ subgroup of the global symmetry
can be gauged without restoring SUSY.

$\bullet$ Non-renormalizable
$SU(N)\times U(1)$ models~\cite{dnns}: The matter content is
(we also give charges under a maximal global
$SU(N-3)$ symmetry)
\begin{eqnarray}
A \sim (\Yasymm, 2-N,1), ~~~~N\sim(M,1,1),~~~~ 
\overline{N}_i \sim (\Ybfund, N-1, \Ybfund),\\ \nonumber
S_i \sim(1,-N,\Yfund)),~~~~ S_{ij}\sim(1,-N,\Yasymm) \ ,
\end{eqnarray}
where $i,j=1,\ldots,N-3$. The superpotential 
\beq
W =\lam_i \overline{N}_i N S_i + \gamma_{ij} A
\overline{N}_i\overline{N}_j S_{ij}
\eeq
lifts all flat directions while preserving a global symmetry.
Note that for $N=4$, $S_{ij}$ does not exist and this 
is just the 4--1 model of Section~\ref{fourone}.

$\bullet$ $SU(N)\times SU(N-1)$~\cite{PST1},
and $SU(N)\times SU(N-2)$~\cite{PST2}:
The $SU(N)\times SU(N-1)$ models were
discussed in Section~\ref{flat}. We discussed SUSY breaking without
R-symmetry in these models in Section~\ref{noR}.
The $SU(N)\times SU(N-2)$ models, which
are similar by construction but have very different dynamics
are discussed in Section \ref{dual}. Both classes
of models have calculable minima with an unbroken global
symmetry 
($SU(N-2)$ and $SP(N-3)$ respectively)~\cite{poptrisigma,amm,shadmi}.

$\bullet$ ITIY models~\cite{IT,IY} and their modifications: 
We discussed the ITIY models in section~\ref{ITIY}. 
They are based on an $SU(N)$ ($SP(N)$) gauge
group with $N$ ($N+1$) flavors of matter in the fundamental
representation coupled to a set of gauge singlet fields in
such a way that all D-flat directions are lifted. Even after
supersymmetry breaking these models posses a flat direction
which is only lifted by (perturbative) corrections to the
\kahler potential. In
\cite{shirmanflat,MA-H,DDGR} it was argued that the
perturbative corrections generate a growing potential for
large vevs along this direction. By gauging a subgroup of
the global symmetry it is possible to obtain a modification of
the model with a calculable local SUSY breaking minimum at
large vevs \cite{murinvert,DDGR}.
These
models can be generalized in the following way. Take any
model with a quantum modified constraint and couple all gauge
invariant operators to singlet fields. Since the quantum
constraint becomes incompatible with the singlet F-term conditions, 
supersymmetry must be broken~\cite{css2}. As an 
example consider an  $SO(7)$ gauge group with five
matter multiplets transforming in the spinor
representation. The theory possesses an $SU(5)$ global
symmetry, and a $U(1)_R$ under which all matter fields are
neutral. The gauge invariant composites transform as
an antisymmetric tensor $A$ and antifundamental $\Fbar$ of the
global symmetry. The quantum constraint is
\beq
A^5 + A\Fbar^4 = \Lam^{10} \ .
\eeq
Coupling all gauge invariants to gauge singlets $\tilde A$
and $F$ and implementing the constraint through the Lagrange
multiplier $\lam$ one finds
\beq
W= \tilde A A + F \Fbar + \lam(A^5 + A \Fbar^4 - \Lambda^{10}) \ ,
\eeq
and obviously SUSY is broken. We note that the model is
nonrenormalizable.

$\bullet$ $SU(2)$ with one $I=3/2$ matter field \cite{ISS}:
We discussed this theory in section~\ref{isssection}.
A non-renormalizable tree-level superpotential
lifts all classical flat directions. The theory
confines, no superpotential is generated dynamically,
and supersymmetry is broken since the tree-level superpotential
cannot be extremized in terms of the confined field.

$\bullet$ 
$SU(7)$  with two
symmetric tensors, $S_a$\, , $a=1,2$, six
antifundamentals $\Qbar_i$, $i=1,\ldots,6$,  and the
tree level superpotential 
\beq
W=\sum_{i}^3 S^1 \Qbar_{2i} \Qbar_{2i-1} + S^2
\Qbar_{2i}\Qbar_{2i+1} \ ,
\eeq
where in summing over $i$ we identify $7 \sim
1$~\cite{css2,NT}. 
This is another example of SUSY breaking through
confinement, which we saw in the \cit{ISS} model in
Section~\ref{isssection}. 
The superpotential lifts all classical flat directions
while preserving a global anomaly free $U(1)\times U(1)_R$
symmetry.
$SU(7)$ dynamics leads to confinement,
and generates the nonperturbative superpotential 
\beq
W_{dyn}= \frac{1}{\Lam^{13}} H^2 N^2 \ ,
\eeq
where $H^a_{ij} = S^a \Qbar_i \Qbar_j$ and
$N_i = S^4 \Qbar_i$.
Near  the origin of the moduli space 
the \kahler potential is canonical in terms
of the composite fields. Solving the equations of motion for
$H^a_{ij}$ one finds that at least some of the composite 
fields acquire vevs, breaking the global symmetry, and therefore,
supersymmetry.

$\bullet$ $SO(12)\times U(1)$ and $SU(6)\times U(1)$~\cite{css2}. 
The matter content of the $SO(12)\times U(1)$ model is
$(32,1)$, $(12, -4)$, $1,8)$, $(1,2)$, $(1,6)$.
The matter content of the $SU(6)\times U(1)$   model is
$(20,1)$, $(6, -3)$, $({\bar 6},-3)$, $(1,4)$, $(1,2)$.
These models are constructed 
by starting with a non-chiral theory with a dynamical
superpotential and gauging a global $U(1)$ symmetry (adding
the necessary fields to make the full theory anomaly free) in a way
which makes the theory chiral. Supersymmetry is broken by
the interplay between a dynamically-generated superpotential
and the tree-level superpotential.

\section*{Acknowledgements}

It is a pleasure to thank Michael Dine, Galit Eyal,
Jonathan Feng, Martin Gremm, and Yossi Nir for
useful comments. 
Y.~Shirman is grateful to the Aspen Center for Physics for hospitality
during the early stages of this work. The work of Y.~Shirman was
supported in part by NSF grant PHY-9802484.
The work of Y.~Shadmi was supported in part by
DOE grants \#DF-FC02-94ER40818 and \#DE-FC02-91ER40671,
and by the Koret Foundation.

\appendix \section{Some Results on SUSY Gauge Theories}

In this appendix we shall briefly review some results in 
supersymmetric gauge theories. Our main goal here is to
introduce notations (which will mainly follow those of
\cite{WB}), and to summarize the results necessary to make the
present review self-contained. Much more detailed reviews of
the progress in our understanding of  supersymmetric gauge
theories exist in the literature, \eg 
\cite{ISreview,peskinreview,shifmanreview}.

\subsection{Notations and Superspace Lagrangian}
\label{notations}

We shall consider an effective low-energy theory of light
degrees of freedom well above the possible scale of 
supersymmetry breaking. In this case the effective action
will have linearly realized supersymmetry, and
it is convenient to write an effective supersymmetric
Lagrangian in  ${\cal N}=1$ superspace, where four (bosonic) space-time
coordinates are supplemented by four anti-commuting
(fermionic) coordinates $\theta_\alpha$, and $\overline
\theta^{\dot \alpha}$, $\alpha=1,2$. The light matter fields
combine into chiral superfields\footnote{
We shall usually 
use the same notation for a chiral superfield and its lowest
scalar component.}
\beq
\label{chiral}
\Phi=\phi +\sqrt{2} \theta \psi + \Theta^2 F,
\eeq
while gauge bosons and their superpartners combine into
vector superfields
\beq
\label{vector}
V=-\theta \sigma^\mu \overline \theta A_\mu+
i  \theta^2 \overline \theta \bar \lam - 
i \overline \theta^2 \theta \lam +
\frac{1}{2} \theta^2 \overline \theta^2 D,
\eeq
where we have used Wess-Zumino gauge.

The effective supersymmetric Lagrangian for a theory with
gauge group $G$, and matter fields $\Phi_i$ transforming in
the representation $r$ of the gauge group (with $T^a$ being a
generator of this representation)  can be written as
\beq
\label{lagrangian}
{\cal L}= \int d^4 \theta K(\Phi^\dagger,e^{V\cdot T}\Phi) + 
\frac{1}{g^2} \int d^2 \theta {\cal W}^\alpha {\cal
W_\alpha} + {\rm h.c} + \int d^2 \theta W(\Phi) +{\rm h. c.}
\eeq
The first term in (\ref{lagrangian}) is a \kahler
potential which contains, among others, kinetic terms for the
matter fields.  
The \kahler potential also contributes gauge-interaction terms
to the scalar potential.
The second term in (\ref{lagrangian}) is
the kinetic term for the gauge fields.
In particular 
${\cal W}^\alpha=- \frac{1}{4}\overline {\cal D} \overline
{\cal D}
{\cal D}_\alpha V$, where $\cal D$ is a superspace derivative, 
is a supersymmetric generalization of the gauge field-strength
$F^{\mu\nu}$. 
The last term in (\ref{lagrangian}) is the superpotential. 

The superpotential is a holomorphic function of chiral
superfields and obeys powerful non-renormalization
theorems. In particular, in perturbation theory the
superpotential can only be modified by field rescalings
(which can be absorbed into renormalization of the \kahler
potential). Using holomorphy, symmetries of the theory, and
known weakly coupled limits it is often possible to
determine the superpotential exactly, including all
non-perturbative effects. Similarly, the kinetic term for
the gauge multiplet is a holomorphic function allowing one
to obtain exact results on the renormalization of the gauge
coupling. 

The \kahler potential on the other hand can be a general
real-valued function of $\Phi^\dagger$ and $\Phi$ consistent
with symmetries. Classically it is given by
\beq
K = \Phi^\dagger e^{V\cdot T} \Phi,
\eeq
but quantum mechanically it is renormalized both
perturbatively and non-perturbatively. 

In studying the dynamical behavior of the supersymmetric
theory it is often useful to remember 
that
the Hamiltonian 
is determined by the supersymmetry generators
\beq
\label{appendH}
H= \frac{1}{4} (\Qbar_1 Q_1 +Q_1 \Qbar_1 +\Qbar_2 Q_2 +
Q_2 \Qbar_2 ) \ .
\eeq
>From Eq.~(\ref{appendH}) we see that a supersymmetric vacuum
state (a state annihilated by the supersymmetry charges) has vanishing
energy.
Therefore, a particularly important role (especially in the analysis of
supersymmetry breaking) is played by the scalar
potential of the theory
\beq 
\label{susypotential}
V= \frac{1}{2} g^2 \sum_a \left( D^a \right )^2 +
F^\dagger_i g_{ij}^{-1} F_j,
\eeq
where $g_{ij}=\frac{\partial^2 K}{\partial \Phi^{\dagger i}
\partial \Phi^j}$, and the auxiliary fields $F$, and $D$ are
given 
in terms of the scalar fields by
\beq
\begin{array}{l}
F_i =\frac{\partial}{\partial \Phi_i} W\\
D^a = \sum_i \Phi^{\dagger i} t^a \Phi^i \ ,
\end{array}
\eeq
where in $F_i$ one takes the derivatives of the superpotential
with respect to the different superfields and then keeps
only the lowest component, and in $D^a$, $\Phi^i$ stands for the
scalar field of the $\Phi^i$ supermultiplet.

Typically a supersymmetric gauge theory possesses a set of
directions in field space (called D-flat directions) along which
$D^a=0$ for all $a$. 
Along some or all of these D-flat directions, the $F$-flatness
conditions, $F_i=0$, can also be satisfied. The subspace of
field space where the scalar potential vanishes is called a
moduli space and to a large degree determines the low-energy
dynamics.

The study of non-perturbative effects in SUSY gauge
theories relies heavily on the use of symmetries. An
important role, especially in the applications to dynamical
supersymmetry breaking, is played by an ``R-symmetry''. We
therefore pause to introduce this symmetry. 
Under an R-symmetry, the fermionic coordinates rotate as,
\beq
\theta \ra e^{i\alpha} \theta \, .
\eeq
A chiral field with R-charge $q$ transforms under this
symmetry as follows
\beq
\Phi(x, \theta, \bar \theta) \ra e^{-iq\alpha} 
\Phi(x, e^{i\alpha} \theta, e^{-iq\alpha} \bar \theta) \, .
\eeq
Note that different component fields transform differently
under R-symmetry, and thus it does not commute with
supersymmetry.
On the other hand, the vector superfield $V$ is neutral under
R-symmetry (therefore, the gaugino transforms as $\lambda \ra
e^{-i\alpha} \lambda$). 
Clearly there always exists an assignment of R-symmetry
charges to the superfields such that the \kahler potential
contribution to the action is invariant under
R-symmetry. 
On the other hand the superpotential
contributions to the action explicitly break R-symmetry
unless the superpotential has charge $2$ under R-symmetry.

In the
following subsections we shall discuss methods for 
determining the classical moduli space. We shall also describe
the quantum behavior of supersymmetric QCD (SQCD) with
various choices of the
matter content.  At the end of this section
we shall comment on analogous result for models with
different gauge groups and matter content. These results will
provide us with tools needed to analyze supersymmetry breaking.

\subsection{D-flat Directions}
\label{Dflatsection}

Classically, one could set all superpotential couplings to
zero. Then the moduli space of the theory is determined
by D-flatness conditions. Even when tree level
superpotential couplings are turned on but remain small, the
vacuum states of the theory will lie near the solutions of
D-flatness conditions (still in the classical
approximation). 
It is convenient, therefore, to
analyze SUSY gauge theories in two stages. First find a
submanifold in the field space on which the D-terms vanish, and
then analyze the full theory including both tree level and
non-perturbative contributions to the superpotential.

We start by describing a useful technique for finding
the D-flat directions of a theory \cite{adssoten,adsdsb}
with $SU(N)$ gauge symmetry.
Consider the $N\times N$ matrix
\beq
\label{Dmatrix}
D^i_j= \phi^{\dagger l} \left(A^i_j\right)^k_l \phi_k,
\eeq
where $\left(A^i_j\right)^k_l$ are the real generators  of
$GL(N)$. For 
$\phi$ in the fundamental representation 
$\left(A^i_j\right)^k_l=\delta^i_l \delta^k_j$ (the
generalization of $\left(A^i_j\right)$ 
for a general multi-index representation is obvious).
It is easy to see that the vanishing of all $D^a$'s is
equivalent to the requirement that $D^i_j$
be proportional to the unit matrix, $D^i_j \sim \delta^i_j$.
To show this it is sufficient to note that $D^a=D^i_j
\lam^{aj}_i$, where $\lam^a$ are generators of $SU(N)$ in
the fundamental representation.

Another way to parameterize the moduli space is by the use of
gauge invariant composite operators. It has been shown that
a complete set of such operators is in one to one
correspondence with the space of D-flat 
directions~\cite{lutytaylor}. 
An important feature
of this latter parameterization of the moduli space is that
in some cases there exist gauge invariant operators which
vanish identically due to the Bose statistic of the
superfields. For this reason they do not have counterparts
in the ``elementary'' parameterization of the moduli
space. However, due to quantum effects these operators 
typically describe light (composite) degrees of freedom of
the low-energy theory and play an important role in the
dynamics.

\subsection{Pure Supersymmetric $\sun$ Theory}
\label{psun}

The Lagrangian of a pure supersymmetric Yang-Mills (SYM) theory can be
written as
\beq
\label{SYM}
{\cal L} = \frac{1}{4 g_W^2} \int d^2 \theta {\cal W^\alpha}
{\cal W_\alpha} + {\rm h.~c.} \ .
\eeq
The Wilsonian coupling constant in the Lagrangian can be
promoted to a vev of the background chiral superfield 
$\frac{1}{g_W^2} \ra S=\frac{1}{g_W^2}-i\frac{\Theta}{8 \pi^2}$.
Since the physics is independent of
shifts in $\Theta$, the
Wilsonian gauge coupling in Eq. (\ref{SYM}) receives
corrections only at one loop. 
On the other hand, the gauge
coupling constant in the 1PI action receives contributions at
all orders in perturbation theory. These two coupling constants can
be related by field redefinitions \cite{SV1,SV2}. In the
following we shall always use the Wilsonian action and work
with Wilsonian coupling constants. We shall use functional
knowledge of the exact beta functions only to establish
the scaling dimensions of the composite operators in our
discussion of duality in Section \ref{duality}.

SYM is a strongly interacting non-abelian theory
very much like QCD. In particular it is believed that it
confines and develops a mass gap. By using symmetry arguments
it is possible to show that if the gaugino condensate
develops it has the form
\beq
\label{condensate}
\vev{\lam \lam} = {\rm const} \times \Lam_{SYM}^3 = {\rm
const} \times \mu^3 e^{-\frac{8 \pi^2}{N_c g^2}} \ .
\eeq
In fact the constant can be exactly calculated~\cite{NSVZ,SV3}.
The theory has $N_c$ supersymmetric vacuum states.

\subsection{$N_f<N_c$: Affleck-Dine-Seiberg Superpotential}
\label{adsW}

As a next step one can consider an $\sun$ gauge theory 
with $N_f$ ($<N_c$) flavors of matter fields in the fundamental
$Q$ and antifundamental $\Qbar$ representations. 
This theory possesses a large non-anomalous 
global symmetry under which matter fields
transform as follows: 
\beq
\label{sunsymmetries}
\begin{array}{cccccccccc}  
&SU(N_f)_L & \times & SU(N_f)_R&\times&U(1)_B&\times&U(1)_R 
\\
Q&  N_f & &1&&1&&{N_f-N_c \over N_f} \\
\Qbar& 1 & &\overline N_f&&-1&&{N_f-N_c \over N_f} \\
\end{array}
\eeq
Classically there are D-flat directions
along which the scalar potential vanishes. Using the techniques
described above we can parametrize these flat directions 
(up to symmetry transformations) by
\beq
\label{Dflat}
 Q=\pmatrix{v_1 &&& \cr & v_2
&& \cr 
&&\ldots&\cr
&&&v_{N_f }\cr
\dots & \dots & \dots & \dots \cr} = \Qbar \ .
\eeq

These flat directions can also be parameterized by the vev's of the gauge
invariant operators $M_{ij} = Q_i \Qbar_j$. 
These composite degrees of freedom 
give a better (weakly coupled) description near the origin
of moduli space where the theory is in a confined regime.

In this model a unique nonperturbative superpotential
is allowed by the symmetries \cite{adsqcd}
\beq
\label{sqcd}
W_{dyn} =\left ({\Lam^{3 N_c -N_f} \over \det (Q \Qbar)}\right)
^{1\over N_c -N_f} \ ,
\eeq
where $\Lam$ is the renormalization group invariant scale
of the theory. It was shown \cite{adsqcd,cordes} 
that this superpotential is in fact
generated by  instanton effects
for $N_f=N_c-1$. 
It is generated by gaugino condensation in
all other cases.

Before proving this last statement, let us pause for a moment
and discuss the relation between the renormalization group
invariant scales of the microscopic and effective theories.
Suppose the microscopic theory is $SU(N_c)$ with $N_f$ flavors.
As has been mentioned above 
the {\it Wilsonian} coupling 
\cite{SV1,SV2}
of the theory runs only at one loop
\beq
\label{effectivefullg}
{1\over g^2(\mu)} = 
{1 \over g^2(M)} + 
{b_0 \over 16 \pi^2} \ln \left( {\mu\over M}\right) \ .
\eeq
Suppose also that at a scale $v$ some fields in the theory
become massive, and the physics below this scale is described
by an $SU(N_c^\prime)$ gauge group with $N_f^\prime$ flavors.
The Wilsonian coupling of the effective theory is
\beq
\label{effectivelowg}
{1\over g_L^2(\mu)} = 
{1 \over g_L^2(M)} + {\tilde b_0 \over 16 \pi^2} 
\ln \left( {\mu\over M}\right) \ .
\eeq
In equations (\ref{effectivefullg}) and (\ref{effectivelowg})
$b_0=3N_c -N_f$ and 
$\tilde b_0 = 3N_c^\prime -N_f^\prime$
are the $\beta$-function coefficients. But the couplings should be equal
at the scale $\mu = v$. This allows us to derive scale matching conditions.
For example, take an $SU(N_c)$ theory with $N_f$ flavors. Its renormalization
group invariant scale is given by
\beq
\label{examplescalehigh}
\Lam^{3 N_c - N_f} = \mu^{3 N_c -N_f} \exp\left(-{8 \pi^2
\over g^2}\right) \ .
\eeq
If one of the
matter fields is massive with mass $m \gg \Lam$,
the effective theory has $N_f-1$ flavors and its scale is
\beq
\label{examplescalelow}
\Lam_L^{3 N_c - N_f+1} = 
\mu^{3 N_c -N_f+1} \exp\left(-{8 \pi^2 \over g^2}\right) \ . 
\eeq
Requiring equality of couplings at the scale of the mass
we find
\beq
\label{examplematching}
\Lam_L^{3N_c-N_f+1} = m \Lam^{3N_c -N_f} \ .
\eeq
In a general case the equation above becomes
\beq
\label{scalematching}
\Lam_L^{\tilde b_0} = v^{\tilde b_0 - b_0} \Lam^{b_0} \ ,
\eeq
where $v$ represents a generic vev and/or mass in the theory.

Now, in the $N_f=N_c-1$ theory with small masses $m \ll \Lam$
an instanton calculation \cite{adsqcd,cordes} is reliable and 
gives (\ref{sqcd}). Due to the
holomorphicity of the superpotential the result can be extrapolated
into the region of moduli space where one flavor, say $N_f$'th, 
is heavy, $m_{N_fN_f} \gg \Lam$. It decouples from the low energy 
effective theory. Solving the equations of motion for the heavy field
and using the scale matching condition (\ref{scalematching})
one finds the superpotential (\ref{sqcd}) for the effective theory
with $N_f=N_c-2$ flavors. In the low energy effective theory
this superpotential can be interpreted as 
arising from gaugino condensation.
One can continue this procedure by induction
and not only derive the superpotential for arbitrary
$N_f<N_c$ but also fix the numerical coefficient in front of the
superpotential. (This coefficient can be absorbed into the definition of 
$\Lam$, and we shall set it to $1$ most of the time.)

Even though the classical flat directions are lifted in the massless theory
by the superpotential (\ref{sqcd}), 
the scalar potential 
\beq
\label{sqcdscalar}
V = \sum_i \left(\abs{\partial W \over \partial Q_i}^2 + 
\abs{\partial W \over \partial \Qbar}^2\right)
\eeq
tends to zero as 
$Q=\Qbar \ra \infty$ and as a result theory does not possess 
a stable vacuum state. In our discussion of DSB we discuss
examples where flat directions
may be not only lifted but stabilized 
due to nonperturbative effects (see Section \ref{flat}).

One could lift classical flat directions 
by adding a mass term to the superpotential
\beq
\label{sqcdtree}W_{tree} = m_{ij} Q_i \Qbar_j \ .
\eeq
Note that this superpotential explicitly breaks the $U(1)_R$ symmetry.
In Section \ref{global} we argue that 
this is often a signal of unbroken supersymmetry.
It is easy to find the supersymmetric vacua in this model. In terms of
the meson fields they are given by
\beq
\label{sqcdsolution}
M_{ij} =\left( \det(m) \Lam^{3N_c - N_f} \right)^{1/N_c}
\left({1 \over m_{ij}}\right) \ .
\eeq
It is also worth noting that if some number of matter 
fields are  massive they decouple
from the low energy theory and can be integrated out. Solving the equations 
of motion for the massive fields one can find the 
superpotential of Eq.~(\ref{sqcd}) and the solution
(\ref{sqcdsolution}) for the vev's 
of 
the remaining light fields.

\subsection{$N_f=N_c:$ Quantum Moduli Space}

Additional flat directions exist for $N_f=N_c$. 
The most general expression for the flat directions is
\beq
\label{baryonflat}
Q=\pmatrix{a_1 &&&& \dots \cr & a_2
&&& \dots \cr 
&&\dots&&\dots\cr
&&& a_{N_c} & \dots \cr} ~~~~~
 \Qbar = \pmatrix{b_1 &&&& \dots \cr  & b_2
&&& \dots \cr 
&&\dots&& \dots \cr
&&& b_{N_c} & \dots \cr} 
\eeq
subject to the condition
\beq
\label{condition}
\abs{a_i}^2-\abs{b_i}^2 = v^2 \ .
\eeq
As was mentioned above, the flat directions can be parameterized by vevs of 
gauge invariant polynomials. In this case new flat directions
can be represented by fields with the quantum 
numbers of baryons\footnote{Summation over both
color and flavor indices is implied.} $B = Q^N$
and antibaryons $\Bbar = \Qbar^N$.
Note, however, that due to the Bose statistics of the superfields, 
the gauge
invariant polynomials obey the constraint, classically,
\beq
\label{classicalconstraint}\det M - B \Bbar = 0 \ .
\eeq
\citename{seibergexact} (\citeyear{seibergexact,seibergduality}) showed
that this constraint is modified quantum mechanically,
\beq
\label{quantumconstraint}
\det(M) - B \Bbar = \Lam^{2N} \ .
\eeq
We refer the reader to 
\cite{seibergexact,ISreview,peskinreview,shifmanreview} for
a detailed explanation of this result.

It is often convenient to enforce this 
quantum mechanical constraint by introducing
a Lagrange multiplier term in the 
superpotential
\beq
\label{lagrange}
W=A(\det(M) - B \Bbar - \Lam^{2N}) + m_{ij} M_{ij} \ .
\eeq
Once again the validity of this
superpotential can be verified in the limit that some of the matter 
fields are heavy and decouple from the low energy theory. Integrating
them out leads to the superpotential (\ref{sqcd})
for the light matter. 

Naively, the \kahler potential of the $N_f=N_c$ theory is singular
at the origin. This corresponds to the fact that
at the origin, the full gauge group $SU(N_c)$ is
restored and additional degrees of freedom
become massless. This singular point,
however, does not belong to the quantum moduli space.
$SU(N_c)$ cannot be restored because of the constraint
(\ref{quantumconstraint}) and the \kahler potential in terms
of composite degrees of freedom is non-singular. 
In the infrared mesons and baryons represent a good description
of the theory. One of many non-trivial tests they pass is
't Hooft anomaly matching conditions~\cite{thooft}.
Far from the origin
the quantum moduli space is very close to classical one and
the elementary degrees of freedom should represent a good
(weakly coupled) description of the theory.

\subsection{$N_f=N_c+1$}
\label{nplusone}

In this case there are $N_f$ bary\-ons and antibary\-ons 
transforming under the global $SU(N_f)_L \times SU(N_f)_R$
as $(N_f, 1)$ and $(1, \bar N_f)$ respectively. 
Classically, the gauge invariants obey the constraints
\beq
\label{constraintn+1}
\begin{array}{l}
\det (M) - B_i M_{ij} \Bbar_j = 0 \ , \\
B_i M_{ij}=M_{ij}\Bbar_{j}=0 \ .\\
\end{array}
\eeq
These constraints are not modified quantum mechanically. 
One can easily see this by adding a tree level superpotential
$W_{tree}=\sum_{ij} m_{ij} M_{ij}$. 
Holomorphy guarantees that meson vevs are 
given by Eq. (\ref{sqcdsolution}).
Taking various limits of the mass matrix
one can see that the mesons $M_{ij}$ can have any values
on the moduli space. This can also be shown for the baryons.

In terms
of the elementary fields the \kahler potential is singular 
at the origin reflecting
the fact that $SU(N_c)$ is restored there and 
additional degrees of freedom become massless.
In terms of composite degrees of freedom the \kahler potential 
is regular, and they represent 
a suitable infrared description of the theory. 
As in the case $N_f=N_c$,
't~Hooft anomaly matching conditions are satisfied by the effective
description.
In this model, the constraints can be implemented by the superpotential
\beq
\label{sun+1super}
W={1 \over \Lam^{2N_c-1}}\left(B_i M_{ij} \Bbar_j - \det
M\right) \ .
\eeq
Adding mass for one flavor correctly leads to the $N_f=N_c$ model.

\subsection{$N_f>N_c+1:$ Dual Descriptions of the
Infrared Physics}
\label{duality}

We shall start from the case ${3 \over 2} N_c < N_f < 3 N_c$. 
This theory flows to an infrared fixed 
point~\cite{seibergduality}. 
\cit{seibergduality} suggested that 
in the vicinity of the infrared fixed point the theory
admits a dual, ``magnetic'', description with the same global 
symmetries but in terms of a 
theory with a different gauge group.
This theory is based on the gauge group $SU(N_f - N_c)$ with $N_f$ 
flavors of $q$ and $\qbar$ 
transforming as fundamentals and antifundamentals respectively, 
as well as 
gauge-singlet fields $M$, corresponding to the mesons of
the original (``electric'') theory. 
The global-symmetry charges are given by
\beq
\label{dualsymmetries}
\begin{array}{cccccccccc}  
&SU(N_f)_L & \times & SU(N_f)_R&\times&U(1)_B&\times&U(1)_R 
\\
q& \overline N_f&&1&&{N_c\over N_f-N_c}&&{N_c\over N_f}\\
\qbar& 1&&N_f&&-{N_c\over N_f-N_c}&&{N_c\over N_f}\\
M & N_f && \overline N_f&&  0 && 2{N_f-N_c \over N_c}\\
\end{array}
\eeq
The magnetic theory also flows to a fixed point. 
However, in the magnetic
theory a tree level superpotential is allowed by symmetries
\beq
\label{magneticW}
W = Mq\qbar.
\eeq
In the presence of this superpotential the theory flows
to a new fixed point which is identical to the fixed point
of the ``electric" theory.

At the fixed point the superconformal symmetry can be used to
understand the behavior of the theory.
For example the scaling dimensions of the gauge invariant operators are
known. The exact beta function for the coupling in 1PI
action (in the electric description) is
given by \cite{NSVZ,SV1,SV2}
\begin{eqnarray}
\label{exactbeta}
\beta(g) &=& {g^2\over 16 \pi^2} {3N_c - N_f+N_f \gamma(g^2)
\over 1 - N_c{g^2\over 8 \pi^2}} 
\\ \nonumber
\atop
\gamma(g) &=& - \frac{g^2}{16 \pi^2} \frac {N_c^2-1}{N_c} 
+ 
{\cal O}( g^4) \ .
\end{eqnarray}
At the zero of the $\beta$-function the anomalous dimension is 
$\gamma = -3 N_c/N_f + 1$, and one finds
\beq
\label{mesondimension}
D(Q\Qbar) = 2+\gamma = 3{N_f-N_c \over N_f}\ .
\eeq
The dimension of the baryon operators can be determined by
exploiting the R symmetry \cite{seibergduality}
\beq
\label{baryondimension}
D(B)=D(\Bbar)={3N_c(N_f-N_c)\over 2 N_f}\ .
\eeq
This allows one to determine the scaling of the \kahler potential
near the fixed point both in the electric and in the magnetic description
\begin{eqnarray}
\label{kahlerscaling}
K_e &\sim& (Q \Qbar)^{2N_f \over 3(N_f - N_c)} \ ,\\ \nonumber 
K_m &\sim& (q \qbar)^{2N_f \over 3 N_c}  \ .
\end{eqnarray}

Let us summarize the correspondence
between the electric 
and magnetic theories:
\beq
\label{mapping}
\begin{array}{rcl}
M_{ij}=Q_i\Qbar_j &\ra& M_{ij} \ , \\ 
W = m_{ij} M_{ij} &\ra& W = m_{ij} M_{ij} + 
M_{ij} q_i \qbar_j \ , \\ 
b, \bbar &\ra & B, \Bbar \ .\\
\end{array}
\eeq
By performing a second duality transformation one can verify that in fact
the magnetic meson is identified with the composite electric meson through
the equations of motion.

The scales of the electric and the magnetic theories are related by
\beq
\label{dualscale}
\Lam^{3N_c-N_f} \tilde \Lam^{3(N_f-N_c)-N_f} = 
(-1)^{N_f-N_c}\mu^{N_f} \ ,
\eeq
where the scale $\mu$ 
is needed to map the composite electric meson
$Q\Qbar$ into an elementary magnetic meson $M$. These fields
have the same dimension at the infrared fixed point, but different
dimensions in the ultraviolet.

If the number of flavors is $N_c+1 < N_f < {3 \over 2} N_c$ one could 
construct a dual  description in a similar way.
In that case only the electric description is asymptotically free and
makes sense in the ultraviolet.

\subsection{Other Models}
\label{othermodels}

There are numerous generalizations of the results presented
in previous subsections to theories with different gauge
groups and matter fields. Here we mention some of the
generalizations which will be useful for our discussion of
\susy breaking.

Results analogous to those for SQCD  can be found for $SP(N)$ 
theories with $N_f$ 
flavors 
of matter fields transforming
in the fundamental representation\footnote{In our notation
$SP(1)=SU(2)$, and $N_f$
flavors correspond to $2N_f$ fields.} \cite{sptheory}. 
The one loop $\beta$-function 
coefficient is given by
\beq
\label{SP}
b_0=3 (N_c+1) - N_f\ .
\eeq
This can be compared to the one loop $\beta$-function
of the $SU(N)$ models given in Eq. (\ref{effectivefullg})
In fact, one can find many of the results for $SP(N)$ theories by making
the substitution $N_c \ra N_c+1$ in the expressions for $SU(N)$ models
(and rewriting determinants as Pfaffians). 
The theory does not have baryons
for any number of flavors. The supersymmetric vacuum is given by
\beq
\label{SPsolution}
M_{ij}=Q_iQ_j=\left(Pf(m) \Lam^{3(N_c+1)}\right)^{1 \over
N_c+1}\left(1 \over m_{ij}\right).
\eeq
We can easily see that for $N_f=N_c+1$, the 
quantum constraint is different from the classical one,
\beq
\label{SPquantum}
Pf(M)=\Lam^{2(N_c+1)}\ .
\eeq
If the number of flavors is $N_f>N_c+2$ there is a dual description 
analogous to the one for the $SU(N)$ theories. 

We conclude this section by mentioning several other classes
of models which are useful in the study of the
supersymmetry breaking. \cit{PT1} studied quantum moduli
space and exact superpotentials in $SU(N)$ gauge
theories with matter in the antisymmetric tensor,
fundamental, and antifundamental  
representations. In the case without fundamental fields
\cit{pouliot} has constructed a dual by using 
\possessivecite{berkooz} deconfining trick.
In \cite{PST1} duality was studied in
the product group theories and it was shown that dual models
can be constructed by using single group duality.


\def\np#1#2{Nucl. Phys. {\bf B #1}, #2}
\def\pl#1#2{Phys. Lett. {\bf B #1}, #2}
\def\plb#1#2{Phys. Lett. {\bf B #1}, #2}
\def\prl#1#2{Phys. Rev. Lett. {\bf #1}, #2}
\def\physrev#1#2{Phys. Rev. {\bf D #1}, #2}
\def\prd#1#2{Phys. Rev. {\bf D #1}, #2}
\def\ap#1#2{Ann. Phys. {\bf #1}, #2}
\def\prep#1#2{Phys. Rep. {\bf #1}, #2}
\def\rmp#1#2{Rev. Mod. Phys. {\bf #1}, #2}
\def\cmp#1#2{Comm. Math. Phys. {\bf #1}, #2}
\def\cqg#1#2{Class. Quant. Grav. {\bf #1}, #2}
\def\mpl#1#2{Mod. Phys. Lett. {\bf A #1}, #2}
\def\ptphys#1#2{Prog. Theor. Phys. {\bf #1}, #2}
\def\prpnphys#1#2{Prog. Part. Nucl. Phys. {\bf #1}, #2}
\def\jhep#1#2{J. High Ener. Phys. {\bf #1}, #2}
\def\ijmp#1#2{Int. J. Mod. Phys. {\bf A #1}, #2}

\end{document}